\pdfoutput=1
\documentclass[a4paper,11pt]{article}

\usepackage{jheppub}

\usepackage{preamble}
\usepackage[nameinlink]{cleveref}
\usepackage{subcaption}
\usepackage{lineno}
\newcolumntype{d}[1]{D{.}{.}{#1}}

\title{Explainable Equivariant Neural Networks for Particle Physics: PELICAN}

\author[b]{Alexander Bogatskiy}
\author[a]{Timothy Hoffman}
\author[a,c]{David W. Miller}
\author[a]{Jan T. Offermann}
\author[a]{Xiaoyang Liu}
\affiliation[a]{Department of Physics, University of Chicago, Chicago, IL, U.S.A.}
\affiliation[b]{Center for Computational Mathematics, Flatiron Institute, New York, NY, U.S.A.}
\affiliation[c]{Enrico Fermi Institute, Chicago, IL, U.S.A.}

\emailAdd{abogatskiy@flatironinstitute.org, hoffmant@uchicago.edu, jano@uchicago.edu}

\abstract{PELICAN is a novel permutation equivariant and Lorentz invariant or covariant aggregator network designed to overcome common limitations found in architectures applied to particle physics problems. Compared to many approaches that use non-specialized architectures that neglect underlying physics principles and require very large numbers of parameters, PELICAN employs a fundamentally symmetry group-based architecture that demonstrates benefits in terms of reduced complexity, increased interpretability, and raw performance. We present a comprehensive study of the PELICAN algorithm architecture in the context of both tagging (classification) and reconstructing (regression) Lorentz-boosted top quarks, including the difficult task of specifically identifying and measuring the $W$-boson inside the dense environment of the Lorentz-boosted top-quark hadronic final state. We also extend the application of PELICAN to the tasks of identifying quark-initiated vs.~gluon-initiated jets, and a multi-class identification across five separate target categories of jets. When tested on the standard task of Lorentz-boosted top-quark tagging, PELICAN outperforms existing competitors with much lower model complexity and high sample efficiency. On the less common and more complex task of 4-momentum regression, PELICAN also outperforms hand-crafted, non-machine learning algorithms. We discuss the implications of symmetry-restricted architectures for the wider field of machine learning for physics.}

\begin{document}
\maketitle
\flushbottom

%========================
% INTRODUCTION
%========================
\section{Introduction}
\label{Introduction}

Identifying, reconstructing, and measuring the properties and dynamics of high-energy, short-distance particle phenomena are inherently inference tasks, since direct access to the fundamental processes is often impossible due to the time and length scales at which they occur. The suite of detection techniques, pattern recognition algorithms, and measurement approaches used to perform such tasks inevitably imposes constraints on both the nature of the information used as well as on the form and structure of the results. Such constraints play a crucial role in the context of jet substructure measurements, in which detailed analysis is performed on the long-distance features of Lorentz-boosted particle decays, parton showering, and radiation patterns found in the collimated sprays of particles that form the jets themselves. We present a comprehensive analysis of a new approach to multiple jet substructure-based inference tasks using a machine learning (ML) architecture that fundamentally  respects permutation and Lorentz-group symmetries: PELICAN, the permutation equivariant and Lorentz invariant or covariant aggregator network. Our approach imposes explicit physics-informed symmetry constraints on the architecture and consequently yields new insights and capabilities.

Decades of jet substructure research have yielded a wide range of approaches to performing inference tasks such as: distinguishing quark-initiated from gluon-initiated jets~\cite{gallicchio_quark_2013,Larkoski:2014pca,Komiske:2016rsd,10.3389/frai.2022.852970,10.21468/SciPostPhys.6.6.069}; discriminating jets formed from Lorentz-boosted top quarks, Higgs and $W$-bosons, from the continuum background of jets formed from light-quarks and gluons~\cite{Butterworth:2008iy,Kaplan:2008ie,KasiePlehn19,Thaler:2008ju}; dissecting and measuring the parton-shower structure of light-quark and gluon jets themselves~\cite{Soper:2012pb,Feige:2012vc,Chien:2014nsa,Marzani:2017kqd,Dreyer:2018nbf,kogler_advances_2021}. Many approaches have been adopted to perform these tasks, including the direct use of discriminating high-level observables and multi-variate methods~\cite{Thaler:2010tr,CMS:2011usu,EFP}, as well as a growing number of ML architectures using a variety of latent-space representations. For a comprehensive overview of jet substructure measurements and approaches, see refs.~\cite{Kogler:2018hem,Marzani:2019hun}, as well as ref.~\cite{hepmllivingreview} for a general review of ML methods in high-energy physics (including substructure measurements). As the model complexity has grown, so too have questions regarding the relationship of both the methods and the constraints that they impose on the fundamental physical processes that they are used to model. In particular, the use of observables, architectures, and latent space representations that adhere closely to the structure and dynamics of the physics processes under study have been found to provide not only enhanced performance, but also significant insights and improvements in interpreting the results~\cite{Butter:2017cot,Erdmann:2018shi,EFP}. Imbuing these models with knowledge of, or even fundamental respect for, the symmetries inherent to the system under study has thus become increasingly impactful in the study of jet substructure, especially in the context of ML models and various neural network (NN) architectures~\cite{EFN,Bogatskiy:2020tje,LorentzNet22}.

There are several common approaches to enforcing continuous symmetries in NNs. Data augmentation can be used to train a model to have a particular sparsity structure and become approximately symmetric. However, when model complexity and interpretability are of concern, as is the case in many particle physics applications, a different approach is helpful. Similar issues arise when using data preprocessing or normalization, which often come with inherent ambiguities and discontinuities that can be detrimental for more complex tasks. 

Traditionally, ML algorithms are evaluated based on basic performance metrics such as accuracy and computational cost. However, in contexts where the trained algorithms are treated not only as predictors or generators, but as actual models for physical processes -- which is especially true in scientific applications -- other metrics of model quality are valuable. Model complexity (e.g.~the number of parameters), explainability and interpretability are also important for using an ML algorithm as a viable physics model. Furthermore, certain problem-specific properties such as symmetries can be critical as well. Symmetries in ML are known to produce less complex models which respect basic geometrical rules and arguably provide more opportunities for interpretability and explainability (e.g.~convolutional neural network (CNN) kernels are often interpreted as visual features). Even in realistic settings where the symmetries are merely approximate, symmetry-constrained architectures often outperform more general architectures in terms of pure accuracy (see e.g. \secref{toptagging}), but even in cases when that is not true, symmetric architectures should not be discounted due to their other benefits. For these reasons, as advocated for in ref.~\cite{Snowmass21}, we have adopted the approach of building all symmetries directly into the PELICAN network architecture itself, similar to the inherent translational symmetry of CNNs. 

\subsection*{Summary of results}

In \secref{equivariance} we discuss equivariance in jet physics and introduce the tools required to build an efficient equivariant architecture. In \secref{architecture} we describe the architectures of PELICAN classifiers and regressors. We briefly summarize the main results presented in this work, corresponding to \secrange{toptagging}{irc}.

\paragraph{Top-tagging with a PELICAN classifier}
We train PELICAN top taggers using a public benchmark dataset, to distinguish between top quark jets (signal), and light quark and gluon jets (background). These taggers achieve state-of-the-art performance on the benchmark with fewer learnable parameters than the previous highest-performing network. PELICAN top taggers with as few as $11\text{k}$ parameters outperform all non-equivariant networks in the benchmark. See \secref{toptagging} for details.

\paragraph{Quark-vs-gluon-initiated jet tagging with a PELICAN classifier}
We extend the study of a PELICAN classifier to the task of distinguishing jets produced by light-quarks (signal) from those produced by gluons (background). As in the task of top-tagging, PELICAN demonstrates state-of-the-art performance on this benchmark task, with fewer learnable parameters than any other similarly performant network architecture. See \secref{qgtagging} for details.

\paragraph{Multi-class jet tagging with a PELICAN classifier}
The final classification benchmark task involves identifying 5 different categories of jets (gluon jets, light quark jets, $W$-boson, $Z$-boson, and top quark jets). PELICAN achieves state-of-the-art classification performance across each of these five categories and once again demonstrates a more efficient utilization of its parameters that any rival architecture. See \secref{hlstagging} for details.

\paragraph{$W$-boson 4-momentum reconstruction with PELICAN}
We train a PELICAN model using a custom dataset~\cite{btW6} of fully-hadronic top-quark decays to reconstruct the full 4-momentum of the intermediate $W$-bosons using only the list of 4-momenta of the top quark jet constituents as inputs. PELICAN performs favorably in reconstructing the full $W$-boson momentum when compared with the Johns Hopkins (JH) top tagger~\cite{Kaplan:2008ie}, which identifies $W$-boson candidates for the subset of jets that pass its tagging. PELICAN achieves better transverse momentum ($p_T$), mass, and angular resolutions on JH top-tagged jets and achieves comparable resolutions to the JH tagger even when evaluated on the full dataset. 
Additionally, we train a PELICAN model to reconstruct the 4-momentum of only the products of the $W\rightarrow q q'$ decay which are contained within the jet. We discuss differences in performance and effects of this choice in reconstruction targets in \secref{Wreco}.

\paragraph{$W$-boson mass reconstruction with PELICAN}
Particle mass reconstruction is a common particle physics analysis task, and any reconstruction algorithm should be robust and relatively free of bias. In \secref{Wmass} we discuss the nuances of PELICAN mass reconstruction targeting the $W$-bosons in the above-mentioned dataset~\cite{btW6} as an example. The results show that eliminating bias in the underlying dataset is required to produce an unbiased final algorithm. In the case of $W$-boson mass ($m_W$) reconstruction, this is achieved by training PELICAN on a dataset with a range of values of $m_W$.

\paragraph{Explaining PELICAN 4-momentum reconstruction}
PELICAN's respect of the particle permutation and Lorentz symmetries inherent to particle datasets provides it with explainability and interpretability rarely found in particle physics ML applications. In \secref{Weights} we investigate the rich penultimate layer of PELICAN and its discriminatory and explanatory capabilities. In particular, we discuss interpretations of PELICAN as a form of soft-clustering and detector-unfolding algorithm. By \textit{explainability} of a network we mean the ability to identify physical features that highly correlate with the output. PELICAN's unique regression architecture, constrained by the simultaneous imposition of Lorentz and permutation symmetries, produces intermediate outputs (``PELICAN weights'') which we directly interpret as jet clustering coefficients and demonstrate the correctness of this interpretation. Despite the task being a simple regression on a single 4-vector, PELICAN reconstructs a much more complex set of physical features in the form of the labels of all input jet constituents by parent type.

\paragraph{IRC-safety and PELICAN}
In particle physics, so-called \textit{IRC-safety} is an algorithmic concern which restricts tools to be robust with respect to soft-particle emissions (infrared -- IR) and collinear (C) splittings which proliferate due to divergences in perturbative quantum chromodynamics (QCD). In \secref{irc} we introduce a simple IRC-safe modification of PELICAN and display its state-of-the-art performance.

\section{Equivariance and jet physics}
\label{equivariance}
%========================
% LORENTZ GROUP
%========================

This section aims to establish a clear connection between the group theory that underlies the PELICAN architecture and the implementation of this approach for both classification and regression, as described in Section~\ref{architecture}. 

In general, given a symmetry group $G$ and two sets $X,Y$ on which an action of $G$ is defined, a mapping $F:X\to Y$ is called $G$\textit{-equivariant} if $F(g\cdot x)=g\cdot F(x)$ for any $x\in X$ and $g\in G$. In particular, if the action of $G$ on $Y$ happens to be trivial (i.e. $g\cdot y=y$ for all $g,y$), then $F$ is called \textit{invariant}. In relativistic physics, equivariant maps are typically represented by tensors with equivariant spacetime indices treated via Einstein notation. For instance, the electromagnetic field tensor $F^{\mu\nu}$ can be viewed as a Lorentz-equivariant mapping from covariant vector fields to contravariant ones. In this work we will be interested in tasks from particle physics that can be reduced to learning a Lorentz-equivariant map. In this section we review several aspects of Lorentz symmetry in the context of such tasks.

%------------------------
% LORENTZ SYMMETRY
%------------------------
\subsection{Lorentz symmetry and jets}

The Lorentz symmetry is one of the fundamental symmetries of the Standard Model of particle physics. The full Lorentz group $\mathrm{O}(1,3)$ can be defined as the set of linear transformations of the 4-dimensional spacetime that preserve the Minkowski metric $\eta=\mathrm{diag}(1,-1,-1,-1)$. However, in this work we will restrict ourselves to the \textit{proper orthochronous} subgroup $\SO^+(1,3)$ that preserves spatial and temporal orientations. Lorentz invariance is the mathematical encapsulation of the fact that the outcomes of physical phenomena don't depend on the inertial frame of the observer. In the context of particle accelerators, this boils down to the observation that all initial and final states of a particle interaction are the same in all inertial frames. This is formally reflected in the fact that the Standard Model of particle physics is Lorentz-invariant, and therefore any model of any physically relevant processes described by the Standard Model can be as well.

Several subtle points are worth addressing before applying Lorentz symmetry to experimental tasks in jet physics. Neither the actual particle detectors nor the software simulating particle decays and their detection are Lorentz-invariant. Reasons for this include: non-invariant corrections to perturbative computations in quantum chromodynamics (QCD); non-invariance of jet clustering algorithms; practical limitations of detectors such as finite spatial and temporal resolutions, as well as energy and momentum thresholds. Nevertheless, it is still valid to learn Lorentz-invariant models from data obtained this way. Firstly, QCD is globally Lorentz-invariant and boosting the \textit{entire} event does not change the outcome of the decay process. As long as inference is performed on data obtained in conditions similar to the conditions of the perturbative simulation, corrections from effects such as the running of the couplings with varying momentum scales are not a concern either. The same applies to jet clustering algorithms and finite detector resolution: as long as the data used for inference was obtained in the same reference frame as the data used for training, the inference is valid and the outputs are expected to be Lorentz-equivariant. Finally, the fact that the detector itself introduces a fixed reference frame can be fully addressed without breaking the symmetry of the model by including detector geometry among its inputs. This is discussed further in  \secref{beams}.

While the imprecisions in our measurements of 4-momenta in themselves do not undermine the validity of Lorentz-invariant models, further investigation of the relative robustness to systematic measurement biases of different architectures is warranted. One step in this direction, via dataset augmentation, was recently taken in ref.~\cite{whiteson23}.

%------------------------
% LORENTZ INVARIANCE
%------------------------
\subsection{Lorentz invariance}\label{LI}

The classification tasks considered in this work are exactly Lorentz invariant. The physical implications of this statement are discussed below, but may be stated mathematically in the following way: if the inputs to the network are a collection of 4-vectors (energy-momentum vectors in our case) $p_1,\ldots,p_N$, the output is $F(p_1,\ldots,p_N)$, and $\Lambda\in\SO^+(1,3)$ is a Lorentz transformation, then
\[F\left(\Lambda p_1,\ldots,\Lambda p_N\right)=F\left( p_1,\ldots, p_N\right).\]
There are multiple approaches to constructing an ML model that satisfies such a constraint. The simplest one is to hand-pick a set of invariant observables (such as particle masses, jet masses, relative masses, particle identification labels and charge) and use them as input to a generic NN architecture. 

Another approach inspired by convolutional networks is to preserve group-equivariant latent representations in the hidden layers. In this case, the neuron nonlinearity must be a Lorentz-equivariant operation, and examples of this can be found in both the Lorentz Group Network (LGN)~\cite{Bogatskiy:2020tje} and LorentzNet~\cite{LorentzNet22} architectures. Equivariance with respect to the part of the Lorentz group that fixes the proton beam axis was also used for regression problems in ref.~\cite{CPT}. As in traditional CNN's used in image processing, equivariant latent representations, as opposed to invariant ones, can regularize the network via efficient weight-sharing and improve training.

The PELICAN design adopts a slightly different approach. Given a set of 4-vector inputs $p_1,\ldots,p_N$, we compute a \textit{complete} set of Lorentz invariants on that set. For classical groups, including the Lorentz group, the space of invariants constructed out of a collection of vectors in the fundamental representation consists of functions of only the pairwise invariant dot products (using the appropriate invariant quadratic form for the given symmetry group) and of square determinants (e.g.~of 4 column-vectors for the Lorentz group)~\cite{Weyl46}. Furthermore, if the invariant is required to be symmetric in the vector inputs, then it is \textit{only} a function of the dot products (see also the discussion in ref.~\cite{Gripaios2021}). In short, all totally symmetric Lorentz invariants can be written in the following form:
\[I(p_1,\ldots,p_N)=f\left(\{p_i\cdot p_j\}_{i,j}\right).\label{eq}\]

This is the first key idea used in the PELICAN architecture. The first step performed by the input layer is the computation of the $N\times N$ array of dot products between the input 4-momenta (also known as the Gram matrix). The $N(N-1)/2$ components of the Gram matrix $\{p_i\cdot p_j\}$ cannot be independent, which is apparent from dimension counting. The physical manifold inside this high-dimensional space is defined by the set of constraints $\det M_5=0$ for \textit{every} 5-minor $M_5$ of the Gram matrix (that is, any matrix obtained from the original one by crossing out $N-5$ rows and $N-5$ columns). Moreover, a causally related set of points such as a particle jet will always satisfy $p_i\cdot p_j\geq 0$ for all $i,j$. Therefore a neural network whose input is an $N\times N$ matrix will learn the task only on this $(4N-6)$-dimensional submanifold of $\mathbb{R}^{N^2}$. The outputs of the trained model on the rest of the space will be uncontrollable and physically meaningless. As a result, PELICAN has the complexity of $\mathcal{O}(N^2)$, similar to any message passing architecture on a fully connected graph. An explicit set of coordinates for this manifold would allow for complete Lorentz-invariant networks of the superior $\mathcal{O}(N)$ complexity, however such coordinates, as of this time, are unknown for $N>5$ \cite{Gripaios2021}.

%------------------------
% PREMUTATION INVARIANCE
%------------------------
\subsection{Permutation equivariance}\label{EUA}

Particle data are often interpreted as a point cloud since there is no natural ordering on the vectors. For such problems it makes sense to use a permutation-invariant or equivariant architecture. One of the simplest approaches is called \textit{Deep Sets}~\cite{ZaKoRaPoSS17}, which has been applied to jet tagging~\cite{EFN} and even heavy-flavor tagging~\cite{ATLAS:2020jip}. The fundamental fact used in Deep Sets is that any permutation-invariant continuous mapping of inputs $x_1,\ldots,x_N$ can be written in the form $\psi\left(\sum_i\varphi(x_i)\right)$, where $\psi$ and $\varphi$ can be approximated by neural networks. 

The main limitation of permutation-invariant architectures such as Deep Sets is the difficulty of training. Since aggregation (summation over the particle index) happens only once, the Deep Sets architecture can struggle with modeling complex higher-order interactions between the particles, as shown rigorously in ref.~\cite{wagstaff22}. The network representing $\psi$ is forced to be a relatively wide fully connected network, which presents difficulties in training.

An alternative to permutation-invariant architectures is provided by permutation-\textit{equivariant} ones. Given a symmetry group $G$ (e.g.~the group of permutations), a representation $(V,\rho)$ is a tuple where $V$ is a set and $\rho:G\times V\to V$ is a map that becomes a bijection $\rho_g=\rho(g,\cdot):V\to V$ for any fixed value of the first argument, $\rho_e=\mathrm{id}$, and $\rho_{g^{-1}}=\rho_g^{-1}$. Given two representations $(V,\rho)$ and $(V',\rho')$ of a group $G$, a map $F:V\to V'$ is called equivariant if it \textit{intertwines} the two representations, that is:
\[
F(\rho_g(v))=\rho_g'\left(F(v)\right),\quad v\in V,\; g\in G.
\]
Equivariance is a key property of all convolutional networks -- for example, in CNN's the convolution operation is inherently equivariant with respect to translations (up to edge effects). 

Similarly, Graph Neural Networks (GNNs) use permutation equivariance with respect to the reordering of the rows and columns of the adjacency matrix for problems where the inputs can be naturally represented by a graph data structure. In this context, we review the standard definition of a message passing layer where the particles are treated as nodes in a graph (e.g.~a fully connected graph), and every layer of the network only updates the activation at every node. If we denote by $f_i$ the data assigned to node $i$, then the message passing layer will typically construct ``messages'' $m_{ij}=m(f_i,f_j)$ and update each node by aggregating the messages from all neighbors of that node and combining the result with the original state of the node: $f_i'=\psi(f_i, \sum_j m_{ji})$. Sometimes the graph also possesses ``edge data'' $D_{ij}$ that can be incorporated into the message-forming stage.

Message passing architectures have been successfully applied to jet tagging, most prominently in refs.~\cite{Bogatskiy:2020tje, LorentzNet22}. Closely related permutation-equivariant transformer architectures were also applied to particle physics in \cite{ParT,CPT}. However, attempts to combine message passing with Lorentz invariance reduced at the input stage as described above run into a major obstacle: the new inputs to the network consist of \textit{nothing but} edge data $d_{ij}=p_i\cdot p_j$. Traditional message passing would require a reduction of this set of inputs to a point cloud (with only one particle index), potentially restricting the set of possible higher-order interactions between the points or adding redundancy to the network. To avoid making these unnecessary choices, we employ the general permutation-equivariant layers suggested in refs.~\cite{Maron18, KondorPan}.

In the general setting, permutation equivariance is a constraint on mappings $F$ between arrays $T_{i_1i_2\cdots i_r}$ of any rank $r$, every index $i_k\in \{1,\ldots,N\}$ referring to a particle label, whereby permutations of the particles commute with the map:
\[F\left(\pi\circ T_{i_1i_2\cdots i_r}\right) =\pi\circ F\left(T_{i_1i_2\cdots i_s}\right), \quad \pi\in S_N.\]
Here, the action of permutations is effectively diagonal: $\pi\circ T_{i_1i_2\cdots i_r} =T_{\pi(i_1)\ldots \pi(i_r)}$. GNNs explicitly implement this constraint for rank 1 arrays (node information). A higher-order generalization of the message passing layer can be defined as
\[\text{\bf Equivariant Layer:}\quad T^{(\ell+1)}=\textsc{Agg}\circ \textsc{Msg}\left(T^{(\ell)}\right).\]
In this nomenclature, $\textsc{Msg}$ is a node-wise nonlinear map (\textit{message forming}) shared between all nodes, and $\textsc{Agg}$ is a general permutation-equivariant linear mapping (\textit{aggregation}) acting on the particle indices of $T$. Note that whether $\textsc{Msg}$ is node-wise and whether $\textsc{Agg}$ is linear is somewhat ambiguous based on how one separates the mappings into their components, which is why, in particular, the traditional formulation of message passing allows messages to be functions of pairs of nodes. In practice, our aggregation block will permit nonlinear aggregation functions such as max-pooling and in fact contain an additional nonlinear activation function.

%------------------------
% PREMUTATION INVARIANCE
%------------------------
\subsection{Elementary equivariant aggregators}\label{EEA}
%\paragraph{Elementary Equivariant Aggregators}\label{EEA}

The exact structure of the equivariant aggregation layers defined above must still be specified. Since the general case is presented in refs.~\cite{Aggregators, KondorPan}, here we will only present the layers that are required for jet physics tasks. Since the input is an array of rank $2$, the main equivariant layer for us is one that transforms arrays of rank $2$ to other arrays of the same rank: $T_{ij}\mapsto T_{ij}'$. The space of all linear maps of this type turns out to be 15-dimensional. The basis elements of this space can be conveniently illustrated using binary arrays of rank $4$. There are 15 such arrays $B^{a}_{ijkl}, a=1,\ldots,15$, and the action of the equivariant layer can be written as
\[T^{\prime a}_{ij}=\sum_{k,l=1}^N B^{a}_{ijkl} T_{kl}.\label{aggregators}\]

\begin{figure}
    \centering
    \includegraphics[page=1,  scale=0.5]{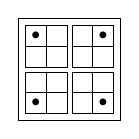}
    \includegraphics[page=2,  scale=0.5]{dots.pdf}
    \includegraphics[page=3,  scale=0.5]{dots.pdf}
    \includegraphics[page=5,  scale=0.5]{dots.pdf}
    \includegraphics[page=4,  scale=0.5]{dots.pdf}\\
    \includegraphics[page=7,  scale=0.5]{dots.pdf}
    \includegraphics[page=9,  scale=0.5]{dots.pdf}
    \includegraphics[page=11, scale=0.5]{dots.pdf}
    \includegraphics[page=6,  scale=0.5]{dots.pdf}
    \includegraphics[page=8,  scale=0.5]{dots.pdf}
    \includegraphics[page=10, scale=0.5]{dots.pdf}
    \includegraphics[page=13, scale=0.5]{dots.pdf}
    \includegraphics[page=12, scale=0.5]{dots.pdf}\\
    \includegraphics[page=15, scale=0.5]{dots.pdf}
    \includegraphics[page=14, scale=0.5]{dots.pdf}
    \caption{The 15 binary arrays of rank $4$ that represent the basis elements of the permutation equivariant aggregators of PELICAN.\label{fig:aggregators}}
\end{figure}

\noindent The 15 aggregators $B^a$ may be visualized as is done in \figref{fig:aggregators} for $N=2$.
%===========================================
%===========================================
The smallest squares represent components of the input $2\times 2$ array, and the larger $2\times 2$ squares represent components of the output array. Dots represent the non-zero components of the binary tensors $B^a$, and every component of the output tensor is the result of aggregation over all inputs marked by the dots. Output components that lack any dots are set to be a fixed constant, by default zero (the affine versions of these mappings include two such parameters: one constant for the diagonal and another for the remaining components). By ``aggregation'' we mean, in general, any symmetric function, but in practice it is usually a sum or mean. For example, the first aggregator is simply the identity map on matrices: the $ij$'th component of the output array is the result of aggregation over only the $ij$'th component of the input. The second aggregator realizes the transposition of arrays $T'_{ij}=T_{ji}$. The following three aggregators represent various ways of embedding the diagonal of the input array in an equivariant way. It is easy to see that simultaneously swapping the two rows and the two columns of the input is equivalent to doing the same to the output, which confirms equivariance. These first 5 aggregators are ``order zero'' in $N$ because they do not actually perform any aggregation. Instead, they can be thought of as permutation-equivariant skip-connections.

The second group of 8 ``order one'' aggregators aggregate over $N$ components of the input by aggregating either over rows, columns, or the diagonal, and then embedding the result into the output array in all possible equivariant ways. Finally, the last 2 aggregators are the ``order two'' aggregators that aggregate over all $N^2$ components of the input.

If we allow aggregators to be nonlinear, then they can take the following form: the binary array $B^a$ selects a subset of the components of the input array, and then a general symmetric function $S^a$ is applied to that subset:
\[T^{\prime a}_{ij}=S^a\left(\{T_{kl}\mid k,l: B^{a}_{ijkl}\neq 0\}\right).\]
In practice we define $S^a$ as the mean of its inputs followed by an additional scaling by a factor of $N^{\alpha_a}/\bar{N}^{\alpha_a}$ with learnable exponents $\alpha_a$, where $\bar{N}$ is a constant representing the typical number of input vectors expected in the dataset, provided to the model as a hyperparameter.

These equivariant layers and their generalizations have been thoroughly studied since their introduction (see e.g.~refs.~\cite{EqGNNs,Sannai, Maron19OnUniversality,Maron19ProvablyPowerful}) and applied to problems in graph learning and chemistry, e.g.~in refs.~\cite{KondorPan,Kondor2023}. However, many important questions about the universality and expressivity of these networks remain open. For example, networks that can use any $\mathrm{Eq}_{r\to s}$ layers were proven to be universal in refs.~\cite{EqGNNs, Maron19OnUniversality}, but the exact value of the maximum rank (which is 2 in PELICAN) required for universality in a given problem is not known.

%------------------------
% SEGUE TO HEP: EQUIVARIANCE & JETS
%------------------------
\subsection{Equivariance and Jet Physics}

There are several reasons for enforcing the full Lorentz symmetry in our ML models. First and foremost, it is a fundamental symmetry of the space to which the inputs belong. Lorentz transformations represent the effect of switching between different inertial frames, and most fundamental processes in physics are independent of the choice of the observer's inertial frame: if a given collection of particles consists of the products of a decay of a top quark for one observer, then the same is true for all other observers.

Nevertheless, some processes involved in generating and observing high-energy collision events break the Lorentz symmetry in some subtle ways. At the fundamental level, the running of the couplings in QCD can cause Lorentz symmetry breaking in the parton shower distribution functions. Even the amount of final decay products depends on the transversal boost of the initial parton-level particles. However, there is no question that both the original protons and the final (asymptotic) decay products are accurately represented by a collection of 4-vectors subject to the spacetime Lorentz symmetry: the asymptotic outcome of a collision event is independent of the observer's reference frame. 

Another reason for symmetry-restricted modeling is that, from the geometric perspective, only some mathematical operations are permissible when working with objects that transform in a certain way under a symmetry group. A non-equivariant neural network effectively neglects the vector nature of the inputs by treating individual components of the input vectors as scalars. While improving network expressivity, non-equivariance fails to deliver physically interpretable models. Ultimately, a statement about equivariance is a statement about what the basic \textit{features} of the data are -- e.g.~vectors are features, but the individual components of those vectors are not.

More relevant to the applications is the fact that both the simulation and the observation of collisions inevitably involves some degree of \textit{clustering}. A particle detector is made of cells (e.g.~calorimeters) of finite size and as such is unable to distinguish between some particles that are collinear or very close to collinear. Similarly, the standard algorithms for collision simulation typically perform \textit{jet clustering} to closely reproduce the detector behavior. Clustering of course is not a Lorentz-invariant procedure: particle tracks that diverge by a small angle in one frame will diverge by a large angle in another highly boosted frame. However, this limitation of Lorentz-invariant architectures is fairly minor. Since clustering is always done in a fixed laboratory frame, it is still reasonable to impose the full Lorentz symmetry on the resulting 4-vector data. So unless the pre-clustering data itself is coming from multiple significantly different inertial frames, clustering is not interfering with the fundamental symmetry. Simply put, however a given set of 4-vectors is obtained and represented in a specific inertial frame, those vectors will respect the Lorentz symmetry. Finally, one important way to address the issue of network sensitivity to the clustering of collinear particles is through enforcing so called IRC-safety, which will be discussed separately in Section~\ref{irc}.

% These issues are part of the issue of Infrared and Collinear (IRC) Safety, which is a statement that the final ``hard jet'' products of a decay must be invariant under additions on light infrared particles (IR-safety) or under splittings of any present particle into two or more collinear particles with the same total energy-momentum (C-safety). Since any particle detector inherently performs some phase-space integration, it can only observe IRC-safe observables. However, other ``observables'' (such as differential scattering amplitudes) in the theory itself need not be IRC-safe. This opens up the debate about the importance of non-IRC-safe observables in computational methods such as machine learning of collision events. For instance, in \cite{EFN} the non-IRC-safe network performed better than an analogous IRC-safe one, indicating that ``there is discrimination power to be found in IRC-unsafe information''. Hence part of this paper is also dedicated to a comparison of an IR-safe instance of our top tagger with the general non-IRC-safe one and quantifying the IRC-safety of the model.

%========================
% NETWORK DESIGN
%========================
\section{PELICAN architecture}
\label{architecture}

The PELICAN architecture is simpler than many previous architectures due to its use of a complete set of Lorentz-invariants at the input stage, namely the set of pairwise dot products between the input 4-momenta (see \secref{equivariance}), and this has significant implications for both the overall architecture as well as the ease of training and interpretability of the network. This section discusses each of the primary components of the network, including the inputs and their embedding, the permutation-equivariant blocks, and the output layers that determine the nature of the task, namely classification or 4-vector regression.

%------------------------
% PELICAN INPUTS
%------------------------
\subsection{Inputs and embeddings}\label{PELICANinputs}

\paragraph{Dot Products and Beams}\label{beams}
On the input side of the architecture, the first step is to compute all pairwise dot products of the input 4-momenta. Appended to the list of these 4-momenta are two auxiliary beam particles with 4-momenta $(1,0,0,\pm 1)$. This is helpful since the datasets we are using are all simulated in a fixed laboratory frame where the original proton-proton collision happens along the $z$-axis, and the auxiliary inputs restore this orientational knowledge. In particular, the dot products between constituents and beams give PELICAN access to the energies and transverse momenta of all constituents.

It is worth emphasizing that introducing beams in this manner allows us to fix a particular spatial orientation of the events without restricting or violating the global Lorentz symmetry inherent in the architecture. Indeed, if one were to treat the auxiliary beams as constant vectors of hyperparameters, then this action would reduce the full Lorentz symmetry to merely rotations in the $xy$-plane and $z$-boosts. However, due to the fact that the beams are fed into the network on equal footing with all other inputs, they are properly treated as full-fledged 4-vectors that should also transform under the global Lorentz symmetry. Thus, counter-intuitively, we let the network access individual energies, transverse momenta and $z$-momenta while still preserving the full Lorentz symmetry and all the computational benefits that come with it (e.g.~the number of learnable parameters remains unchanged, whereas an architecture with a smaller symmetry group would have more degrees of freedom).

\paragraph{Embedding of dot products}
 Next there is an embedding layer that applies the function $f_\alpha(x)=((1+x)^{\alpha^2} -1)/\alpha^2$ to each dot product with several values of the trainable parameter $\alpha$ (initialized to span the interval $[0.05,0.5]$). This is roughly inspired by the logarithmic embedding in ref.~\cite{LorentzNet22} and serves to mollify the extremely heavy-tailed (approximately Pareto) distribution of the input momenta. The addition of the learnable $\alpha$ parameter allows PELICAN to resolve the inputs at multiple momentum scales, providing a slight boost to performance. Such an embedding may be generally useful for inputs which are naturally Pareto distributed (i.e.~positive with density of the form $p(x)\propto x^{-\gamma},\gamma>0$).

\paragraph{Embedding of scalar constituent data} In addition to dot products, we introduce a binary scalar label to distinguish jet constituents from the custom beam vectors. More generally, jet datasets can involve scalar particle data $s_i$ such as particle IDs (PID), charge, color, spin, etc. Let's say each $s_i$ is a column of dimension $C_{\text{scalar}}$. To be able to process such inputs alongside the dot products, they need to be promoted to arrays with two particle indices. One way to do this is to double the number of channels and define the array $s_i\oplus s_j$ of shape $[B,N_{\rm max},N_{\rm max},2C_{\text{scalar}}]$. This array can then be concatenated with the embedded dot products.
 
However, there is a more natural way of doing this within the PELICAN framework. In the next section we will introduce permutation-equivariant $\mathrm{Eq}_{2\to 2}$ blocks that act on arrays with two constituent indices. To embed scalar particle data (``node features'') into the space of such arrays (``edge features''), it is convenient to utilize an analogous equivariant $\mathrm{Eq}_{1\to 2}$ block and apply it directly to $s$, the array of constituent scalars. This way we can produce a flexible number $C_{\text{scalar}}'$ of scalar channels. By choosing the dimensionality of the dot product embedding to be $C^0-C_{\text{scalar}}'$ and then concatenating it with the array of promoted scalars, we get a tensor of shape $[B,N_{\rm max},N_{\rm max}, C^0]$, where the feature vector for each particle pair is $C^0$-dimensional and has the form $\left(f_{\alpha_1}(d_{ij}), \ldots, f_{\alpha_{C^0-C_{\text{scalar}}'}}(d_{ij})\right)\oplus \mathrm{Eq}_{1\to 2}(s)_{ij}$. This approach allows for any number of Lorentz invariants per constituent, but in all of the tasks considered in this paper we have $C_{\text{scalar}}=1$.

%------------------------
% PERMUTATION EQUIV BLOCKS
%------------------------
\subsection{Permutation equivariant blocks}\label{permutationequiv}
%\paragraph{Permutation Equivariant Blocks}

\begin{figure}[t]
    \begin{center}
    \includegraphics[width=0.3\textwidth]{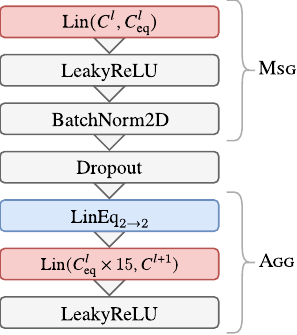}
    \caption{The PELICAN equivariant block updating square arrays. \label{fig:PELICANblock}}
    \end{center}
\end{figure}

The main element of the equivariant architecture is the permutation-equivariant block transforming arrays of rank 2, represented schematically in \figref{fig:PELICANblock}. Namely, we assume that the input tensor to the block has shape $[B,N_{\mathrm{max}},N_{\mathrm{max}},C^l]$, where $B$ is the batch size, $N_{\rm max}$ is the maximum number of jet constituents per event (with zero padding for events with fewer constituents), and $C^l$ is the number of input channels. We also use a binary mask of shape $[B,N_{\rm max},N_{\rm max}]$ to appropriately exclude the zero padding from operations like \texttt{BatchNorm} and aggregation. The output of the block will be a similar tensor of shape $[B,N_{\mathrm{max}},N_{\mathrm{max}},C^{l+1}]$ with the same mask. 
%===========================================
%===========================================

As outlined above, the equivariant layer consists of a message block and an aggregation block. The message block is chosen to be a dense multilayer perceptron (MLP) acting on the channel dimension with a \texttt{LeakyReLU} activation and \texttt{BatchNorm2D} (normalization over the first three dimensions of the tensor, for each channel separately, followed by an affine transform with two learnable parameters per channel). Here we use a masked implementation of batch normalization so that the variable particle number is respected. The message block is then followed by \texttt{Dropout} that zeroes out each of the $B\times N_{\mathrm{max}}^2\times C^l_{\text{eq}}$ components independently with a certain probability.

The aggregation block applies 15 linear aggregation functions ($\texttt{LinEq}_{2\to 2}$) which, for each component of the output tensor, compute the mean over some subset of the components of the input tensor, as explained in \secref{EEA}. Note that this is a non-parametric transformation performed on each channel separately. Each of the $C_{\text{eq}}^l\times 15$  resulting aggregation values is then independently multiplied by $N^\alpha/\bar{N}^\alpha$ with a trainable exponent $\alpha$ (initialized as a random float in $[0,1]$), where $N$ is the number of particles in the corresponding event. This allows for some flexibility in the aggregation process, for example $\alpha=1$ returns the sum aggregation function, and combining multiple aggregators is known to boost accuracy, see e.g.~ref.~\cite{Aggregators}.

Aggregation is followed by a dense layer that mixes the $C_{\text{eq}}^l\times 15$ aggregators down to $C^{l+1}$ features. Due to the size of this layer, we employ a simple factorization to reduce the number of parameters. Namely the weight tensor $W_{abc}$, where $a$ is the input channel index, $b$ is the basis index ($1$ to $15$), and $c$ is the output channel index, can be replaced by the following combination:
\[W_{abc}^{\text{factorized}}=W^0_{ab}W^1_{ac}+W^2_{cb}W^3_{ac}.\]
Here, the first term first mixes the $15$ aggregators among each other for each output channel, and then mixes the channels. Similarly, the second term first mixes the $15$ aggregators for each input channel, and then mixes the channels. It is technically possible to increase the rank of this factorization by adding together multiple tensors of this form, but in practice we find one to be sufficient -- in our tasks the factorized network performs as well as the unfactorized one (except at very low network widths, in which case the unfactorized network performs better and may even have fewer parameters). The final result is a tensor of shape $[B,N_{\mathrm{max}},N_{\mathrm{max}},C^{l+1}]$, so these equivariant layers can be stacked multiple times.

As already indicated above, a few other equivariant blocks will be useful to us, namely $\mathrm{Eq}_{1\to 2}$, $\mathrm{Eq}_{2\to 1}$, and $\mathrm{Eq}_{2\to 0}$. The only difference in their definition is that they involve just 5, 5, and 2 aggregators, respectively.

%------------------------
% OUTPUTS
%------------------------
\subsection{Classification and 4-vector regression outputs}\label{outputs}

One of the strengths of the PELICAN architecture is the ability to easily switch between serving as a classification tool for jet tagging, to being able to provide 4-vector outputs in tasks such as momentum reconstruction. Here we summarize both architectures.

\paragraph{PELICAN classifier}
To build a classifier, aside from the $\mathrm{Eq}_{2\to 2}$ equivariant layer one needs a $\mathrm{Eq}_{2\to 0}$ layer that reduces the rank 2 array to permutation-invariant scalars, as represented by \equref{Classifierlayer}. This layer involves just 2 aggregation functions instead of $15$ -- the trace and the total sum of the input square array, but is otherwise identical to the equivariant layer described in the last section. The input block concatenates the embeddings of the dot products $d_{ij}$ and the scalar data $s_i$ as described above.
\[
\tikz[overlay]{\draw[draw=black] (1.65,-1.3) rectangle (10.5,1.3);}
\begin{tikzcd}[column sep=small, row sep=small]
\{d_{ij}\}  & \arrow[r]& \text{Emb} \arrow[dr] &&&&\\
&& & \bigoplus \arrow[r] & \left[\mathrm{Eq_{2\to 2}}\right]^L \arrow[r] & \mathrm{Eq_{2\to 0}} \arrow[r] & \mathrm{MLP} \arrow[r] & \,& \hspace{-0.5em} \{w_c\}\\
\{s_i\} & \arrow[r] & \text{Eq}_{1\to 2} \arrow[ur] &&&&
\end{tikzcd}
\label{Classifierlayer}
\]
% \[\small \{d_{ij}\}\to \boxed{\mathrm{Emb} \to \left[\mathrm{Eq_{2\to 2}}\right]^L \to \mathrm{Eq_{2\to 0}}\to \mathrm{MLP}}\to \{w_c\}\]
% \begin{center}{\includesvg[width=0.9\linewidth]{diagram}}\end{center}

From the input block, the tensor is passed through $L$ equivariant $\mathrm{Eq}_{2\to 2}$ layers, and the $\mathrm{Eq}_{2\to 0}$ layer with dropout. This produces a tensor of shape $[B,C_{\mathrm{out}}]$. One final MLP mixes this down to just $n_{\text{classes}}$ classification weights $w_c$ per event. A cross-entropy loss function is then used for optimization.

\paragraph{PELICAN 4-vector regression}
The same architecture can also be easily adapted for 4-vector regression tasks such as momentum reconstruction. Any Lorentz-equivariant map from a collection of 4-vectors $p_1,\ldots,p_N$ to one 4-vector (or several) has the form
\[F(p_1,\ldots,p_N)=\sum_{i=1}^N f_i(p_1,\ldots,p_N)\cdot p_i,\label{equiv}\]
where $f_i$'s are Lorentz-invariant functions, see e.g.~ref.~\cite{Hogg21} for a proof. Combining this with permutation invariance, we conclude that the multi-valued map $(p_1,\ldots,p_N)\mapsto (f_1,\ldots,f_N)$ must also be equivariant with respect to permutations of the inputs.

The only change required to the architecture we've introduced for classification is that  $\mathrm{Eq}_{2\to 0}$ must be replaced with $\mathrm{Eq}_{2\to 1}$ and the final output layer must have only one output channel (assuming we are regressing on a single 4-vector). The $\mathrm{Eq}_{2\to 1}$ layer is again identical to $\mathrm{Eq}_{2\to 2}$ except that it uses only $5$ linear aggregators: taking the diagonal, row sums, column sums, trace, and full sum. The architecture is summarized by the relationship represented in \equref{RegressionLayer}, where we treat $f_i$ as the outputs, and use \equref{equiv} to recover the final predicted vector.
\[
\tikz[overlay]{\draw[draw=black] (1.65,-1.3) rectangle (10.5,1.3);}
\begin{tikzcd}[column sep=small, row sep=small]
\{d_{ij}\}  & \arrow[r]& \text{Emb} \arrow[dr] &&&&\\
&& & \bigoplus \arrow[r] & \left[\mathrm{Eq_{2\to 2}}\right]^L \arrow[r] & \mathrm{Eq_{2\to 1}} \arrow[r] & \mathrm{MLP} \arrow[r] & \,& \hspace{-0.5em} \{f_i\}_{i=1}^N\\
\{s_i\} & \arrow[r] & \text{Eq}_{1\to 2} \arrow[ur] &&&&
\end{tikzcd}
\label{RegressionLayer}
\]
In summary, \textit{the simultaneous imposition of the full Lorentz and permutation symmetries lets us express a Lorentz-equivariant and permutation-invariant vector-valued model via an equivalent Lorentz-invariant and permutation-equivariant scalar-valued one}. To recover the final 4-vector output one needs only aggregate the output using \equref{equiv}. As we will see, in particular tasks the coefficients $f_i$ become identifiable physical features, making the model explainable. Crucially, this would not have been possible with only a partial Lorentz symmetry.
% \[\small \{d_{ij}\}\to \boxed{\mathrm{Emb} \to \left[\mathrm{Eq_{2\to 2}}\right]^L \to \mathrm{Eq_{2\to 1}}\to \mathrm{MLP}}\to \{f_i\}_{i=1}^N\]

%========================
% TOP TAGGING
%========================
\section{Tagging jets from Lorentz boosted top quarks}
\label{toptagging}
This section presents the dataset, training approach, and results of using PELICAN as a classifier in the context of identifying ``Lorentz-boosted'' (high transverse momentum) top quarks. Three different versions of PELICAN are discussed, each with a different size in terms both the width of the network and the number of trainable parameters. Lastly, the dependence of the performance on the size of the training dataset is also presented, providing a quantitative relationship between the size of the network, the training dataset efficiency, and the resulting performance.

%------------------------
% DATASET
%------------------------
\subsection{Top tagging dataset}
%\paragraph{Dataset}

We perform top-tagging on the reference dataset~\cite{KasPleThRu19}, which was also used in ref.~\cite{KasiePlehn19}. This dataset consists of 2M entries, each entry corresponding with a single hadronic top jet or the leading jet from a QCD dijet event. There are 1.2M training entries, 400k validation entries and 400k testing entries. The events were generated with the \PythiaEight event generator~\cite{Bierlich:2022pfr}, and the \Delphes framework~\cite{deFavereau:2013fsa} was used for fast detector simulation in order to incorporate detector effects. For each jet, the 4-momentum of the $200$ leading constituents are stored in Cartesian coordinates $(E,p_x,p_y,p_z)$, in order of decreasing $p_T$. This list is zero-padded, and all jets in the dataset have fewer than 200 constituents. The dataset does not contain any other information on the jet constituents, such as charge or spin.

%------------------------
% TRAINING PROCEDURE
%------------------------
\subsection{Classification training procedure}
%\paragraph{Training} 

The top-tagging model contains five $\mathrm{Eq}_{2\to2}$ blocks of identical shapes (followed by a sixth $\mathrm{Eq}_{2\to1}$ block). We train three different versions of the model with different widths. The widest model has $132$ input and $78$ output channels on every messaging layer (the equivariant layer then produces $132\times 15$ quantities which get mixed down to $78$ channels by a fully connected linear layer). The output MLP is just one layer that mixes $132$ channels down to $2$ classification weights. The number of jet constituents was capped at 80 (no noticeable performance gain was seen beyond that number). The dropout rate was $0.025$, and the model was optimized using the \textsc{AdamW} optimizer \cite{AdamW} with weight decay of $0.005$. The training on the full dataset went on for 35 epochs with the same learning rate schedule as in ref.~\cite{LorentzNet22}: 4 epochs of linear warm-up up to learning rate of $0.001$, followed by 28 epochs of \textsc{CosineAnnealingLR} with $T_0$ of $4$ epochs and $T_{\mathrm{mult}}=2$, and then 3 epochs of exponentially decaying learning rate with exponent $\gamma=0.5$ per epoch. We ensure that each minibatch contains an equal number of signal and background events, and shuffle the order of the minibatches at every epoch. The three models were trained on Nvidia H100 GPU's with batch size of 100, taking $0.43$, $0.17$, or $0.08$ seconds per batch, respectively. Inference took $0.17$, $0.07$, or $0.04$ seconds per batch. Batches were shuffled between epochs.
%===========================================
\begin{table}[t]
    \vspace{-0.\intextsep}
    \centering
    \begin{small}
    \begin{tabular}{l@{\hspace{2mm}}l@{\hspace{2mm}}l@{\hspace{2mm}}l@{\hspace{2mm}}r}
    \toprule
    Architecture    &   Accuracy    &   AUC         &   $1/\epsilon_B$ ($\epsilon_S=0.3$)  &   \# Params \\
    \midrule
    TopoDNN\cite{Pearkes:2017hku}   & 0.916 & 0.972 & 382$\pm$ 5 & 59k \\
    EFN\cite{EFN}    &   0.927       &   0.979       &   729 $\pm$ 13    &   82k     \\
    LGN\cite{Bogatskiy:2020tje}             &   0.929(1)    &   0.964(14)   & 424 $\pm$ 82      &   4.5k    \\
    BIP(XGBoost)\cite{Ortner22}        &   0.929       &   0.978       &   600  $\pm$ 47   & 312       \\
    EFP\cite{EFP}             &   0.932       &   0.980       &   384             & 1k       \\
    BIP(MLP)\cite{Ortner22}        &   0.931       &   0.981       &   853  $\pm$ 68   & 4k       \\
    PFN\cite{EFN}             &   0.932       &   0.982       &   891 $\pm$ 18    &   82k     \\
    DisCo-FFS\cite{Shih22}             &   --       &   0.982       &   1249 $\pm$ 43    &   1.4k     \\
    ResNeXt\cite{KasiePlehn19}         &   0.936       &   0.984       &   1122 $\pm$ 47   &   1.46M   \\
    ParticleNet\cite{ParticleNet}     &   0.938       &   0.985       &   1298 $\pm$ 46   &   498k    \\
    ParT\cite{ParT}      &   0.940       &   0.9858      & 1602 $\pm$ 81    &   2.1M    \\
    LorentzNet\cite{LorentzNet22}      &   0.942       &   0.9868      & 2195 $\pm$ 173    &   220k    \\
    \midrule
    $\text{PELICAN}$        &   0.9426(2)   &   0.9870(1)   & 2250 $\pm$ 75    &   208k     \\ 
    $\text{PELICAN}_{\text{IRC}}$   &   0.9406(2)   &   0.9844(11)   & 1711 $\pm$ 208    &   208k   \\   
    \bottomrule
    \end{tabular}
    \end{small}
    \caption{Comparison of different classifiers trained on the full top-tagging dataset. Note that BIP(XGBoost) and EFP are not neural networks. PELICAN's metrics are averaged and the uncertainties are given by the standard deviation over 5 runs with different values of the random seed. $\text{PELICAN}_{\text{IRC}}$ is an IRC-safe modification detailed below in \secref{irc}. Note that Ref.~\cite{ParT} also includes a higher-performing model pre-trained on a large custom dataset. For the direct comparison, this table refers only to the non-pre-trained ParT model. \label{tab1}}
\end{table}

\begin{table}[t]
    \vspace{-0.\intextsep}
    \centering
    \begin{small}
    \begin{tabular}{l@{\hspace{2mm}}l@{\hspace{2mm}}l@{\hspace{2mm}}l@{\hspace{2mm}}l@{\hspace{2mm}}r}
    \toprule
    Depth $L$ & Width &   Accuracy    &   AUC         &   $1/\epsilon_B$  ($\epsilon_S=0.3$)  &   \# Params \\
    \midrule
      5    &  132/78  &   0.9425(1)   &   0.9870(1)   & 2250 $\pm$ 75    &   208k     \\  % 500_100p
      5    &  60/35  &   0.9424(1)   &   0.9868(1)   & 2148 $\pm$ 125    &   48k     \\   
      5    &  25/15  &   0.9410(3)   &   0.9858(4)   & 1879 $\pm$ 103    &   11k     \\   
    %  5    &  10/6  &   0.9385(3)   &   0.9848(2)   & 1471 $\pm$ 81    &   3k     \\   % 501_100p
      5    &  10/6  &   0.9386(2)   &   0.9850(1)   & 1494 $\pm$ 43    &   3k     \\   % 551_100p
      3    &  6/4  &   0.9358(7)   &   0.9835(2)   & 1145 $\pm$ 74    &   1k     \\   % 548_100p
      2    &  6/3  &   0.9336(5)   &   0.9823(3)   & 901 $\pm$ 59    &   605     \\   % 554_100p
    % 3    &  2/2 &   0.9288(21)   &   0.9799(10)   & 622 $\pm$ 72    &   339   \\   
      1    &  6/3  &  0.9291(5)   &   0.9801(4)   &  669$\pm$41     &   326     \\   % 550_100p
      %1    &  6/3  &  0.9291(9)   &   0.9796(8)   & 637 $\pm$ 56    &   294     \\   % 505_100p
      1    &  -/6    &  0.9258(8)   &   0.9780(6)   & 516 $\pm$ 52    &   248     \\   
    \bottomrule
    \end{tabular}
    \end{small}
    \caption{Comparison of PELICAN classifiers of varying shapes trained on the full top-tagging dataset. The depth $L$ is the number of equivariant $\mathrm{Eq}_{2\to 2}$ blocks, and the width consists of two numbers. E.g.~132/78 means that each messaging block takes in 132 channels and outputs 78 channels (and the reverse for the aggregation block). The input width of the output MLP matches the messaging blocks. The last model has the fully connected messaging layers disabled altogether, so only the shape of the aggregation blocks is given. See also \figref{toptag-params}. \label{tab_pelican_model_size}}
\end{table}

%===========================================
% \WFclear % see here: https://tex.stackexchange.com/a/442369

%------------------------
% RESULTS
%------------------------
\subsection{Top tagging results}
%\paragraph{Results} 

\begin{figure}[t]
    \vspace{-1em}
    \begin{center}
    \includegraphics[width=0.9\textwidth]{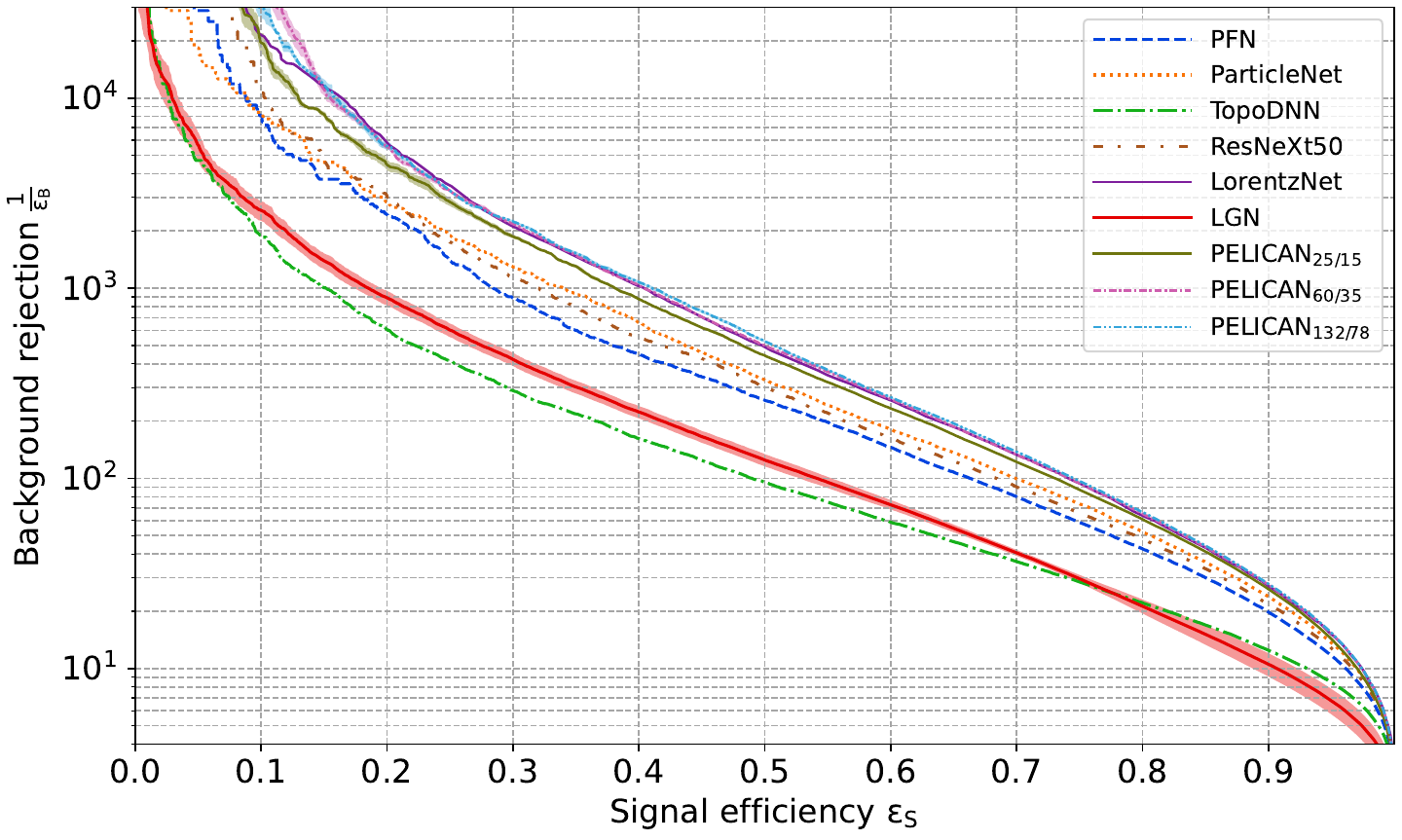}
    \caption{Performance of various ML architectures represented by the background rejection as a function of the signal efficiency.}
    \label{PELICAN-ROC}
    \end{center}
\end{figure}

\begin{figure}[t]
    \vspace{-1em}
    \begin{center}
    \includegraphics[width=0.6\textwidth]{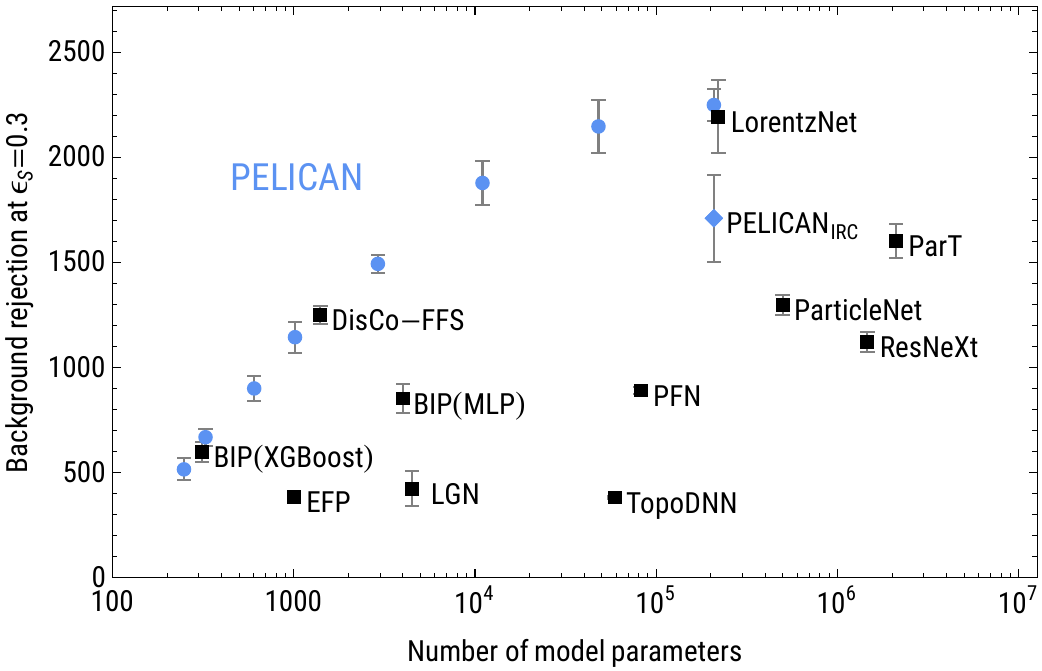}
    \caption{Comparison of top-tagger background rejection performance for fixed signal efficiency ($\epsilon_{S} = 0.3$) as a function of the number of parameters in each model considered, combining data from \tabref{tab1} and \tabref{tab_pelican_model_size}.}
    \label{toptag-params}
    \end{center}
\end{figure}

\Figref{PELICAN-ROC} shows the \textit{receiver operating characteristic}, here represented by the background rejection as a function of the signal efficiency, for the classification performance. In \tabref{tab1} we compare the accuracy, area under the curve (AUC), and background rejection values at $30\%$ signal efficiency between PELICAN and multiple existing ML top-taggers, including the previous state-of-the-art LorentzNet~\cite{LorentzNet22}. We also include two non-ML taggers: the Energy Flow Polynomials~\cite{EFP}, designed to be IRC-safe; and the Boost Invariant Polynomials~\cite{Ortner22}, designed to be partially Lorentz-invariant and used as inputs to XGBoost~\cite{XGBoost}. These two taggers stand out due to their low numbers of parameters and efficient utilization of physical constraints. We trained three PELICAN top-taggers with layers of differing widths, with 208k, 48k, and 11k trainable parameters respectively. The results are averaged over $5$ random initialization seeds, and the uncertainties are given by the standard deviation. The large PELICAN model improves upon the LorentzNet result with a comparable number of parameters, and the medium model roughly matches LorentzNet despite having 5 times fewer parameters. Perhaps most remarkably, the small model with 11k parameters beats every pre-LorentzNet competitor despite having at least several times fewer, and up to 190 times fewer, parameters than other networks. The metrics for these three and a few even smaller models are presented in \tabref{tab_pelican_model_size}, and we visualize this comparison across all models in \figref{toptag-params}.

%===========================================
%===========================================

In addition to different model sizes, we also explore sample efficiency. Each of the three models above was trained on $0.5\%$, $1\%$ and $5\%$ of the training data and compared to the original. For these, the training went on for 70 epochs with 60 epochs of \textsc{CosineAnnealingLR} instead of 28, and 6 epochs of exponential decay instead of 3. The results can be found in \tabref{tab2}. Notice that at lower amounts of training data the differences in performance between models of different width become much less significant, and at $1\%$ and $0.5\%$ of training data all three models fall within each other's uncertainty ranges. These results suggest that the larger PELICAN networks are likely able to learn a greater range of more subtle features from the training data and thus benefit from seeing a larger training dataset. On the other hand, the primary features are already learnable with just a few percent of the data. In particular, with $5\%$ of the training data and only 11k learnable parameters, the $25/15$-channel wide version of the network achieves similar background rejection performance as ResNeXt, which uses 1.46M parameters learning on the full dataset.
%===========================================

%===========================================
%===========================================

\begin{table}[t]
    \centering
    \begin{small}
    \begin{tabular}{l@{\hspace{2mm}}l@{\hspace{2mm}}l@{\hspace{2mm}}l@{\hspace{2mm}}r}
    \toprule
    Width    & \% training data &   Accuracy    &   AUC         &   $1/\epsilon_B\;(\epsilon_S=0.3)$ \\
    \midrule
    $132/78$ & $100\%$  &   0.9425(1)   &   0.9870(1)   & 2250 $\pm$ 75       \\   
                              & $5\%$ &   0.9368(3)   &   0.9843(1)   & 1270 $\pm$ 65 \\   % 547_5p
                              & $1\%$ &   0.9316(6)   &   0.9810(5)   & 789 $\pm$ 49     \\   
                              & $0.5\%$ &   0.9289(11)   &   0.9800(5)   & 633 $\pm$ 28      \\   
    \midrule
    $60/35$  & $100\%$ &   0.9423(1)   &   0.9868(1)   & 2133 $\pm$ 148       \\         % 546_100p
                              & $5\%$ &   0.9368(2)   &   0.9841(1)   & 1148 $\pm$ 49      \\   
                              & $1\%$ &   0.9323(3)   &   0.9813(4)   & 799 $\pm$ 52     \\   
                              & $0.5\%$ &   0.9289(9)   &   0.9795(5)   & 637 $\pm$ 105     \\   
    \midrule
    $25/15$  & $100\%$  &   0.9411(2)   &   0.9863(1)   & 1885 $\pm$ 109      \\           % 545_100p
                              & $5\%$ &   0.9360(2)   &   0.9837(1)   & 1111 $\pm$ 108      \\   
                              & $1\%$ &   0.9316(1)   &   0.9810(5)   & 798 $\pm$ 116       \\   
                              & $0.5\%$ &   0.9286(11)   &   0.9795(6)   & 615 $\pm$ 133      \\   
    \bottomrule
    \end{tabular}
    \end{small}
    \caption{Comparison of PELICAN top-tagging models of different widths (depth is always $L=5$) trained on different fractions of the training data. \label{tab2}}
\end{table}

%========================
% QG TAGGING
%========================
\section{Tagging quark and gluon jets}
\label{qgtagging}
In this section we apply PELICAN to another binary classification task: distinguishing jets produced by gluons from those produced by light quarks. We study three variants of PELICAN for this task: both with and without the use of particle ID labels (PID), as well as a third IRC-safe version. We compare the performance of each of these three variants to other published architectures. We also compare PELICAN models of three different widths and find that even a very small PELICAN model achieves state-of-the-art tagging performance.

%------------------------
% DATASET
%------------------------
\subsection{Quark-gluon jet dataset}

We use the public dataset introduced in ref.~\cite{EFN}. It consists of an equal number of jets produced either by gluons or light quarks $(u,d,s)$. The non-neutrino products are clustered using \textsc{FastJet}~\cite{FastJet} with anti-$k_T$ jet~\cite{Cacciari:2008gp} radius $R=0.4$, and the jets are restricted to $p_T \in [500,550]$ GeV and $|y|<2$. There is no detector simulation, and the particle ID is stored. We select the first 200k events for validation, the following 200k for final testing, and the remaining 1.6M for training. The samples can be downloaded from \cite{QG_Zenodo}.

%===========================================
% \WFclear % see here: https://tex.stackexchange.com/a/442369

%------------------------
% RESULTS
%------------------------

\begin{table}[t]
    \vspace{-0.\intextsep}
    \centering
    \begin{small}
    \begin{tabular}{l@{\hspace{2mm}}l@{\hspace{2mm}}l@{\hspace{2mm}}l@{\hspace{2mm}}l@{\hspace{2mm}}r}
    \toprule
    Architecture                 &   Accuracy    &   AUC   &   $1/\epsilon_B\;(\epsilon_S=0.3)$ & $1/\epsilon_B\;(\epsilon_S=0.5)$ &   \# Params \\
    \midrule
    \textbf{Not IRC-safe, w/ PID}  &&&&&\\
    PFN-ID\cite{EFN}              &   --          &   0.9052(7)   &   --       & 37.4 $\pm$ 0.7 &   82k     \\
    ParticleNet-ID\cite{ParticleNet} &   0.840       &   0.9116       &   98.6 $\pm$ 1.3 &  39.8 $\pm$ 0.2 &   498k   \\
    ABCNet\cite{ABCNet} &   0.840       &   0.9126      & 118.2 $\pm$ 1.5  & -- &   230k    \\
    LorentzNet\cite{LorentzNet22} &   0.844       &   0.9156      & 110.2 $\pm$ 1.3  & 42.4 $\pm$ 0.4&   220k    \\
    $\text{ParT}_{\text{full}}$ \cite{ParT}    &   0.849    &   0.9203   & 129.5 $\pm$ 0.9  &  47.9 $\pm$ 0.5&   2.1M    \\
    % $\text{PELICAN-ID}_{5,25/15}$   &    0.8519(5)  &   0.9212(7) &  131.1  $\pm$ 2.5  &  47.8 $\pm$ 0.5  & 12k  \\   %n=100, qg5
    % $\text{PELICAN-ID}_{5,60/35}$   & 0.8547(2)  &  0.9236(5)  & 133.4 $\pm$ 0.9   &  50.3 $\pm$ 0.8   &   50k  \\  %n=80, qg3
    $\text{PELICAN}_{\text{PID}}$  &  \textbf{0.8551(2)}    &   \textbf{0.9252(1)}   &  \textbf{149.8 $\pm$ 2.4} &  \textbf{52.3 $\pm$ 0.3}  &   209k \\ %n=100, gq17
    \midrule
    \textbf{Not IRC-safe, w/o PID}  &&&&&\\
    PFN\cite{EFN}                 &   --          &   0.8911(8)   &   --       & 30.8 $\pm$ 0.4 &   82k     \\
    ParticleNet\cite{ParticleNet} &   0.828       &   0.9014       &   85.4      &  33.7 &   498k   \\
    $\text{PELICAN}$  &    \textbf{0.8349(4)}  &   \textbf{0.9076(1)}   &  \textbf{94.5 $\pm$ 0.5} &  \textbf{37.5 $\pm$ 0.1}  &   208k     \\    % qg17
    \midrule
    \textbf{IRC-safe}  &&&&&\\
    EFN\cite{EFN}                 &   --          &   0.8824(5)   &   --       & 28.6 $\pm$ 0.3 & 82k        \\
    EFP\cite{EFP}                 &   --          &   0.8919      &   --       & 29.7 & 1k        \\
    EMPN\cite{Spannowsky_IRC_safe}&   --          &   0.8932(6)   &   --       & 30.8 $\pm$ 0.2 & $\sim$110k \\
    $\text{PELICAN}_{\text{IRC}}$  &  \textbf{0.8321(9)}    &   \textbf{0.9056(7)}   &  \textbf{91.7 $\pm$ 0.4}  &  \textbf{36.5 $\pm$ 0.3}  &   9k     \\   % n=100, (qg4)
    \bottomrule
    \end{tabular}
    \end{small}
    \caption{Comparison of different quark-gluon classifiers trained on the quark-gluon tagging dataset \cite{QG_Zenodo}.\label{tab_QG}}
\end{table}

\begin{table}[t]
    \vspace{-0.\intextsep}
    \centering
    \begin{small}
    \begin{tabular}{l@{\hspace{2mm}}l@{\hspace{2mm}}l@{\hspace{2mm}}l@{\hspace{2mm}}l@{\hspace{2mm}}l@{\hspace{2mm}}l@{\hspace{2mm}}r}
    \toprule
    PELICAN variant & $L$ & Width &    Accuracy    &   AUC   &   $1/\epsilon_B\;(\epsilon_S=0.3)$ & $1/\epsilon_B\;(\epsilon_S=0.5)$ &   \# Params \\
    \midrule
    % Not IRC-safe, w/ PID  & 132/78  &  0.8555(2)    &   0.9247(3)   &  134.8 $\pm$ 1.8 &  51.3 $\pm$ 0.7  &   209k \\ %n=100, gq4
    Not IRC-safe, w/ PID  & 5& 132/78  &  \textbf{0.8551(2)}    &   \textbf{0.9252(1)}   &  \textbf{149.8 $\pm$ 2.4} &  \textbf{52.3 $\pm$ 0.3}  &   209k \\ %n=100, gq17
     % & 60/35   & 0.8547(2)  &  0.9236(5)  & 133.4 $\pm$ 0.9   &  50.3 $\pm$ 0.8   &   50k  \\  %n=80, qg3
     % & 60/35   & 0.8538(3)  &  0.9242(2)  & 147.6 $\pm$ 1.4   &  50.7 $\pm$ 0.4   &   50k  \\  %qg12
     &5& 60/35   & 0.8544(5)  &  0.9245(2)  & 148.7 $\pm$ 2.2   &  51.5 $\pm$ 0.2   &   49k  \\  %qg16
     % & 25/15   &    0.8519(5)  &   0.9212(7) &  131.1  $\pm$ 2.5  &  47.8 $\pm$ 0.5  & 12k  \\ %n=100, qg5
     &5& 25/15   &    0.8506(3)  &   0.9216(2) &  134.8  $\pm$ 0.7  &  48.5 $\pm$ 0.4  & 12k  \\ % qg14
    \midrule
%    Not IRC-safe, w/o PID  & 132/78 &    0.8342(2)  &   0.9059(8)   &  88.9 $\pm$ 0.5 &  36.0 $\pm$ 0.2  &   209k     \\    %n=100, qg9
    % Not IRC-safe, w/o PID  &5& 132/78 &   \textbf{0.8347(3)}  &   \textbf{0.9075(1)}   &  \textbf{95.3 $\pm$ 1.0} &  \textbf{37.7 $\pm$ 0.2}  &   208k     \\    %n=100, qg9
    Not IRC-safe, w/o PID  &5& 132/78 &   \textbf{0.8349(4)}  &   \textbf{0.9076(1)}   &  \textbf{94.5 $\pm$ 0.5} &  \textbf{37.5 $\pm$ 0.1}  &   208k     \\    % qg17
     % & 60/35   &    0.8344(2)  &   0.9059(9)   &  89.2  $\pm$ 0.5  &  35.6 $\pm$ 0.4  &   48k \\ %n=100, qg2 (qg2-5-9)
     %& 60/35   &    0.8343(3)  &   0.9073(1)   &  93.9  $\pm$ 0.6  &  37.4 $\pm$ 0.2  &   48k   \\ % qg12
     &5& 60/35   &    0.8344(1)  &   0.9073(1)   &  93.6  $\pm$ 1.1  &  37.4 $\pm$ 0.1  &   48k   \\ % qg16
     % & 25/15   &    0.8324(3)  &   0.9046(5)   &  87.5  $\pm$ 0.7  &  35.0 $\pm$ 0.2  &   11k  \\  %n=120, qg5
     &5& 25/15   &    0.8330(5)  &   0.9063(3)   &  92.8  $\pm$ 0.5  &  36.8 $\pm$ 0.3  &   11k  \\  % qg14
    \midrule
    % IRC-safe  & 132/78  &  0.8299(3)    &   0.8955(18)   &  85.7 $\pm$ 1.2  &  33.8 $\pm$ 0.2  &   204k  \\% n=100, (qg4)
    %IRC-safe  & 132/78  &  0.8292(7)    &   0.9034(3)   &  89.9 $\pm$ 0.8  &  35.7 $\pm$ 0.2  &   204k  \\%qg11
    IRC-safe  & 5&132/78  &  0.8294(3)    &   0.9034(1)   &  90.5 $\pm$ 1.3  &  35.7 $\pm$ 0.2  &   204k  \\%qg17
     % & 60/35   &  0.8293(2)  &  0.8949(24)  &   85.2 $\pm$ 0.7  &  33.5 $\pm$ 0.2   &   48k  \\   %n=100, qg8
     %& 60/35   &  0.8287(4)  &  0.9028(1)  &   89.2 $\pm$ 1.2  &  35.2 $\pm$ 0.2   &   48k  \\   % qg12
     &5& 60/35   &  0.8291(4)  &  0.9030(1)  &   90.5 $\pm$ 1.2  &  35.2 $\pm$ 0.2   &   46k  \\   % qg16
     % & 25/15   &  0.8282(4)  &  0.9005(3)  &   83.4 $\pm$ 0.4 &  33.3 $\pm$ 0.3   &   11k  \\  % n=100, qg5
     &4& 25/15   &  \textbf{0.8321(9)}  &  \textbf{0.9056(7)}  &   \textbf{91.7 $\pm$ 0.4} &  \textbf{36.5 $\pm$ 0.3}   &   9k  \\ % qg13
    \bottomrule
    \end{tabular}
    \end{small}
    \caption{Comparison of PELICAN $q/g$ classifiers of varying widths. The smallest IRC-safe model was found to have a lower optimal depth of $L=4$.\label{tab_QG_size}}
\end{table}

\subsection{Quark-gluon jet tagging results}

The training procedure for the PELICAN quark-gluon tagger is identical to that of the top-tagger described above. The hyperparameters were also chosen to be the same, except here we take up to 100 highest-$p_T$ constituents. As before, we train models of three different shapes and compare their performance. In addition, we train these models both with and without PID information supplied as a set of scalar inputs. Finally, we train IRC-safe PELICAN models so that they can be directly compared to other IRC-safe architectures. For the details of the IRC-safe implementation, see \secref{irc}.

In \tabref{tab_QG} we compare PELICAN to other architectures trained on the quark-gluon dataset. We group the architectures according to whether they use PID inputs, and whether they are IRC-safe, to enable a more direct and fair comparison. Moreover, in \tabref{tab_QG_size} we also compare the performance of PELICAN models of different sizes, analogous to the comparison in \secref{toptagging}. Interestingly, among the IRC-safe models the one with the lowest number of parameters performed the best. This can potentially be explained by the accumulation of floating point errors that we discuss below in \secref{irc}. PELICAN achieves impressive state-of-the-art classification performance with as few as 11k parameters, surpassing architectures of up to 190 times larger sizes. Finally, while \tabref{tab_QG} includes only the non-pre-trained version of ParT to facilitate a direct and fair comparison, it is worth noting that the PELICAN quark-gluon tagger surpasses even the pre-trained ParT tagger as reported in Ref.~\cite{ParT}.

%========================
% QG TAGGING
%========================
\section{Multi-class jet tagging}
\label{hlstagging}

\begin{table}[t]
    \vspace{-0.\intextsep}
    \centering
    \begin{small}
    \begin{tabular}{l@{\hspace{2mm}}l@{\hspace{2mm}}l@{\hspace{2mm}}l@{\hspace{2mm}}l@{\hspace{2mm}}l@{\hspace{2mm}}l@{\hspace{2mm}}r}
    \toprule
    Architecture   & Gluon         &   Light quark & $W$-boson & $Z$-boson & Top quark & \# Params\\
    \midrule
    \multicolumn{7}{c}{\textbf{AUC}} \\
    JEDI-net    &   0.9529     &   0.9301      & 0.9739  & 0.9679 & 0.9683 &   34k    \\
    PCT     &   0.9623    &   0.9414   & 0.9789  &  0.9814 &  0.9757  &  193k          \\
    LorentzNet &   0.9681(3)    &   0.9479(4)   & 0.9837(2)  & 0.9813(3) & 0.9793(3) &   224k    \\
    $\text{PELICAN}$  &   \textbf{0.9693(1)}  &   \textbf{0.9493(1)}   &  \textbf{0.9840(1)} &  \textbf{0.9816(1)} &  \textbf{0.9803(1)} &   208k     \\ 
    \midrule
    \multicolumn{7}{c}{\textbf{TPR at FPR=0.10}} \\
    JEDI-net     &   0.878(1)  &   0.822(1)  & 0.938(1)  & 0.910(1) & 0.930(1)  &   34k  \\
    PCT     &   0.891(1)    &   0.833(1)   & 0.932(1)  &  \textbf{0.946(1)} &  0.941(1)   &  193k        \\
    LorentzNet  &   0.912(1)    &   0.855(1)   & 0.952(1)  & 0.939(1) & 0.949(1)  &   224k   \\
    $\text{PELICAN}$  &   \textbf{0.916(1)}  &   \textbf{0.860(1)}   & \textbf{0.953(1)}  & 0.940(1)  &\textbf{ 0.951(1)}   &   208k   \\ 
    \midrule
    \multicolumn{7}{c}{\textbf{TPR at FPR=0.01}} \\
    JEDI-net      &   0.485(1)  &   0.302(1)  & 0.704(1)  & 0.769(1) & 0.633(1)  &   34k  \\
    PCT    &   0.513(2)    &   0.298(2)   & \textbf{0.834(1)}  &  0.781(1) &  0.700(3)  &  193k       \\
    LorentzNet  &   0.557(4)    &   0.319(2)   &  0.800(3)  & 0.850(3) & 0.753(3)  &   224k  \\
    $\text{PELICAN}$  &   \textbf{0.567(1) } &   \textbf{0.320(1)}   &  0.804(1) & \textbf{0.850(1)}  & \textbf{0.761(1)}   &   208k  \\ 
    \bottomrule
    \end{tabular}
    \end{small} 
    \caption{Three metrics of the receiver-operator curves for each jet category in the HLS4ML Jet dataset (in the ``one-vs-rest'' strategy), compared across architectures: area under the curve and the values of signal efficiencies (true positive rate, or TPR) at background efficiencies (false positive rate, or FPR) of 10\% and 1\%. LorentzNet and PELICAN results are averaged over 5 random initializations and the uncertainties are given by the standard deviation. The results for JEDI-net and PCT were taken from refs.~\cite{JEDI-net,Mikuni21}.\label{tab_hls}}
\end{table}

In this section we consider a final, more advanced jet tagging task. It involves identifying jets that are produced by the decays of five different types of particles: gluons, light quarks, $W$-bosons, $Z$-bosons, and top quarks. We compare the performance of PELICAN models of three different widths to previously published architectures.

\subsection{HLS4ML LHC jet dataset}

The HLS4ML dataset consists of jets produced by simulated $\sqrt{s}=13\,\text{TeV}$ proton-proton collisions using the parameters of a typical LHC detector as detailed in ref.~\cite{Duarte18}. The jets are clustered using the anti-$k_T$ algorithm~\cite{Cacciari:2008gp} with jet radius $R=0.8$, and the jet transverse momentum is required to be around $1\,\text{TeV}$. There are 5 categories of jets: $g,q,W,Z,t$, labeled using a one-hot vector (e.g.~$Z$ is labeled $(0,0,0,1,0)$). We use the version of the dataset that includes up to 100 constituents per jet, which can be downloaded from ref.~\cite{HLS100dataset}. The original dataset includes 16 features per constituent, 4 of them being the Cartesian components of the 4-momentum, and the rest are various representations of the 4-momentum in cylindrical coordinates relative either to the beam axis or the jet axis. In addition, it includes a number of jet-level scalar features. For PELICAN, we need only the 4-momentum of each constituent. This simplified version of the dataset appropriate for our dataloader can be found at ref.~\cite{HLS100datasetConverted}. To avoid any potential bias, the original training set was split into a fully balanced set of 567k jets and an unbalanced validation set of 63k jets (matching the sizes in ref.~\cite{Mikuni21}). The final testing dataset consists of 240k jets.

\begin{table}[t]
    \vspace{-0.\intextsep}
    \centering
    \begin{small}
    \begin{tabular}{l@{\hspace{2mm}}l@{\hspace{2mm}}l@{\hspace{2mm}}l@{\hspace{2mm}}l@{\hspace{2mm}}l@{\hspace{2mm}}l@{\hspace{2mm}}r}
    \toprule
    Width  & Gluon         &   Light quark & $W$-boson & $Z$-boson & Top quark & \# Params\\
    \midrule
    \multicolumn{7}{c}{\textbf{AUC}}\\
    132/78  &   0.9693(1)  &   0.9493(1)   &  0.9840(1) &  0.9816(1) &  0.9803(1) &   208k     \\ 
    60/35  &   0.9683(1)   &   0.9483(1)   & 0.9835(2)  & 0.9811(2)  & 0.9795(2)  &   48k     \\ 
     25/15  &  0.9652(2)   &   0.9441(3)   & 0.9821(2)  & 0.9795(3)  & 0.9769(3)  &   11k     \\ 
    \midrule    
    \multicolumn{7}{c}{\textbf{TPR at FPR=0.10}} \\
    132/78  &   0.916(1)  &   0.860(1)   & 0.953(1)  & 0.940(1)  & 0.951(1)    &   208k  \\ 
     60/35  &   0.912(1)   &   0.856(1)   & 0.952(1)  & 0.938(1)  & 0.949(1)   &   48k \\ 
    25/15  & 0.902(1)  &   0.841(1)   & 0.948(1)  & 0.934(1)  & 0.942(1)       &   11k  \\ 
    \midrule
    \multicolumn{7}{c}{\textbf{TPR at FPR=0.01}} \\
    132/78  &   0.567(1)  &   0.320(1)   &  0.804(1) & 0.850(1)  & 0.761(1)  &   208k  \\ 
    60/35  &   0.562(1)   &   0.316(2)   & 0.801(2)  & 0.849(1)  & 0.750(2)  &   48k \\ 
    25/15  &   0.538(2)   & 0.307(3)  & 0.794(3)  & 0.840(2) & 0.719(2)      &   11k \\ 
    \bottomrule
    \end{tabular}
    \end{small} 
    \caption{Comparison of three PELICAN models of different widths on the multi-class jet tagging task. The width for PELICAN is defined as in~\tabref{tab_pelican_model_size}. \label{tab_hls_sizes}}
\end{table}

\subsection{Multi-class jet tagging results}

We use the same PELICAN classifier architecture as above, but this time producing 5 classification scores. All other hyperparameters and the training procedure were left unchanged. In addition, we trained LorentzNet \cite{LorentzNet22} on this dataset without modifying the default hyperparameters (other than the number of target classes) or the training procedure. In \tabref{tab_hls} we compare the receiver-operator curves of different architectures using three metrics: area under the curve (AUC), and the signal efficiencies (a.k.a.~TPR -- true positive rate) at background efficiencies (FPR -- false positive rate) of 10\% and 1\%. On net, both PELICAN and LorentzNet (the only two Lorentz-invariant architectures applied to this task) provide a significant improvement over older architectures, yet PELICAN surpasses even LorentzNet despite its similar model size.

Finally, we trained three PELICAN models of different widths and compared their performance in \tabref{tab_hls_sizes}. Even the small PELICAN model with only 11k parameters achieves state-of-the-art performance in 4 out of the 5 jet categories. Meanwhile the medium model with 48k parameters is on par with the much larger LorentzNet. We see similar (small) changes in performance as the model size is changed for this task compared to the other tasks described above.

%========================
% REGRESSION
%========================
\section{$W$-boson 4-momentum reconstruction}
\label{Wreco}
%\paragraph{Task} 

To test the equivariant regression architecture described in \secref{architecture} we chose a task where the aim is to reconstruct (or \textit{predict}) the full 4-momentum of the $W$-boson within the Lorentz-boosted top-quark decay products. Specifically, we consider the same hadronic top-quark decay process that constitutes the signal in the top-tagging dataset, which uses the $t\to bW\to bqq$ two-step decay, followed by hadronization, showering, and detection. Our aim is to reconstruct the true 4-momentum of the $W$-boson given the full set of observed final state particles of the top-quark decay, as represented by the jet constituents. The work most closely related to this task is the transformer-based reconstruction of the top quark momentum in ref.~\cite{CPT}, and we employ similar evaluation criteria for our task.

%------------------------
% DATASET
%------------------------
\subsection{Regression dataset}

The dataset used for the regression task consists of 1.5M $t\bar{t}$ events simulated with \PythiaEight, via the HEPData4ML package~\cite{Offermann_HEPData4ML}, consisting of 700k events for training, 200k events for validation, and 500k events for testing (with an additional 100k events set aside in a second testing set). From each event, we cluster anti-$k_T$ jets with $R=0.8$ using \Fastjet and we select the jet nearest to the truth-level top quark in $\left( \eta,\phi \right)$, requiring the distance between the top quark and the jet to satisfy $\Delta R \left(\mathrm{top\, quark},\mathrm{jet} \right) < 0.8$. This jet clustering is done both at truth-level, and using calorimeter tower objects produced by running the event through \Delphes fast detector simulation using the ATLAS detector card. Thus, each \textit{event} in the dataset corresponds to a single jet, and includes information for truth-level particles such as the truth-level top quark -- we may therefore use the terms \textit{jet} and \textit{event} interchangeably below with the understanding that each ``event'' in this dataset has one and only one jet recorded.
This dataset is fully reproducible since it contains the full set of parameters used for its own generation via the HEPData4ML package. It is publicly available via Zenodo~\cite{btW6}, where a full description of the various data fields is provided. Here we provide only an overview of some key features:
%===========================================
\begin{enumerate}
    \item There are two versions of the dataset, corresponding with truth- and reconstruction-level (\Delphes) jets. The events are the same between versions, so the two can be compared event-by-event to study the effects of detector reconstruction on network training and performance.
    \item The input data for the network are the 4-momenta of the $200$ leading jet constituents. For use as possible regression targets and for defining jet containment (explained below), each event contains
    \begin{enumerate}
        \item the truth-level top quark that initiated the jet,
        \item the bottom quark from top-quark decay,
        \item the $W$-boson from top-quark decay,
        \item the two quarks from subsequent $W$-boson decay ($W\rightarrow q q'$),
    \end{enumerate}
    In addition, the event contains the stable $W$-boson daughter particles. These are the truth-level, final state particles that are traced back to the $W$-boson by \Pythia.
    \item Each jet is tagged with the Johns Hopkins top tagger~\cite{Kaplan:2008ie} (JH), as implemented in \Fastjet. This allows us to define a subpopulation of JH-tagged events, which we shall sometimes refer to as \textit{JH events}. For jets that it tags as top-quark jets, JH reconstructs a $W$-boson candidate from subjets.
    \item Each jet is also tagged as whether or not it is \textit{fully-contained} (FC). We define FC events as those where the $b$-quark, as well as the two quarks from $W\rightarrow q q'$ decay, are within $\Delta R < 0.8$ of the jet centroid (i.e.~within the jet radius). In such cases almost all of the $W$-daughters are contained within the jet and we can expect a good reconstruction of the $W$ momentum. FC events comprise $75\%$ of the dataset.
\end{enumerate}
%===========================================

%------------------------
% TRAINING PROCEDURE
%------------------------
\subsection{Regression training procedure}

Our model has 4 equivariant $\mathrm{Eq}_{2\to 2}$ blocks. Each messaging layer takes in 132 channels and outputs 78 channels. Conversely, each equivariant aggregation layer has 78 input channels and outputs 132 channels. The $\mathrm{Eq}_{2\to 1}$ block has the same shape, and the final fully-connected layer has the shape $1\times 132$. There are 210k parameters in total. Assuming $N$ non-zero input jet constituents, this produces $N$ scalar coefficients $c_i$ with zero-padding, which are the Lorentz invariants as described by \equref{equiv}. The reconstructed 4-momentum is then computed via 
\[
    p_{\mathrm{reco}}=\sum_i c_i p_i \label{def pelican weights}.
\]
The training regimen for this task is essentially identical to the one for top-tagging: \textsc{AdamW} optimizer with weight decay of $0.01$, 35 epochs in total with 4 epochs of warm-up and exponential learning rate decay for the last 3 epochs. All matching hyperparameters were copied from the top-tagging model with no extra optimization. The main difference is in the choice of the loss function $L(p_{\mathrm{reco}},p_{\mathrm{target}})$. Spacetime geometry allows for many choices of this function, which in turn will affect the shape of the landscape near $p_{\mathrm{target}}$ and in turn the precision of various reconstructed features of the vector, such as the mass, energy, spatial momentum, transverse momentum, and direction. It is even possible to construct Lorentz-invariant loss functions to make the training process itself equivariant. Nevertheless, for the purpose of simultaneous reconstruction of the direction and the mass $m_W$ of the $W$-boson, we found 
%------------------------
\begin{equation}
    \text{Loss}(p_{\text{reco}},p_{\text{target}})=0.01\Vert\boldsymbol{p}_{\mathrm{reco}}-\boldsymbol{p}_{\mathrm{target}}\rVert+0.05|m_{\mathrm{reco}}-m_{\mathrm{target}}| \label{loss-regression}
\end{equation}
to be very effective. It uses all 4 components of the target vector and strikes a good balance between the precision of the reconstructed mass and spatial momentum. The coefficients are chosen such that each term in the loss function is order unity for our datasets.
%------------------------

Aside from the loss function, another rarely discussed feature of this task is the choice of the target vector $p_{\mathrm{target}}$. Even though our ultimate inference target is the true $W$ momentum $p^W_{\mathrm{true}}$, it is not necessarily the best training target given the nature of the dataset. Detection and jet clustering include multiple energy, momentum, and spatial cuts that exclude some decay products from the final jet. For instance, one of the three quarks in $t\to bqq$ might fall outside of the $R=0.8$ radius of the jet clustering algorithm, in which case most of the decay products of that quark are likely to be absent from the event record. If many of the decay products of the $W$-boson are missing, then we lack the information necessary to make an accurate estimate of its true momentum, or even to identify which of the jet constituents belong to the $W$-boson. This effect is often referred to as an \textit{acceptance} issue due to the finite purview of the final state reconstruction. %This is best seen from the mass spectrum of this target, shown in \figref{targetm}.

%===========================================
\begin{figure}[t]
    \vspace{-0\intextsep}
    \begin{center}
    \includegraphics[width=\linewidth]{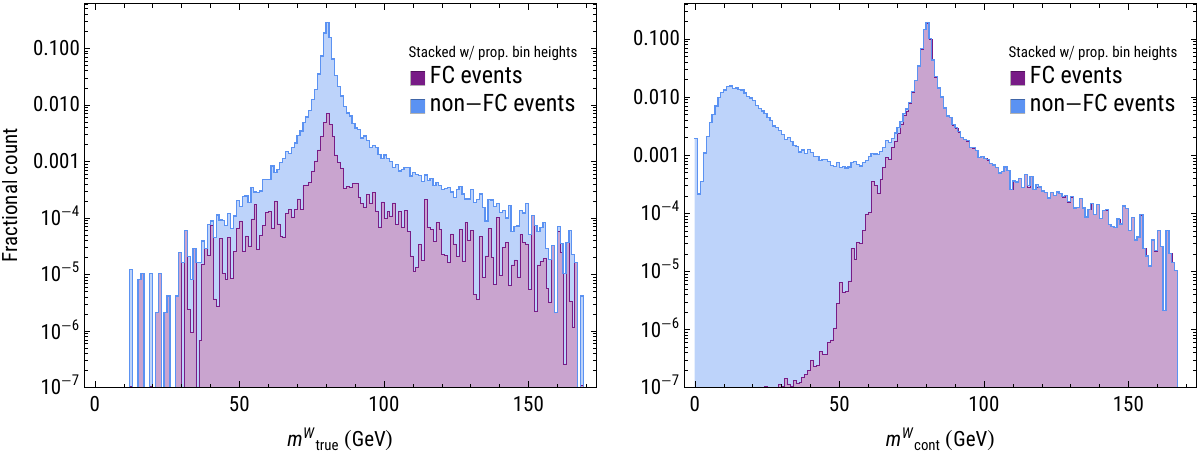}
    \caption{Stacked histogram \textit{with proportional bin heights} showing the mass spectrum of the two targets, the true $W$ momentum $p^W_{\mathrm{true}}$, and the contained true $W$ momentum $p^W_{\mathrm{cont}}$. The left figure shows a stacked histogram of the true $W$ mass spectrum comprised of the FC events (purple) and non-FC events (blue). The right figure shows a stacked histogram of mass spectrum for the contained $W$ mass, similarly comprised of both FC events (purple) and non-FC events (blue). In each case, the bin contents are scaled linearly relative to the total number of events, i.e.~the fraction of FC events in a given bin is given by the apparent height of the FC curve divided by the total height of the bin (heights are measured from the $x$-axis). The two mass spectra of FC events, in fact, match.}
    \label{targetm}
    \end{center}
\end{figure}
%===========================================

To alleviate this issue and provide better control over the inference stage, we propose an alternative target 4-vector that we call the \textit{contained true $W$ momentum} $p^W_{\mathrm{cont}}$, equal to the total 4-momentum of the \textit{truth-level $W$ decay products} that fall within the radius of the final reconstructed top jet. In the truth-level dataset, this is simply $p^W_{\mathrm{cont}}=\sum_k p_{i_k}$ where $i_k$ are the indices of the constituents whose parent is the $W$-boson and not the $b$-quark. In the \Delphes dataset, however, there is no simple analytic relationship between $p^W_{\mathrm{cont}}$ and the jet constituents $p_i$. That is to say that the mapping of the truth-level information to the detector-level reconstruction is highly nonlinear. Nonetheless, in either dataset this vector more accurately reflects the available information about the $W$-boson and allows us to make inferences not only about the $W$-boson itself, but also about the containment qualities of the event. This will be discussed further in \secref{disc-Wreco} below.
For reference, the true mass spectra of both $p^W_{\mathrm{true}}$ and $p^W_{\mathrm{cont}}$ are shown in \figref{targetm}. For fully-contained (FC) events, the mass spectra are similar between the true and the contained $W$ mass as expected. Non-FC events are mostly confined to a clear second peak at 13 GeV corresponding to $qb$ and $q$ jets (where one of the quarks from $W\to qq$ fell outside the jet), and a minor peak at $m^W_{\mathrm{cont}}=0$ corresponding to $b$ jets.

Given the above observations, we prepared two PELICAN models, one trained to reconstruct $p^W_{\mathrm{true}}$, and another trained to reconstruct $p^W_{\mathrm{cont}}$. Otherwise the two models are identical and are trained in the same way and with the same loss function. We then compare the outputs of each model to $p^W_{\mathrm{true}}$ and analyze the benefits of the two choices of the target.

%------------------------
% RESULTS
%------------------------
\subsection{Regression results for $p^W_{\mathrm{true}}$ reconstruction}

\renewcommand{\arraystretch}{1.05} % to get the vertical text to not collide with dividers, we slightly stretch the table vertically
%===========================================
\begin{table}[t]
    % \vspace{-1\intextsep}
    \centering
    \begin{small}
        \begin{tabular}{ccS[table-format=3.2]<{\%}S[table-format=3.2]<{\%}S[table-format=3.3]}
        \toprule
        & Method &  \multicolumn{1}{c}{$\sigma_{p_T}$ (\%)} & \multicolumn{1}{c}{$\sigma_{m}$ (\%)} & \multicolumn{1}{c}{$\sigma_{\Delta R}$ (centirad)}\\
        \midrule
        \multirow{3}{*}{\rotatebox[origin=c]{90}{\parbox{1.3cm}{\centering Without\\ \Delphes}}} 
        & JH               & 0.66    & 1.26     & 0.216   \\
        & PELICAN$\mid$JH  & 0.26    & 0.57     & 0.113   \\
        & PELICAN$\mid$FC  & 0.30    & 0.71     & 0.139   \\
        & PELICAN          & 0.79    & 1.12     & 0.473   \\
        \midrule
        \multirow{3}{*}{\rotatebox[origin=c]{90}{\parbox{1.3cm}{\centering With\\ \Delphes}}} 
        & JH               & 9.8   & 8.3   & 9.6      \\
        & PELICAN$\mid$JH  & 3.5    & 2.6   & 2.8      \\
        & PELICAN$\mid$FC  & 4.0    & 2.9   & 3.1      \\
        & PELICAN          & 5.1    & 3.0   & 4.7      \\
        \bottomrule
        \end{tabular}
    \caption{Momentum reconstruction results for JH and PELICAN trained to reconstruct $p^W_{\mathrm{true}}$. We report the relative $p_T$ and mass resolutions, and the interquantile range for the angle $\Delta R$ between predicted and true momenta. "PELICAN$\mid$JH" refers to PELICAN evaluated only on JH-tagged jets, and "PELICAN$\mid$FC" to PELICAN evaluated only on FC events. PELICAN uncertainties are within the last significant digit.\label{btW6_W_table}}

    % Momentum reconstruction results for the Johns Hopkins (JH) tagger and PELICAN for relative $p_T$ resolution ($\sigma_{p_T}$) and relative mass resolution ($\sigma_{m}$). Uncertainties are given by the central 68th interquantile range (IQR), except for $\psi$ where it is the lower 68th IQR. $\psi$ is in radians.}
    \end{small}
    \vspace{-0.5\intextsep}
\end{table}
%===========================================

The results are summarized in \tabref{btW6_W_table}. We quantify the precision of the reconstruction using the resolution of transverse momentum \pT\footnote{$\pt=\sqrt{p_x^2+p_y^2}$.} and mass as a metric, given by half of the central $68^\text{th}$ interquantile range of $(x_{\mathrm{predict}}-x_{\mathrm{true}})/x_{\mathrm{true}}$, where $x$ is $m$ or $p_T$. In addition we report the lower $68^\text{th}$ interquantile range for $\Delta R$, the $z$-boost-invariant spatial angle between predicted and true momenta\footnote{$\Delta R=\sqrt{(\Delta \phi)^2+(\Delta \ln \tan \theta/2)^2}$.}. 

Since there are no pre-existing ML-based methods for this task, we use the $W$-boson identification of JH for the baseline comparison. JH has a $36\%$ efficiency on the truth-level dataset and $31\%$ on the \Delphes one. It can only identify $W$-boson candidates for jets it tags, so we report PELICAN results both on the JH-tagged jets only (PELICAN$\mid$JH) and on the full dataset (PELICAN). Moreover, we evaluate PELICAN on the population of FC events (PELICAN$\mid$FC). More than $99.9\%$ of JH-tagged events contain all three true quarks $bqq$ within the jet radius, so this population represents an especially restricted and ``ideal'' type of event. The results were evaluated over 5 training runs initialized with different random seeds, and the resolutions reported in \tabref{btW6_W_table} are consistent across the runs.

There are significant differences in PELICAN's performance on the different sub-populations of events. In the direct comparison with the JH tagger, PELICAN$|$JH is 2-4 times more precise. However, even on the much larger class of FC events, PELICAN produces predictions with almost the same precision. The highest loss of precision happens on non-FC events where many of the $W$ decay products are missing from the jet, leading to lower average precision on the entire dataset. As discussed in \secref{Wmass}, this result can be \textit{explained} by interrogating the PELICAN weights and kinematic information directly.

%===========================================
\begin{figure}[t]
    % \vspace{-2\intextsep}
    \begin{center}
    \includegraphics[width=\linewidth]{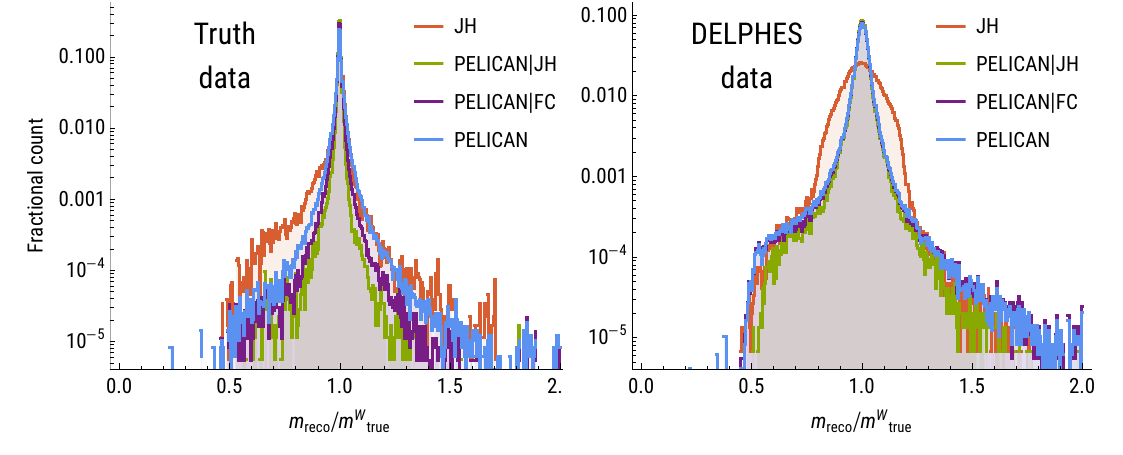}
    \caption{Reconstructed $W$ mass relative to true $W$ mass for the PELICAN model trained on truth (left) or \Delphes (right) data, and targeting $p^W_{\mathrm{true}}$.}
    \label{btW6WW_res_m}
    \end{center}
\end{figure}
%===========================================
In \figref{btW6WW_res_m} we show the relative reconstructed $W$ masses for two of the models, one trained on truth data, and one on \Delphes data. The results also include the curve for the JH tagger's reconstruction, as well as PELICAN$\mid$JH and PELICAN$\mid$FC. The $68^\text{th}$ interquantile ranges of these curves match the numbers in the $\sigma_m$ column of \tabref{btW6_W_table}. See \secref{Weights} for further details on the causes of performance degradation in the \Delphes case. For the complete set of results see \appref{appendix_plots}.

%===========================================
\begin{figure}[t]
    % \vspace{-2\intextsep}
    \begin{center}
    \includegraphics[width=\linewidth]{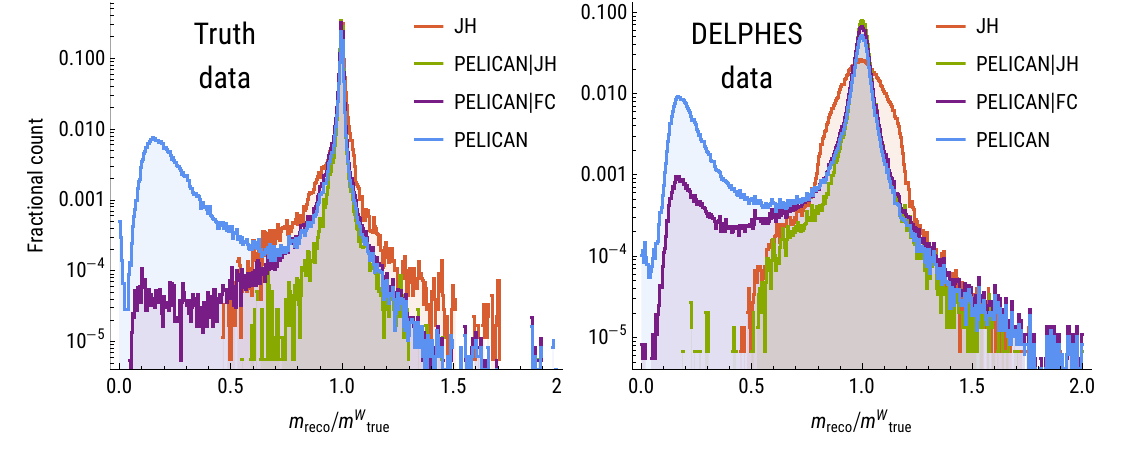}
    \caption{Reconstructed $W$ mass relative to true $W$ mass for the PELICAN model trained (on truth or \Delphes data) targeting $p^W_{\mathrm{true}}$.}
    \label{btW6DW_res_m}
    \end{center}
\end{figure}
%===========================================

%------------------------
% RESULTS
%------------------------
\subsection{Regression results for $p^W_{\mathrm{cont}}$ reconstruction}

Now we train new models with the target vector set to the contained true $W$ momentum $p^W_{\mathrm{cont}}$, evaluate their precision by comparing the outputs to the true $W$ momentum $p^W_{\mathrm{true}}$, and compare the results to \tabref{btW6_W_table}. As shown in \tabref{btW6_D_table}, the resolutions for these models on JH-tagged and FC events are slightly worse than the first set of models, in the \Delphes case by 5-15\%. The largest change is in non-FC events, leading to poor average resolutions on the whole dataset. Despite this, as we will now show, these models can in fact be better suited for real-world applications.

%===========================================
\begin{table}[t]
    % \vspace{-1\intextsep}
    \centering
    \begin{small}
        \begin{tabular}{ccS[table-format=3.2]<{\%}S[table-format=3.2]<{\%}S[table-format=3.3]}
        \toprule
        & Method &  \multicolumn{1}{c}{$\sigma_{p_T}$ (\%)} & \multicolumn{1}{c}{$\sigma_{m}$ (\%)} & \multicolumn{1}{c}{$\sigma_{\Delta R}$ (centirad)}\\
        \midrule
        \multirow{3}{*}{\rotatebox[origin=c]{90}{\parbox{1.3cm}{\centering Without\\ \Delphes}}} 
        & JH               & 0.66    & 1.26     & 0.216   \\
        & PELICAN$\mid$JH  & 0.27    & 0.62     & 0.113   \\
        & PELICAN$\mid$FC  & 0.34    & 0.86     & 0.142   \\
        & PELICAN          & 2.37    & 38.93    & 0.681   \\
        \midrule
        \multirow{3}{*}{\rotatebox[origin=c]{90}{\parbox{1.3cm}{\centering With\\ \Delphes}}} 
        & JH               & 9.8    & 8.3   & 9.6      \\
        & PELICAN$\mid$JH  & 3.6   & 2.8  & 3.1      \\
        & PELICAN$\mid$FC  & 4.2    & 3.6   & 3.4      \\
        & PELICAN          & 6.2    & 39.6   & 5.6      \\
        \bottomrule
        \end{tabular}
    \end{small}
    \caption{PELICAN resolutions for models trained to reconstruct $p^W_{\mathrm{cont}}$. Resolutions are still obtained by comparing the model predictions to $p^W_{\mathrm{true}}$.\label{btW6_D_table}}
    \vspace{-0.5\intextsep}
\end{table}
%===========================================

%------------------------
% DISCUSSION
%------------------------
\subsection{Discussion \label{disc-Wreco}}

To see the main benefit of this model, we present the behavior of the relative reconstructed mass shown in \figref{btW6DW_res_m}. PELICAN-reconstructed masses within the range of true $W$ masses are almost as precise on the full dataset as they are on FC events (see \figref{btW6DW_res_m} near the peak at 1). The most prominent feature obvious from these results is that, despite the slightly lower accuracies on FC events (at fixed width and depth of the network), the model trained to reconstruct $p^W_{\mathrm{cont}}$ accurately reproduces the mass spectrum of $m^W_{\mathrm{cont}}$ in \figref{targetm} and therefore discriminates between FC and non-FC events, allowing us to perform post-inference event selections. 

For instance, in the \Delphes case, choosing a 55 GeV cutoff, $97\%$ of all FC events have $m_{\mathrm{reco}}>55 \text{ GeV}$, and vice versa, $97\%$ of all events with $m_{\mathrm{reco}}>55\text{ GeV}$ are FC. In this manner we can significantly improve the accuracy of the reconstruction without accessing truth-level information that is needed to identify FC events. Notably, this so called ``parton labeling'' via regression networks was recently studied in ref.~\cite{CPT23}. This filtering of events comes at the cost of a modest reduction in signal efficiency -- from the ostensible $100\%$ down to $75\%$. Note that in the \Delphes case, the set of FC events is contaminated with a small number of events with significant losses of $W$ decay products due to detector effects, but it can be refined by reducing the jet radius used in the definition of full containment. Consequently, we propose the following simple routine for real-world applications of these models. First, use the model trained targeting $p^W_{\mathrm{cont}}$ as an FC-tagger to refine the data. Then, apply the model targeting $p^W_{\mathrm{true}}$ to reconstruct the $W$-boson.

We conclude that $p^W_{\mathrm{cont}}$ is the better target for many common reconstruction tasks where one is willing to sacrifice some signal efficiency -- or to only fully measure the 4-momentum on a sub-sample of the identified events -- to gain improved accuracy. In the following sections we will not present models trained on both targets, however a complete set of metrics and figures can be found in \appref{appendix_plots}.

% \begin{figure}[h]
%     \vspace{-0\intextsep}
%     \begin{center}
%     \includegraphics[width=0.85\linewidth]{figures/m_corr/btW6DD_m_corr.pdf}
%     \caption{2D histograms of target vs.~reconstructed masses for models (top: truth data; bottom: DELPHES data) trained targeting $p^W_{\mathrm{cont}}$, broken up into two populations based on containment (left: non-FC events; right: FC events).}
%     \label{m_corr_DW}
%     \end{center}
% \end{figure}

%========================
% MASS MEASUREMENT
%========================
\section{$W$-boson mass measurement}
\label{Wmass}
As we saw above, PELICAN is able to reconstruct the mass of the $W$-boson, $m_W$, found within the dense environment of the complete decay products of a top quark jet. For truth-level datasets, the resolution of this reconstruction is below the natural width ~\cite{ParticleDataGroup:2022pth} of the mass spectrum, $\Gamma_W/m_W\approx 2.59\%$. In the \Delphes case, the resolution is too wide to produce any substantial correlation between the true and reconstructed masses (see \appref{appendix_plots} for figures that demonstrate this). We would like to eliminate the possibility that the reason that the true masses are highly concentrated around $80$ GeV is due in part to the potential for PELICAN to effectively \textit{memorize} a single number: the $W$ mass. In this section we examine a more realistic reconstruction task, where the true mass of the target particle is unknown, and the dataset uniformly covers a wide range of its masses.

The reconstruction task is still identical to that of \secref{Wreco}. Even though we could use an invariant scalar-valued version of PELICAN to target the mass of the $W$-boson, the accuracy of that reconstruction would in fact suffer in comparison with the equivariant 4-vector-valued model. This is simply due to the fact that the 4-momentum contains more relevant information than the mass alone, since the direction and the energy of the particle are, in part, correlated with the mass. Thus the only new element in this experiment will be the dataset, which will now involve $W$-bosons of varying masses uniformly covering the range $m_W\in [64,96]$~GeV.
%
%===========================================
\begin{figure}[t]
    \vspace{-0\intextsep}
    \begin{center}
    \includegraphics[width=\linewidth]{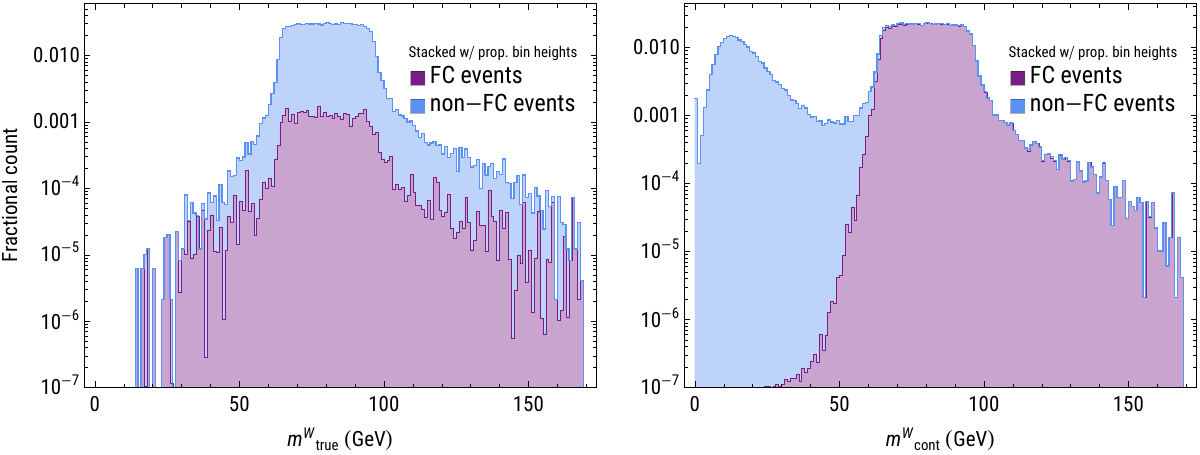}
    \caption{Stacked histogram with proportional bin heights (see description in \figref{targetm}) showing the mass spectrum of the two targets, $p^W_{\mathrm{true}}$ and $p^W_{\mathrm{cont}}$, in the variable $W$ mass dataset.}
    \label{m_targetm}
    \end{center}
\end{figure}
%===========================================
%
%------------------------
% DATASET
%------------------------
%\subsection{Regression dataset} 
%
The dataset is also identical to that used in \secref{Wreco}, including the number of events, except that the $W$ mass is set to be variable. This is achieved by combining multiple independently-produced datasets where the generator-level value of $m_W$ was modified from its default value. \Figref{m_targetm} shows the resulting distribution of $W$ masses, as well as that of the sum of $W$-daughters contained within each jet.

%===========================================
\begin{table}[t]
    % \vspace{-1\intextsep}
    \centering
    \begin{small}
        \begin{tabular}{ccS[table-format=3.2]<{\%}S[table-format=3.2]<{\%}S[table-format=3.3]}
        \toprule
        & Method &  \multicolumn{1}{c}{$\sigma_{p^T}$ (\%)} & \multicolumn{1}{c}{$\sigma_{m}$ (\%)} & \multicolumn{1}{c}{$\sigma_{\Delta R}$ (centirad)}\\
        \midrule
        \multirow{3}{*}{\rotatebox[origin=c]{90}{\parbox{1.3cm}{\centering Without\\ \Delphes}}} 
        & JH               & 7.98    & 4.75     & 22.180   \\
        & PELICAN$\mid$JH  & 0.27    & 0.63     & 0.111   \\
        & PELICAN$\mid$FC  & 0.35    & 0.89     & 0.143   \\
        & PELICAN          & 2.64    & 39.00    & 0.744   \\
        \midrule
        \multirow{3}{*}{\rotatebox[origin=c]{90}{\parbox{1.3cm}{\centering With\\ \Delphes}}} 
        & JH               & 16.0    & 12.0   & 25.4      \\
        & PELICAN$\mid$JH  & 4.2     & 6.5  & 3.4      \\
        & PELICAN$\mid$FC  & 4.9     & 8.0   & 3.8      \\
        & PELICAN          & 7.3     & 40.7   & 6.7      \\
        \bottomrule
        \end{tabular}
    \end{small}
    \caption{PELICAN resolutions for models trained to reconstruct $p^W_{\mathrm{cont}}$ with variable $m_W$. Resolutions are still obtained by comparing the model predictions to $p^W_{\mathrm{true}}$.\label{btW6m_D_table}}
    \vspace{-0.5\intextsep}
\end{table}
%===========================================

%------------------------
% RESULTS
%------------------------
\subsection{Regression results for $m_W$ reconstruction}
%\paragraph{Results} 

\begin{figure}[t]
    \vspace{-0\intextsep}
    \begin{center}
    \includegraphics[width=0.85\linewidth]{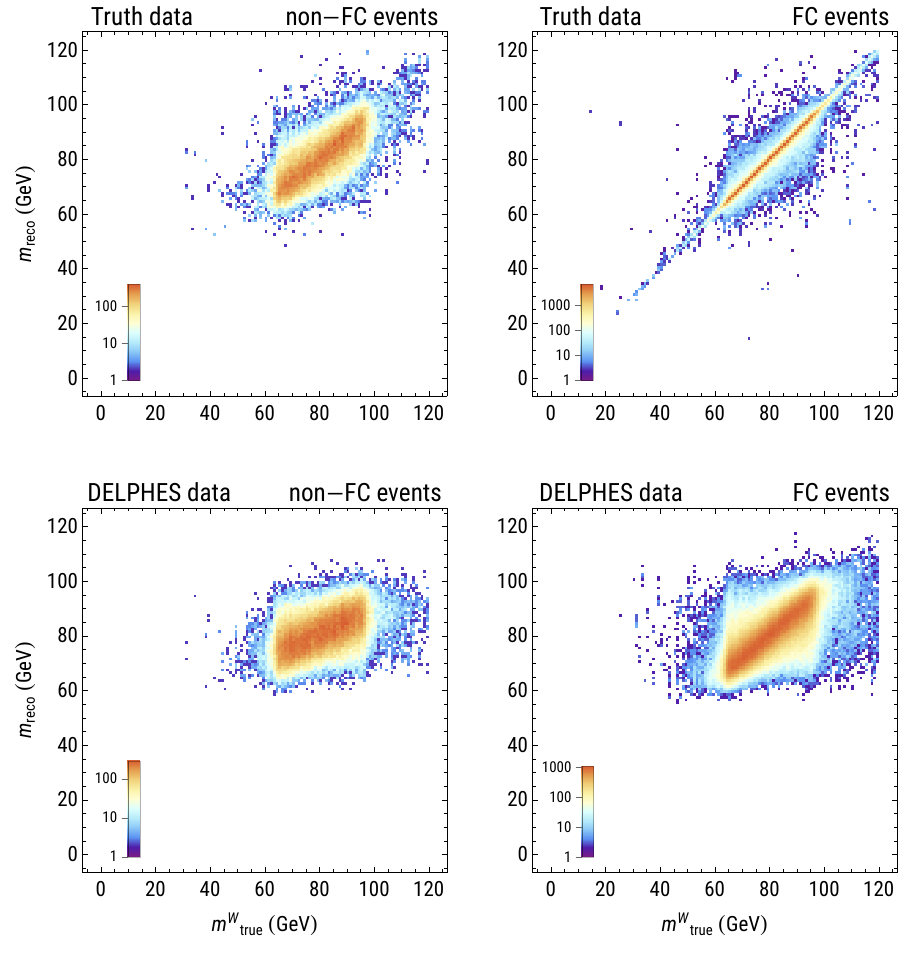}
    \caption{2D histograms of true vs.~reconstructed masses for models trained on the variable mass dataset targeting $p^W_{\mathrm{true}}$ (top: truth data; bottom: \Delphes data), broken up into two populations based on jet containment (left: non-FC events; right: FC events). Top right has a correlation of 88\%, and bottom right 35\%.}
    \label{m_corr_mWW}
    \end{center}
\end{figure}

The hyperparameters and the training regime used here are the same as in \secref{Wreco}. Here we focus on the model trained to reconstruct the contained momentum $p^W_{\mathrm{cont}}$ (see \appref{appendix_plots} to find the results for the model targeting $p^W_{\mathrm{true}}$). The outputs are then compared to the true $W$-boson $p^W_{\mathrm{true}}$. The accuracies for the full 4-vector reconstruction are presented in \tabref{btW6m_D_table}. The largest loss of accuracy relative to \secref{Wreco} is, unsurprisingly, in the mass column. However, since the true mass now covers a much wider range, while the number of training samples remained the same, this still presents a significant improvement in the mass reconstruction capability. To demonstrate this better, we show the 2D correlations between target and reconstructed masses in \figsref{m_corr_mWW}{m_corr_mDD} for the models trained targeting $p^W_{\mathrm{true}}$ and $p^W_{\mathrm{cont}}$, respectively. We also differentiate between non-FC (left) and FC (right) events in the two sides of each of the panels in each figure. 

%===========================================
%===========================================

%===========================================
\begin{figure}[t]
    \vspace{-0\intextsep}
    \begin{center}
    \includegraphics[width=0.85\linewidth]{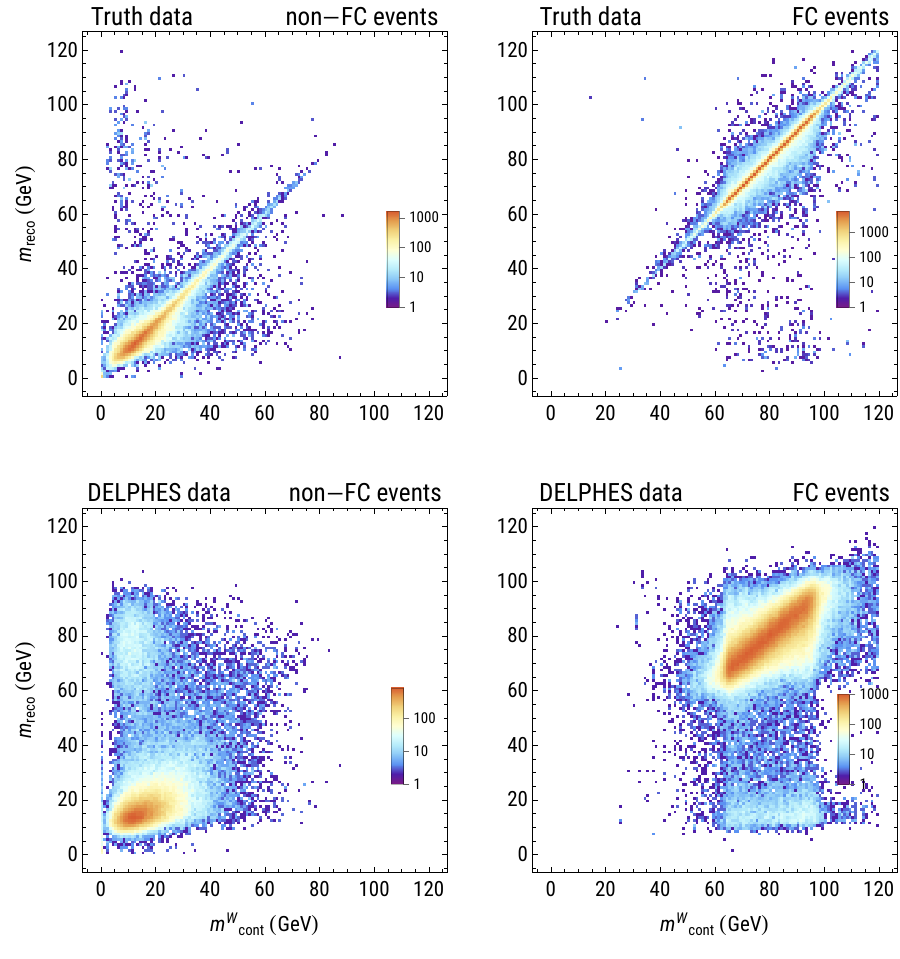}
    \caption{2D histograms of target vs.~reconstructed masses for models trained targeting $p^W_{\mathrm{cont}}$ (top: truth data; bottom: \Delphes data), broken up into two populations based on jet containment (left: non-FC events; right: FC events). Top right has a correlation of 95\%, and bottom right 65\%.}
    \label{m_corr_mDD}
    \end{center}
\end{figure}
%===========================================

%------------------------
% MODEL COMPLEXITY
%------------------------
\subsection{Model complexity}

The model examined above has 210k trainable parameters, however even significantly smaller models achieve good accuracy. As an illustration, we compare the resolutions of three PELICAN models trained on the variable mass dataset targeting $p^W_{\mathrm{true}}$. They are obtained from the original model by a proportional rescaling of the widths of all layers. The first model is the 210k parameter one, with 132/78 channels, i.e.\ each messaging layer has 132 input and 78 output channels. The second model has 60/35 channels and 49k parameters. The third model has 25/15 channels and 11k parameters. The resolutions over the \Delphes test dataset are reported in \tabref{model_size_table}, and we observe that even the 11k-parameter model handily beats the JH method.

%===========================================
\begin{table}[t]
    % \vspace{-1\intextsep}
    \centering
    \begin{small}
        \begin{tabular}{c *{3}{S[table-format=3.1]<{\%}} c}
        \toprule
        PELICAN width &  \multicolumn{1}{c}{$\sigma_{p_T}$ (\%)} & \multicolumn{1}{c}{$\sigma_{m}$ (\%)} & \multicolumn{1}{c}{$\sigma_{\Delta R}$ (centirad)} & \# Params\\
        \midrule
        132/78 & 6.1    & 8.2   & 2.8  & 210k    \\
        60/35 & 6.5     & 8.6  & 3.2   & 49k     \\
        25/15 & 7.4     & 9.5   & 3.8  & 11k    \\       
        \bottomrule
        \end{tabular}
    \end{small}
    \caption{Comparison of PELICAN models of three different widths trained to reconstruct $p^W_{\mathrm{true}}$ with variable $W$ mass. Width is defined as in \tabref{tab_pelican_model_size}. Trained and tested on \Delphes data.\label{model_size_table}}
    \vspace{-0.5\intextsep}
\end{table}
%===========================================

%------------------------
% DISCUSSION
%------------------------
\subsection{Discussion}

\begin{figure}[t]
    \begin{center}
    \includegraphics[width=\linewidth]{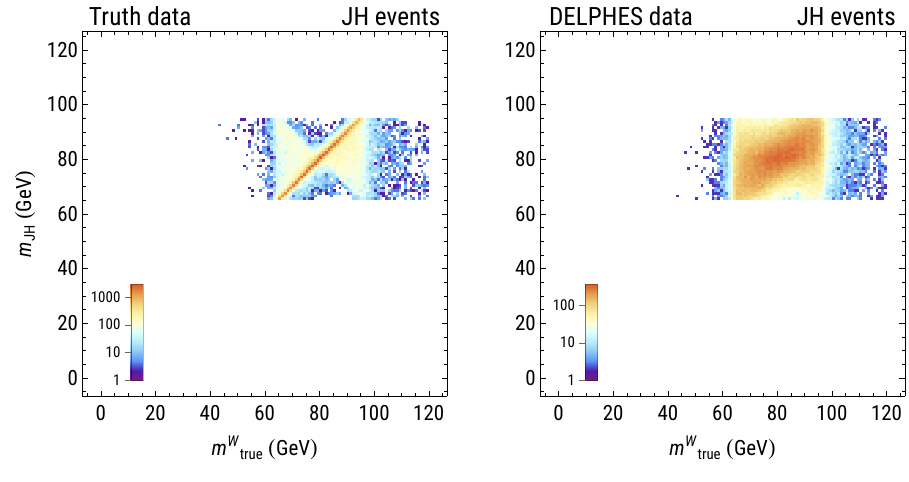}
    \caption{JH tagger's reconstruction of the $W$ mass on the variable $W$ mass dataset (only JH-tagged events), truth-level and \Delphes versions. The correlation values are $47\%$ and $25\%$, correspondingly.}
    \label{m_corr_mJH}
    \end{center}
\end{figure}

In the \Delphes dataset, we observe that for non-FC events (bottom left pane of \figref{m_corr_mDD}), the reconstructed contained mass is only weakly correlated with the true contained mass (or with the true $W$ mass, as shown in \figref{m_corr_DW} in \appref{appendix_plots}). However, in the quadrant where both masses exceed $55$ GeV, we find a $65\%$ correlation on FC events in the \Delphes case. The most important type of error PELICAN makes here is when a non-FC event gets assigned a high reconstructed mass, that is a mass near that of the true $W$-boson was assigned to a jet with few of the $W$ decay products in it. Among all events with $m_{\mathrm{reco}}>55\text{ GeV}$, $3.6\%$ are non-FC, and they bring the correlation among that population down to $51\%$ ($p_T$, mass, and angular resolutions on this population closely track those of PELICAN$\mid$FC above). But since in practice we're interested in $m^W_{\mathrm{true}}$, the correlation between that and $m_{\mathrm{reco}}$ is higher, at $59\%$ among events with $m_{\mathrm{reco}}>55\text{ GeV}$. This is a significant improvement over the model trained on the original $m^W_{\mathrm{true}}\sim 80\text{ GeV}$ \Delphes dataset, and especially over non-ML methods such as the JH tagger (see \figref{m_corr_mJH}). 

Therefore a workflow that guarantees both high background rejection and high reconstruction quality would involve first using a model trained on $p^W_{\text{cont}}$ as a classifier to filter out well-contained events, and then using a model trained on $p^W_{\text{true}}$ to obtain a high-precision reconstruction on that population. However, even a model trained on \Delphes data to reconstruct $p^W_{\mathrm{true}}$, in fact, achieves a $40\%$ correlation with $m^W_{\mathrm{true}}$ on non-FC events (see \figref{m_corr_mWW}), so FC-tagging may not always be necessary. Overall, PELICAN provides a viable method for estimating Lorentz-invariant particle masses.

%===========================================
%===========================================

%========================
% PELICAN WEIGHTS
%========================
\section{PELICAN explainability}
\label{Weights}
Despite the output of the PELICAN regression model ostensibly being a 4-vector (or multiple 4-vectors), the richer and more natural object to treat as the output are the PELICAN weights $\{c_i\}$ introduced in \equref{def pelican weights}. Each $c_i$ is attached to its corresponding input constituent $p_i$ due to permutation equivariance and therefore encodes \textit{a scalar feature of that particle within the event}. As we will show in this section, the behavior of these weights is key to the unique explainability and visualization features of the PELICAN architecture.  

In essence, PELICAN is able to take a set of $N$ input 4-vectors and assign $N$ scalar features to them (of course there can be several features per input as well) in a Lorentz-invariant way. This can be powerful in a variety of applications, but in the context of particle reconstruction the problem of finding the right values of the weights is similar to a soft clustering problem. Assuming an idealized dataset with perfect information about the decay products, the model should identify the decay products of the $W$-boson, assign $c_i=1$ to them, and zero to all other constituents. This is analogous to what taggers like the Johns Hopkins top-tagger aim to do via jet clustering. However, since any five 4-vectors are linearly dependent, there is a continuum family of solutions $\{c_i\}$ and it is not clear that PELICAN will prefer the clustering solution. This section is dedicated to analyzing how well this intuition agrees with the actual behavior of the weights, and to explaining the deviations from this na\"ive picture caused by detector effects.

%===========================================
\begin{figure}[t]
    \begin{center}
    \includegraphics[width=\linewidth]{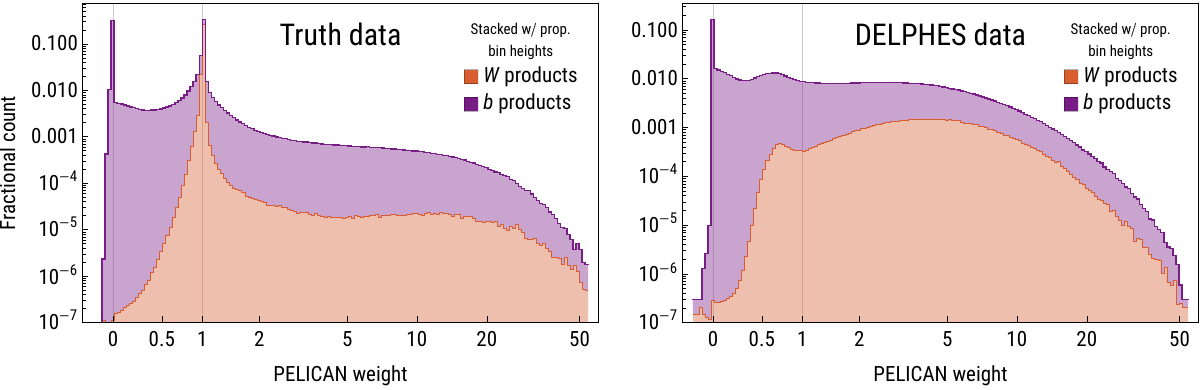}
    \caption{Stacked histograms with proportional bin heights of all PELICAN weights computed over the testing dataset for the 4-vector reconstruction task from \secref{Wreco} using models trained targeting $p^W_{\mathrm{true}}$. Broken up into two populations -- $W$-boson products and $b$-quark products. In the \Delphes case, a constituent is considered a $W$-boson product if the corresponding calorimeter cell detected at least one true $W$-daughter.}
    \label{weights_parent_6W}
    \end{center}
\end{figure}
%===========================================

%------------------------
% MODEL COMPLEXITY
%------------------------
\subsection{Distributions of PELICAN weights} 

In \figref{weights_parent_6W} we display the distributions of all PELICAN weights for models from \secref{Wreco} trained targeting $p^W_{\mathrm{true}}$. We also mark each constituent as either a $W$- or a $b$-daughter. This yields several observations. 

Firstly, nearly all weights are either non-negative or very slightly negative (e.g.~above $-0.1$) with a very sharp peak at zero (the peak is entirely to the left of zero to very high precision\footnote{The bin $[-10^{-6},0)$ contains about 100 times more constituents than the bin $[0,10^{-6})$.}). This is the first feature that justifies the interpretation of PELICAN as a \textit{soft clustering} method. Since our inputs represent realistic events, all input 4-vectors in them are timelike, as is the target vector. This implies that no linear combination of these vectors with positive coefficients can produce a zero vector. The distributions, therefore, show that PELICAN weights assigned to $b$-daughters are not ``contaminated'' with these degenerate combinations.

Secondly, the truth-level distribution is highly concentrated at $0$ and $1$ and very closely matches the binary clustering solution. That is, almost all constituents assigned weight $0$ are $b$-daughters, and almost all of those assigned $1$ are $W$-daughters. Nevertheless, $30\%$ of $b$-daughters are assigned positive weights, prompting further investigation. Moreover, the distribution of $W$-daughter weights in the \Delphes case is so spread out that it becomes difficult to explain it via a mere analogy with clustering.

%===========================================
\begin{figure}[t]
    \begin{center}
    \includegraphics[width=\linewidth]{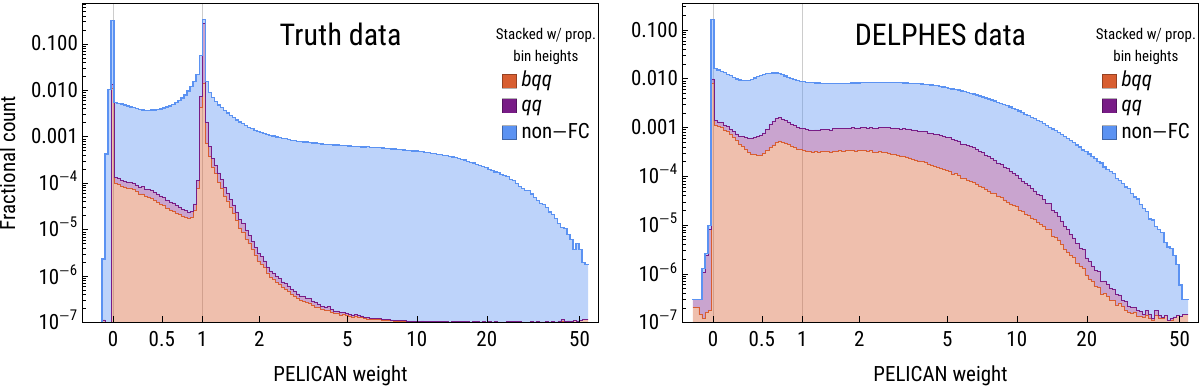}
    \caption{Stacked histograms with proportional bin heights of all PELICAN weights computed over the testing dataset for the 4-vector reconstruction task from \secref{Wreco} using models trained targeting $p^W_{\mathrm{true}}$. Broken up into three populations by jet containment: $bqq$ events (all three truth-level quarks from the $t\to bW\to bqq$ process fall within the jet clustering radius); $qq$ events (only the $b$-quark fell outside of the jet); and non-FC events, which include $bq$, $b$, and $q$ events.}
    \label{weights_event_6W}
    \end{center}
\end{figure}
%===========================================

%===========================================
\begin{figure}[t]
    \begin{center}
    \includegraphics[width=\linewidth]{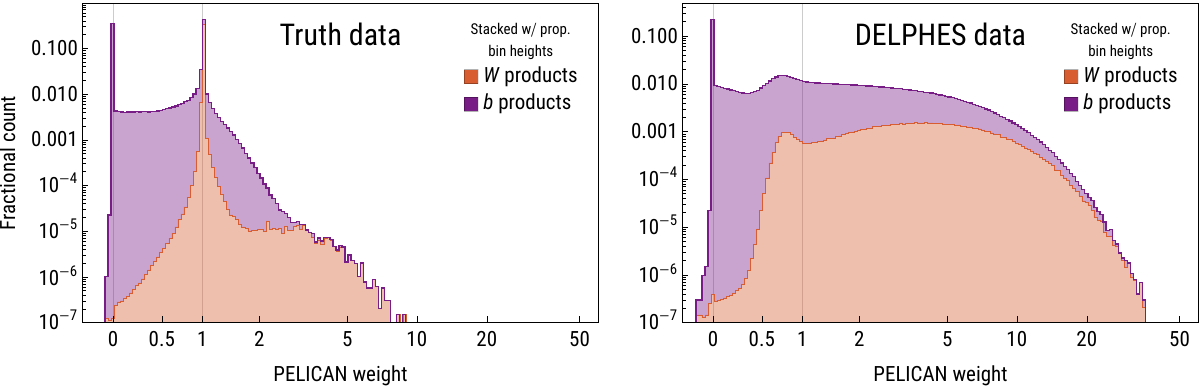}
    \caption{Stacked histograms with proportional bin heights of all PELICAN weights for the 4-vector reconstruction task from  \secref{Wreco} using models trained targeting $p^W_{\mathrm{cont}}$. Broken up into two populations by parent type.}
    \label{weights_parent_6D}
    \end{center}
\end{figure}
%===========================================

%===========================================
\begin{figure}[t]
    \begin{center}
    \includegraphics[width=\linewidth]{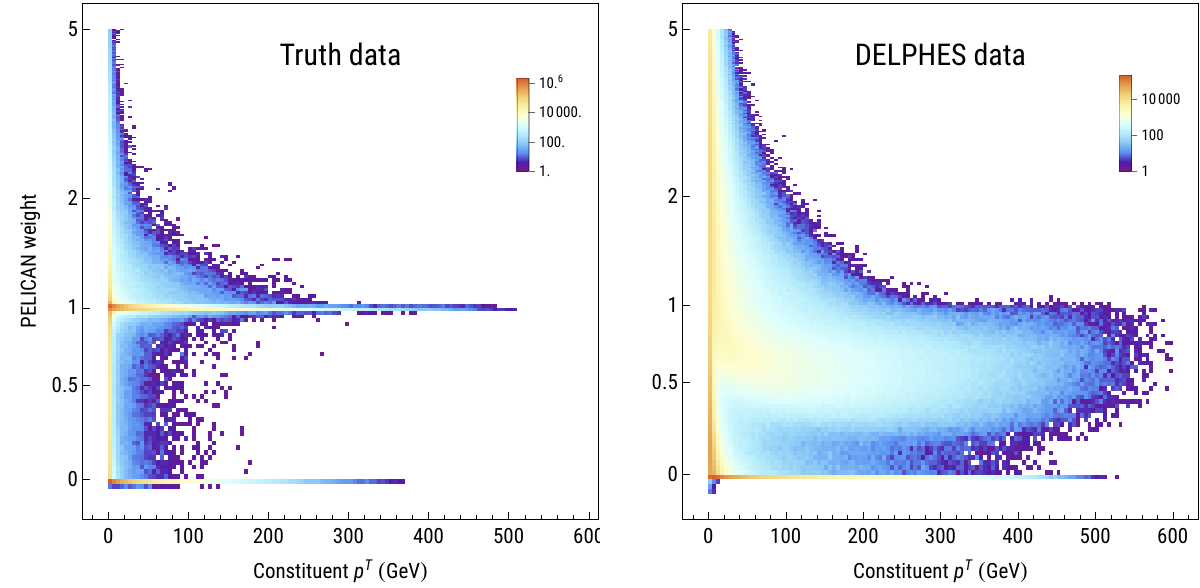}
    \caption{2D histogram of PELICAN weights vs constituent transverse momentum for the 4-vector reconstruction task from \secref{Wreco} using models trained targeting $p^W_{\mathrm{cont}}$. Only FC events shown here.}
    \label{weights_pT_6W}
    \end{center}
\end{figure}
%===========================================

We can delve more deeply into the weight distribution by separating the sub-populations of weights based on jet containment. \Figref{weights_event_6W} shows the distributions of weights for $bqq$, $qq$, and non-FC events. The majority of constituents at the high end of the weight scale belong to non-FC events. Similarly, the weights produced by the models trained targeting $p^W_{\mathrm{cont}}$, shown in \figref{weights_parent_6D}, are more highly concentrated at $0$ and $1$, and have much lower and shorter ``tails'' on the right, especially among $b$-daughters. This is the first indication that PELICAN tends to upweight some constituents in events where it doesn't have enough information for an accurate reconstruction.

This approach allows us to characterize the constituents that are being upweighted. \Figref{weights_pT_6W} shows the constituent weight as a function of the constituent's $p_T$. The main observation here is that among high-energy (``hard'') constituents with $p_T>100\text{ GeV}$ the weight distribution is bimodal, and the vast majority of constituents with weights away from the two peaks are soft, below $20\text{ GeV}$. In the \Delphes case PELICAN appears to downweight high-energy $W$-daughters and upweight soft constituents. Once again, loss of information in the form of detector effects appears to lead to PELICAN upweighting soft constituents.

%------------------------
% DETECTOR EFFECTS
%------------------------
\subsection{Detector effects on PELICAN weights} 

While the truth-level PELICAN models reliably converge to a binary clustering solution, the weights in the \Delphes case do not permit such a straightforward interpretation. To better understand their behavior, we ran additional experiments using custom datasets that exclude different components of the \Delphes detector simulation one by one. \Delphes performs the following steps: simulate the effect of the magnetic field $B_z$ on charged final-state particles; aggregate truth-level particle energies and apply configured cuts within each electromagnetic calorimeter (ECAL) and hadronic calorimeter (HCAL) detector cell; apply energy smearing by sampling a lognormal distribution; unify the ECAL and HCAL cells; apply spatial smearing by picking a uniformly random point within the detector cell; scale the magnitude of the spatial momentum so that the resulting 4-vector, which represents a detector cell, is massless. We found that while each of these steps contributes to smearing out the truth-level distribution of PELICAN weights and shifting the peak downwards, the magnetic field is responsible for almost all of the differences between truth and \Delphes results.

%===========================================
\begin{figure}[t]
    % \vspace{-1\intextsep}
    \begin{center}
    \includegraphics[width=0.5\linewidth]{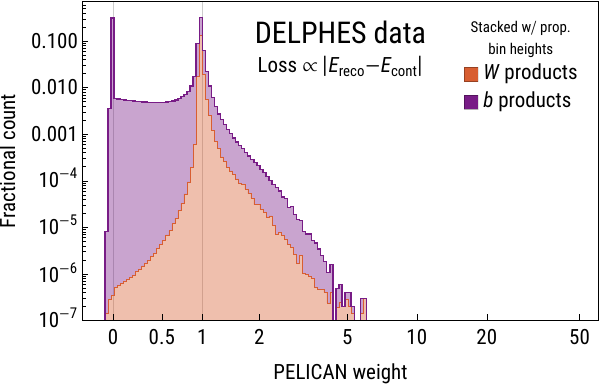}
    \caption{Same as \figref{weights_parent_6D} but this time the model is trained using a single-term loss function proportional to $|E_{\mathrm{reco}}-E_{\mathrm{cont}}|$. We conclude that the shape of the \Delphes distribution in \figref{weights_parent_6D} is overwhelmingly due to \textit{spatial} detector effects.}
    \label{lossE_weights}
    \end{center}
\end{figure}
%===========================================

The simulated magnetic field is able to deflect charged particles very significantly, enough to account for most of the error in PELICAN's reconstruction relative to the truth-level reconstruction. Our hypothesis for why this leads to lower PELICAN weights for hard constituents is the following: Deflected hard particles produce large errors in the direction but not the energy of the reconstruction, and therefore one can downweight them and compensate for the energy deficit using softer constituents. Moreover, by upweighting softer constituents PELICAN can in fact partially correct the error in the direction since the deflections of positively-charged particles can be partially cancelled out by those of negatively-charged particles.

%===========================================
\begin{figure}[t]
    % \vspace{-2\intextsep}
    \begin{center}
    \includegraphics[width=0.5\linewidth]{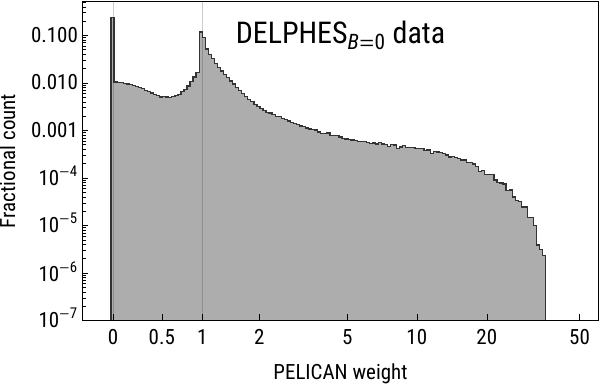}
    \caption{Distribution of PELICAN weights for a model trained on a dataset that differs from a true \Delphes detector simulation only in that the simulated magnetic field is disabled. The model was trained to reconstruct $p^W_{\mathrm{true}}$, so it should be compared to \figref{weights_parent_6W}.}
    \label{noB_weights}
    \end{center}
\end{figure}
%===========================================

%===========================================
\begin{figure}[t]
    \begin{center}
    \includegraphics[trim={0 430 0 0},clip,width=\linewidth]{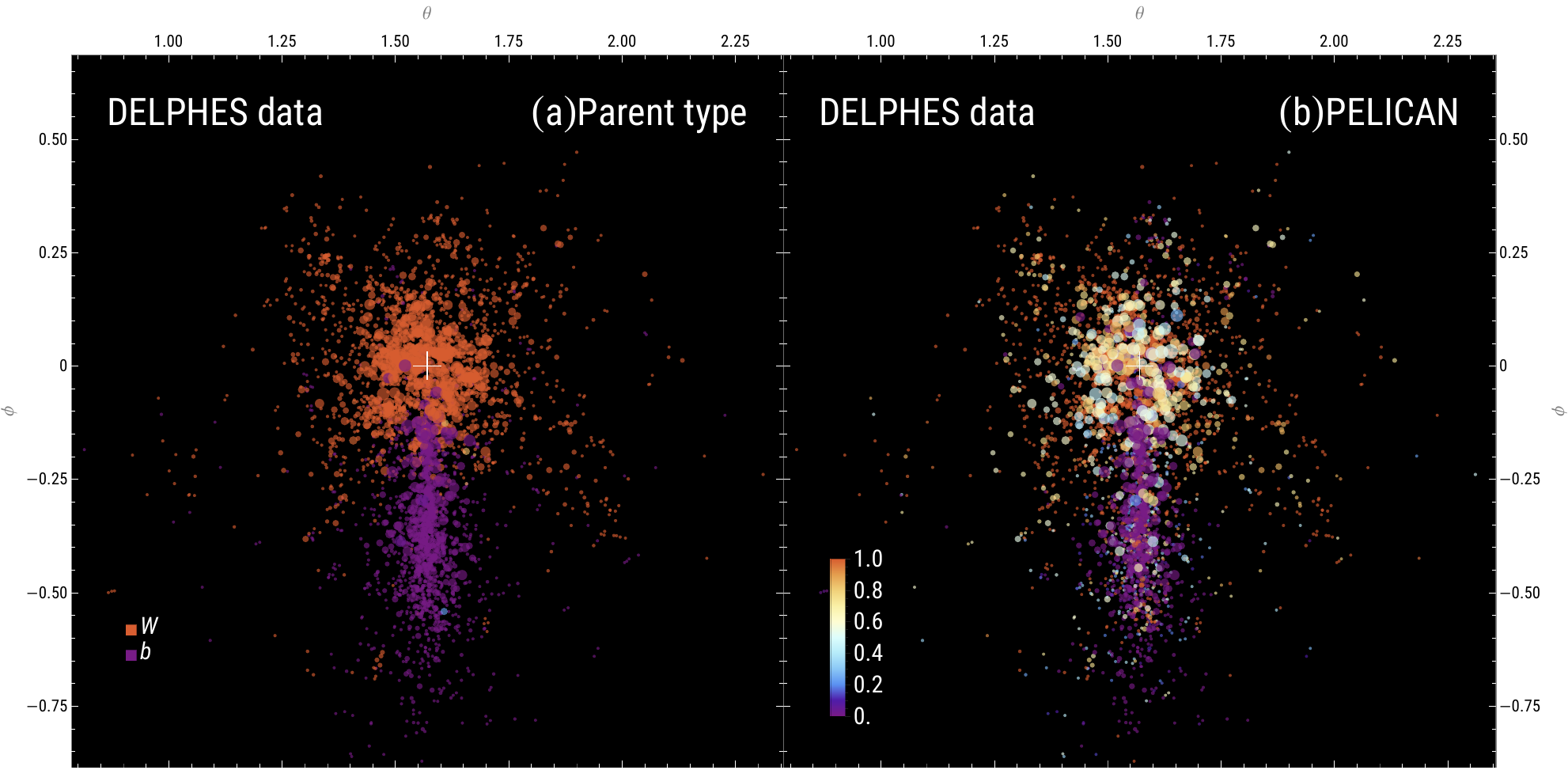}
    \caption{Composite event display of 200 events from the \Delphes dataset from \secref{Wreco}. Each event is transformed using a 3D rotation matrix such that the true $W$-boson ends up at $(\theta,\phi)=(\pi/2,0)$ (white cross), and the true $b$-quark is directly below. Each dot is a \Delphes constituent and the dot size increases logarithmically with constituent energy. (a) Color reflects parent type: constituents that are fully derived from $W$-daughters are orange and those from $b$-daughters are purple; in the rare cases when the fraction of $W$-derived energy in a given calorimeter cell is between $0$ and $1$, the corresponding color is taken from the color scale in the right pane. (b) Color reflects the value of the PELICAN weight, clipped to the interval $[0,1]$, as shown in the legend. Note how the hardest $W$-boson constituents (largest dots) tend to have PELICAN weights between $0.5$ and $1$.}
    \label{multi_display}
    \end{center}
\end{figure}
%===========================================

%===========================================
\begin{figure}[t]
    % \vspace{-2\intextsep}
    \begin{center}
    \includegraphics[trim={0 0 41 33},width=0.7\linewidth]{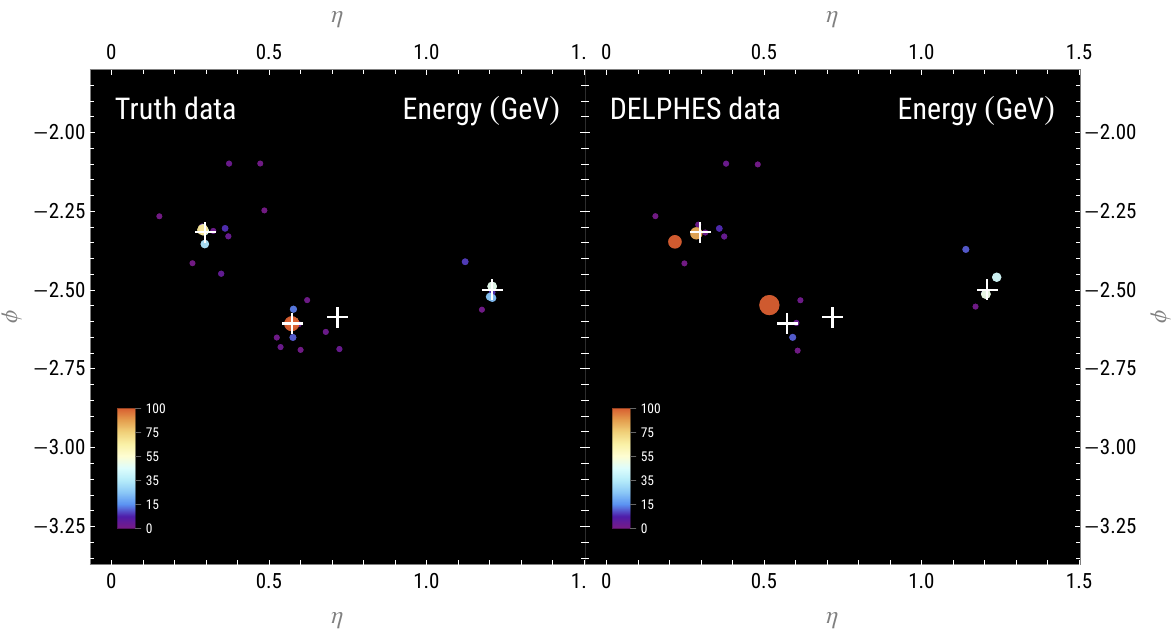}
    \caption{A single event viewed in the $\eta,\phi$ plane with color and size dependent on energy. The central cross marks the true $W$-boson, and the other three crosses mark the three true quarks from the process $t\to bqq$.}
    \label{display_energy}
    \end{center}
\end{figure}
%===========================================

An extra piece of evidence in support of this hypothesis can be found by modifying the loss function. If we re-train the model on the same \Delphes dataset using a loss function consisting of a single energy term $|E_{\mathrm{reco}}-E_{\mathrm{true}}|$, we find a distribution of weights (see \figref{lossE_weights}) nearly as bimodal as the original one trained on truth-level data (see \figref{weights_parent_6D}). This indicates that the source of the error in PELICAN's reconstruction on \Delphes data is overwhelmingly \textit{spatial}. Out of all the steps that \Delphes performs, only two are purely spatial: momentum smearing within one cell, and the simulated magnetic field. However, the detector cells (approximately $0.02\times 0.02$ in $(\eta,\phi)$) are much smaller than the magnitude of PELICAN's typical angular error, and thus smearing alone cannot explain the error.

For an ultimate confirmation of the hypothesis that the reduced precision of the \Delphes model is overwhelmingly caused by the magnetic field, we generated another dataset that differs from a true \Delphes dataset only in that the simulated magnetic field is set to zero. Training the same architecture on this dataset to reconstruct $p^W_{\mathrm{true}}$ produces a model with overall resolutions $\sigma_{p_T}=3.1\%$, $\sigma_m=2.6\%$, and $\sigma_{\Delta R}=1.6$ centirad. These are significantly better than the \Delphes model in \tabref{btW6_W_table}, particularly the angular resolution. However, the largest difference can be seen in the distribution of the PELICAN weights, shown in \figref{noB_weights}. It is much closer to the truth-level model in \figref{weights_parent_6W} than the \Delphes one, which proves that the vast majority of the smearing of the PELICAN weights is caused by the magnetic field. The effect of the magnetic field can likely be mitigated by adding charge information to the input in the \Delphes' energy-flow scheme, which we intend to explore in a future study.

%------------------------
% EVENT VISUALIZATION
%------------------------
\subsection{Event visualization} 
%\paragraph{Event visualization} 

As we discussed above, despite being a single-vector regression model, PELICAN produces one feature \textit{per input constituent} (namely the weight $c_i$), and these features become interpretable by virtue of \equref{def pelican weights}. This gives us a unique opportunity to make event-level visualizations that provide insight into how PELICAN treats jet topology and how it compares to conventional methods such as the JH tagger's jet clustering.

In \figref{multi_display} we show an amalgamation of 200 events from the \Delphes dataset from \secref{Wreco} projected onto the unit sphere. Each event was spatially rotated so that the position of the true $W$-boson within the image is fixed and the true $b$-quark is located in the negative $\phi$ direction. Due to the rotational invariance of PELICAN weights, this normalization does not affect the inference. In one display the constituents are colored according to their parent being either the $W$-boson or the $b$-quark, and in the other they're colored based on their assigned PELICAN weight. The correlation between the two images is clear: $b$-daughters tend to be correctly assigned zero weight, whereas $W$-daughters have positive weights with the hardest constituents having weights between $0.4$ and $0.8$.

\begin{figure}[t]
    \begin{center}
    \includegraphics[width=\linewidth]{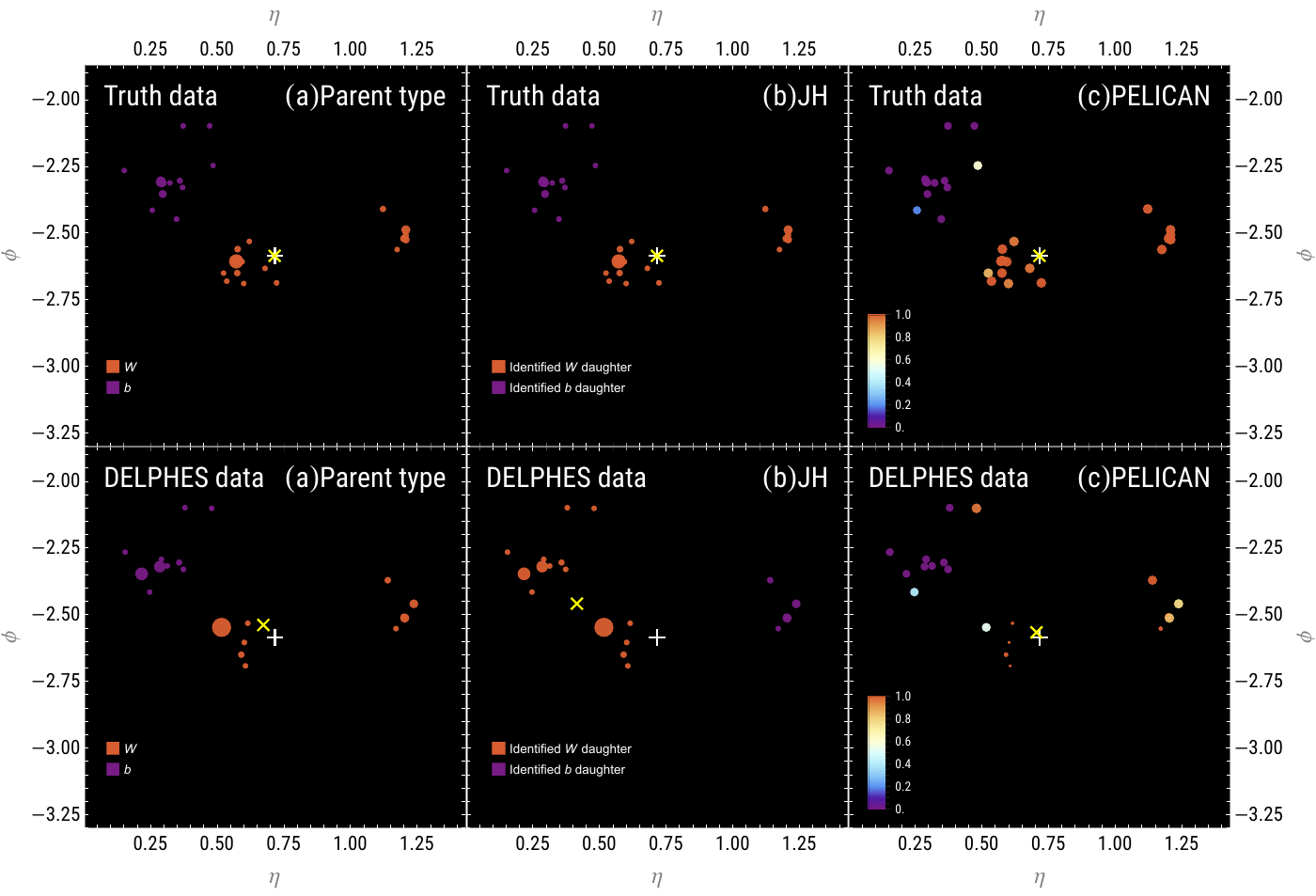}
    \caption{The same event as in \figref{display_energy} in the $(\eta,\phi)$ plane. (a) Constituents are colored according to the actual parent type; size increases nonlinearly with energy; the yellow cross marks the reconstruction obtained by summing all of the constituents that belong to the $W$-boson. (b) Constituents are colored according to how they are tagged by the JH-tagger as either $W$-daughters or not; size increases with energy; the yellow cross marks the JH-reconstructed $W$-boson. (c) Constituents are colored according to their PELICAN weight clipped to the interval $[0,1]$; size increases as the weight goes from $0$ to $1$ and decreases after that. Note how the soft \Delphes $W$-constituents get assigned high PELICAN weights.}
    \label{event_display}
    \end{center}
\end{figure}

In \figref{display_energy} we show a single event in the $(\eta,\phi)$ plane, with dot color and size dependent on the constituent energy. Note the reduced number of constituents in the \Delphes display, and how some of the constituents get strongly deflected by the simulated magnetic field. The same event can be visualized in three more helpful ways. In addition to parent type and PELICAN visualizations introduced in \figref{multi_display}, we can also extract the list of constituents that the JH tagger identifies as belonging to the $W$-boson and highlight them. \Figref{event_display} displays the same single event in all three ways. In addition, we add a special marker for the direction of the reconstructed $W$-boson. In the parent type pane, this reconstruction is defined as $\sum_{i=1}^N r_i p_i$ where $r_i$ is the energy of the true $W$-daughters within that constituent divided by the actual energy of the constituent. In the JH and PELICAN panes, the yellow marker corresponds to the respective reconstructions obtained by those methods.

%===========================================
%===========================================

\subsection{Explainability}

We claim that the interpretation of the PELICAN weights as clustering coefficients constitutes explainability of the PELICAN 4-vector regression model. We are able to fully understand the way PELICAN produces its 4-vector output in terms of a set of physical features of the input constituents which cannot be inferred from the training target -- namely, their identities as $W$-daughters or $b$-daughters. This is possible only due to the combination of the \textit{full} Lorentz symmetry and permutation symmetry, which guarantees the universality of \equref{equiv}. This approach can be contrasted with that of ref.~\cite{CPT23}, where unmatched top quarks are assigned to jets by averaging attention weights in a regression network trained to reconstruct kinematic properties of the top quarks across all layers.

%========================
% IRC-safe PELICAN
%========================
\section{IRC-safety and PELICAN}
\label{irc}
\subsection{Definitions}

Perturbative computations in QCD suffer from a divergence caused by two types of processes: soft emission and collinear splittings. As a consequence, meaningful observables in this theory need to be insensitive to such processes, and this requirement is known as IRC-safety. In this section we provide a precise definition, give a characterization of IRC-safe Lorentz-invariant observables (see details in \appref{appendix_irc}), and describe modifications to the PELICAN architecture that make it IR-safe or IRC-safe.

Infrared safety (IR-safety) guarantees insensitivity to soft emissions, i.e.~particles with relatively low energies and momenta. A family of continuous symmetric observables $f^{(N)}(p_1,\ldots,p_N)$ is said to define an IR-safe observable $f$ if 
\[\lim_{\epsilon\to 0}f^{(N+1)}(p_1,\ldots,p_N,\epsilon p)=f^{(N)}(p_1,\ldots,p_N)\]
for any $N$ and any $p_1,\ldots,p_N,p$, where $\epsilon$ controls how infinitesimally small the considered soft emission $p$ is.

Collinear safety (C-safety) is a restriction on observables in perturbative QCD that arises from the divergent contributions of collinear emissions of gluons. Positive gluon mass would prevent such divergences, which is why C-safety concerns only massless particles. We can define C-safety formally as follows: an observable $f(p_1,\ldots,p_N)$ is C-safe if, whenever two massless 4-momenta $p_1$ and $p_2$  become collinear (which happens for massless particles iff $p_1\cdot p_2=0$), the value of $f$ depends only on the total momentum $p_1+p_2$. Expressed even more explicitly, C-safety says that setting $p_1=\lambda p$ and $p_2=(1-\lambda)p$ with some 4-vector $p$ such that $p^2=0$ must lead to the same output regardless of the value of $\lambda\in (0,1)$, i.e.
\[C_{12}(p)f=\partial_\lambda f(\lambda p,(1-\lambda)p, p_3,\ldots,p_N)=0,\quad \lambda\in (0,1).\label{25773}\]

In \appref{appendix_irc} we characterize a certain class of IRC-safe Lorentz-invariant observables in terms of polynomial bases, but the following summary will suffice for the purpose of designing an IRC-safe version of PELICAN. First, a Lorentz-invariant observable (assumed to be consistently defined for any finite number $N$ of 4-vector inputs) is IR-safe if and only if it has no explicit dependence on the multiplicity $N$. More precisely, adding the zero 4-vector to the list of inputs should leave the output value invariant. Second, an IRC-safe Lorentz-invariant observable is one that is IR-safe and moreover depends on any of its massless inputs in such a way that when any number of them become collinear, the observable ends up depending only on the sum of their energies. E.g.~if $p_1,p_2,p_3$ are fixed to be massless, then $f(p_1,p_2,p_3,p_4,\ldots)$ must depend only on $p_1+p_2+p_3$ when $p_1\parallel p_2\parallel p_3$. Note that such an observable is still completely unrestricted away from the massless manifold.

It is instructive to reflect on the interplay between IR-safety and C-safety. C-safety allows us to exchange momentum between collinear particles without affecting the values of C-safe observables, whereas IR-safety allows us to omit vanishing 4-momenta from the list of inputs. C-safety in itself doesn't change the number of input 4-vectors -- it only requires $f$ to be invariant under a certain transformation that mixes collinear massless inputs. Meanwhile IR-safety effectively requires $f$ to encode an infinite family of observables $f^{(N)}(p_1,\ldots,p_N)$ for any possible number of inputs with the compatibility condition $f^{(N)}(p_1,\ldots,p_{N-1},0)=f^{(N-1)}(p_1,\ldots,p_{N-1})$.

The original PELICAN architecture as introduced above is neither IR- nor C-safe. Below we modify the architecture to make it exactly IRC-safe and evaluate the implications.

\subsection{IRC-safe PELICAN}

% Adding C-safety to the architecture is much simpler. As stated above, the necessary requirement is that the output depend on massless inputs only through their sum. In PELICAN this can be achieved by inserting a linear permutation-equivariant layer with a mass-based soft mask immediately at the input (any nonlinear embedding has to be done later). Consider a case where $p_1,p_2$ are massless and the dot product matrix $\{d_{ij}\}$ is fed into such an equivariant layer. Most of the aggregators will compute sums over rows or columns, thus immediately producing C-safe quantities. However, several of the aggregators, including the identity, will preserve individual information about each $p_i$, therefore their output rows and columns corresponding to $p_1$ and $p_2$ need to be thrown out. This can be done by a soft mask that turns to zero as the mass of any input goes to zero. This mask is defined in the same way as the IR mask except using $m_i^2$ instead of $J\cdot p_i$. It needs to be applied only to the first 2 order zero and the first 7 order one aggregators. 

% Coincidentally, this soft mask can also be used in place of an IR mask, which means that we only need the C-safe soft mask to make a fully IRC-safe PELICAN architecture. Altogether it gets applied to all equivariant aggregators except the third one (which extracts the diagonal and is thus IRC-safe by definition).

To enforce IRC-safety in PELICAN, we follow a strategy almost identical to that of ref.~\cite{Spannowsky_IRC_safe}, based on the original idea from ref.~\cite{EFN}, with the correction that our equivariant layers are a generalization of message passing and that we need to take extra care to preserve Lorentz symmetry. The basic idea is that if $F_i$ are a permutation-equivariant set of IRC-safe observables of our inputs $\{p_i\}$, then so is any observable of the form
\[f\left(\sum_{i=1}^N z_i F_i\right),\label{irc-recursion}\]
where $f$ is any scalar function and $z_i$ are appropriately picked energy-dependent weights, commonly chosen to be either fractional energy $E_i/\sum_j E_j$ or fractional transverse energy $p_i^T/\sum_j p_j^T$. IR-safety is guaranteed due to the fact that any soft constituent has $z\to 0$ and the expression above is invariant under the insertion of such terms. C-safety is slightly more involved: if, say, $p_1$ and $p_2$ are massless and collinear, then due to the C-safety and permutation-equivariance of $F_i$, we have $F_1=F_2$, and all $F_j$ depend on these two particles only through $z_1+z_2$ and their common spatial direction. \Equref{irc-recursion} then guarantees that the new observable also depends on the magnitudes of these two vectors only through the sum $z_1+z_2$. All of these arguments obviously also apply to higher numbers of collinear particles. It is also important for us to note that this property applies to permutation-equivariant observables as well.

To apply this idea to PELICAN, we first need a Lorentz-invariant analog of the energy weights $z_i$ and the corresponding normalized vectors $\hat{p}_i$. Since the only permutation-invariant inertial frame that can be defined based on the list of the constituents $p_i$ is the frame of the jet $J=\sum_i p_i$, it is natural to define the jet-frame energies
\[\mathcal{E}_i=\frac{p_i\cdot  J}{m_J}=\frac{\sum_{j=1}^N d_{ij}}{\sqrt{\sum_{j,k=1}^N d_{jk}}},\label{energy-def}\]
where $m_J$ is the invariant jet mass. The energy weights are then simply the fractional jet-frame energies, and the de-dimensionalized 4-momenta are rescaled by these energies:
\[z_i = \frac{\mathcal{E}_i}{\sum_{j=1}^N \mathcal{E}_j},\quad\quad \hat{p}_i=\frac{p_i}{\mathcal{E}_i},\quad\quad \hat{d}_{ij}=\hat{p}_i \cdot \hat{p}_j=\frac{d_{ij}}{\mathcal{E}_i\mathcal{E}_j}.\label{angle-def}\]

Note that $\hat{d}_{ij}$\footnote{If the inputs are massless, $\hat{d}_{ij}=1-\cos\Theta_{ij}$, where $\Theta_{ij}$ are the pairwise spatial angles in the jet frame.} are Lorentz-invariant and constitute the natural input to the IRC-safe version of PELICAN. We are finally ready to modify the architecture to make it IRC-safe. Combining the idea from \equref{irc-recursion} with the definition of equivariant aggregators in \equref{aggregators}, we come to the following prescription for the $\mathrm{Eq}_{2\to 2}$ blocks:
\begin{itemize}
    \item The 5 order zero aggregators don't need to be modified since the application of any nonlinear function to an IRC-safe observable preserves its IRC-safety.
    \item The 8 order one aggregators involve summation over one equivariant input index, $\sum_i \bullet_i$, and all of these need to be modified by including the Lorentz-invariant energy weights: $\sum_i z_i \bullet_i$.
    \item The 2 order two aggregators are similarly adjusted with two energy weights: $\sum_{i,j} z_i z_j \bullet_{i,j}$.
\end{itemize}
Note that the aggregation function \textit{has to be} based on summation, since max/min pooling or any other nonlinear aggregation will immediately fail C-safety. The way the outputs of these aggregations are embedded into the output $N\times N$ array is unchanged since it doesn't affect IRC-safety. Since the inputs $\hat{d}_{ij}$ are IRC-safe and since \equref{irc-recursion} guarantees recursive IRC-safety, \textit{every individual component} of the output array of every IRC-safe equivariant block is IRC-safe.

The same kind of prescription can also be used to make the $\mathrm{Eq}_{2\to 1}$ (for 4-vector regression), $\mathrm{Eq}_{1\to 2}$ (for scalar inputs), and $\mathrm{Eq}_{2\to 0}$ (for classification) layers IRC-safe. As a consequence, the PELICAN weights $c_i$ become IRC-safe. The regression case deserves special attention, since strictly speaking the PELICAN weights $c_i$ themselves don't have to be IRC-safe for the combination $p_{\mathrm{reco}}=\sum_i c_i p_i$ to be IRC-safe. In particular, whereas the IR-safety of $p_{\mathrm{reco}}$ necessarily makes it possible to find IR-safe $c_i$, there is no such criterion for C-safety.

However, under the interpretation of $c_i$'s as ``soft clustering'' coefficients (as discussed in \secref{Weights}) it does make sense to require IRC-safety from them. In that case the values of weights corresponding to collinear inputs necessarily match due to permutation symmetry. Indeed, IRC-safety implies that if, say, $p_1$ and $p_2$ are collinear, then both $c_1$ and $c_2$ are functions of only their spatial direction and of $E_1+E_2$. Since the direction and the total energy are permutation-invariant but $c_1$ and $c_2$ are permutation-\textit{equivariant}, i.e.\ $c_1(p_1,p_2,\ldots)=c_2(p_2,p_1,\ldots)$, we conclude that $c_1=c_2$. This in turn also guarantees the IRC-safety of the reconstructed 4-vector. Nonetheless, whether IRC-safe PELICAN can approximate \textit{any} IRC-safe vector-valued observable, and whether there is a more general way of generating such observables, is an open problem deserving future investigation.

Finally, since the multiplicity $N$ is not IR-safe, the aggregation function cannot be defined using means. However, there exist IRC-safe analogs of $N$ such as the Soft Drop multiplicity defined in ref.~\cite{Larkoski_SoftDrop}.\footnote{The Soft Drop multiplicity actually measures the depth of the Cambridge-Aachen branching tree along the hard core of the jet. Its definition also involves several new hyperparameters denoted $z_{\mathrm{cut}}$, $\theta_{\mathrm{cut}}$ and $\beta$, see ref.~\cite{Larkoski_SoftDrop}.} Nevertheless, even they can't be directly used in PELICAN due to their non-Lorentz-invariance. For PELICAN tests, we have modified the Soft Drop algorithm by executing it \textit{in the jet frame}, making it manifestly Lorentz-invariant, and defined the Lorentz-invariant Soft Drop multiplicity $n_{\mathrm{SD}}$ that way. This is also equivalent to replacing energies and angles in the original definitions of the Cambridge/Aachen \cite{CA-algorithm-1,CA-algorithm-2} and Soft Drop algorithms with their Lorentz-invariant analogs defined in \equref{energy-def} and \equref{angle-def}. With this, it is safe to use aggregators of the form $n_{\mathrm{SD}}^\alpha\sum_{i=1}^N \bullet_i$.

\subsection{Testing IRC-safe PELICAN models} 
%\paragraph{Testing IR/C-safe PELICAN models} 

First we quantify the deviation in PELICAN's outputs that occurs under soft and collinear splittings and observe how training affects them. We define an IR-splitting as adding a zero 4-vector to the list of input constituents. Then PELICAN's output on IR-split data is directly compared to the original output. Defining a C-splitting is more difficult since realistic events never contain any exactly collinear constituents, and we want to avoid changing the number of particles so as to make this test independent of IR-safety. Therefore we prepare the data by inserting two copies of the vector $(1,0,0,1)$ to each event. Then the C-splitting will amount to replacing these two vectors with $(1.5,0,0,1.5)$ and $(0.5,0,0,0.5)$ (recall that C-safety does not apply when $\lambda=1$ in \equref{25773} because the zero vector is ``collinear'' with anything). The outputs on the same event prepared in these two ways can be directly compared.

To compare two outputs $p_{\mathrm{reco}}, p_{\mathrm{reco}}'$ we compute the relative deviation $|(p_{\mathrm{reco}}'-p_{\mathrm{reco}})/p_{\mathrm{reco}}|$, where the division is component-wise. To estimate the effect of an IR- or C-splitting on PELICAN's predictions, we take the median value of this deviation over a batch of events. The same can also be done with PELICAN weights as the outputs. The splittings are applied to 100-event batches of events from one of our datasets and the relative deviations are averaged over 300 batches. We test 5 randomly-initialized models and 5 models trained on the full variable $W$ mass dataset from \secref{Wmass}.

We find that a randomly-initialized PELICAN regression model's output 4-vector deviates by 0.5\%-14\% (measured by the deviation of each of the 4 Cartesian components) under an IR-split, and the PELICAN weights deviate by up to 9\%. After training on one of our datasets the range of these deviations doesn't appear to change. Under the C-safety test defined above, the 4-vector prediction of a randomly initialized model deviates by an absolute amount of up to $15\,\mathrm{GeV}$, and the median absolute deviation of the PELICAN weights can be as high as $0.03$ (depending on the random seed). With IRC-safety enabled these go down to $10^{-4}\,\mathrm{GeV}$ and $10^{-7}$, respectively. The main reason these deviations aren't even smaller under IRC-safety is the accumulation of numerical errors after repeated aggregation with the energy weights $z_i$. The values of these weights on realistic data are highly concentrated near zero, and overall they can span up to 8 orders of magnitude. After several layers of aggregation with weighting by $z_i z_j$ the spread of the values goes way beyond the size of the standard floating point types, which limits the precision of IRC-safety. However, for practical purposes the current precision of PELICAN is likely to be sufficient, seeing as it is still dominated by the uncertainties originating from the random initialization.

The resolutions $\sigma_{p_T}, \sigma_m$, and $\sigma_{\Delta R}$ of the IRC-safe \Delphes models (using  aggregations weighted by Soft Drop multiplicities) trained on the variable $W$ mass dataset targeting $p^W_{\mathrm{true}}$ are, respectively, 6.8\%, 8.5\%, and 3.1 centirad, which is only about 10\%, 3\%, and 8\% worse (higher) than the resolutions of the original non-IRC-safe models as reported in \tabref{btW6m_W_table}.
Here is how these deviations in the values of the outputs reflect on the overall quality of the predictions. Under an IR-splitting (adding a zero vector to every event) the resolutions of trained non-IRC-safe models get worse by between 0.3\% and 2\% depending on the initialization. Under an ``IRC-splitting'' (splitting the first beam $(1,0,0,1)$ into two constituents $(0.5,0,0,0.5)$)  the resolutions get worse by  between 10\% and 25\%. Therefore non-IRC-safe PELICAN models can be quite sensitive to IRC splittings, especially collinear ones.

Aside from regression, we also trained IRC-safe PELICAN classifiers for top-tagging and quark-gluon tagging. These results are included alongside the non-safe models in \tabref{tab1} and \tabref{tab_QG}. In both cases PELICAN provides state-of-the-art performance among IRC-safe models. In quark-gluon tagging, IRC-safe classifiers perform almost as well as the non-IRC-safe analogs, which confirms the conclusions of the recent study in ref.~\cite{Larkoski23}. Notably, the IRC-safe PELICAN top-tagger used the Lorentz-invariant Soft Drop multiplicity in place of $N$ in its aggregators, whereas the quark-gluon tagger only used simple summation aggregators (i.e.~$N$ was replaced with a fixed constant). This makes the comparison in \tabref{tab_QG} fair, but at the same time we expect even better performance from a model that uses Soft Drop multiplicity. 

%========================
% CONCLUSION
%========================
\section{Conclusion}
\label{conclusion}
We have presented a full description of PELICAN: a network that respects particle permutation and Lorentz symmetries important in particle physics. PELICAN is a general network which is performant at 4-vector regression and provides state-of-the-art performance in the tasks of top-tagging and quark-gluon tagging. The IRC-safe modification of the network is similarly leading among other such architectures. PELICAN also achieves state-of-the-art performance in the more difficult task of multi-class jet identification. Due to the equivariant architecture, all of this is made possible despite PELICAN's relatively compact model size.

To demonstrate PELICAN's regression capabilities, we chose the reconstruction of the $W$-boson's 4-momentum from a full top quark jet, and to our knowledge PELICAN is the first ML method applied to this problem. Even within these tasks there is room to improve PELICAN's performance by introducing additional scalar information such as particle charges, which would allow the network to account for the simulated collider's magnetic field. 

PELICAN's most unique features, however, have to do with its relatively low complexity (in terms of the number of parameters) and explainability. For example, it provides highly competitive top-tagging performance with only several hundred parameters. At the same time, in regression tasks the ``PELICAN weights'' represent a new kind of output that not only has a direct physical interpretation, but also allows one to perform unprecedented particle-level ML-based analysis of particle jets. PELICAN's architecture, its flexibility, and generalizability may also allow for future applications to charged-particle track reconstruction, pile-up identification, and full-event reconstruction. Being a general architecture, PELICAN is not limited to top quark decays or even jets. This network inherently provides powerful tools for investigating its own behavior due to the combination of Lorentz and permutation equivariance and shows promise as a tool which can be thoroughly investigated if deployed in real world scenarios.

%========================
% ACKNOWLEDGEMENTS        
%========================
\section{Acknowledgements}
The authors would like to thank the Data Science Institute at the University of Chicago for its generous support of this research. TH is supported by the Department of Physics at the University of Chicago. DWM and JTO are supported by the National Science Foundation under Grant PHY-2013010.  The computations in this work were, in part, run at facilities supported by the Scientific Computing Core at the Flatiron Institute, a division of the Simons Foundation. The Center for Computational Mathematics at the Flatiron Institute is supported by the Simons Foundation. In addition, we are grateful to Andrew Larkoski for insightful comments.
\label{ack}

%\PRLsep

%\setlength\bibhang{0pt} % remove left margin in citations
%\setlength\bibitemsep{0pt} %set empty space between bib items
%\raggedright %release right end of bib entries
%\vspace*{-3em}
%\setstretch{0.85} %reduce line spacing
%\setquotestyle{english}
%\printbibliography[heading=bibintoc]
%\setstretch{1} %restore line spacing

%========================
% BIBLIOGRAPHY        
%========================
\bibliographystyle{JHEP}
\bibliography{bibliography.bib}

% ========================
% APPENDIX        
% ========================
\appendix
\newpage
\section{Additional results and plots}\label{appendix_plots}

\subsection{$W$-boson 4-momentum reconstruction}

\begin{figure}[h!]
    \begin{center}
    \includegraphics[width=\linewidth]{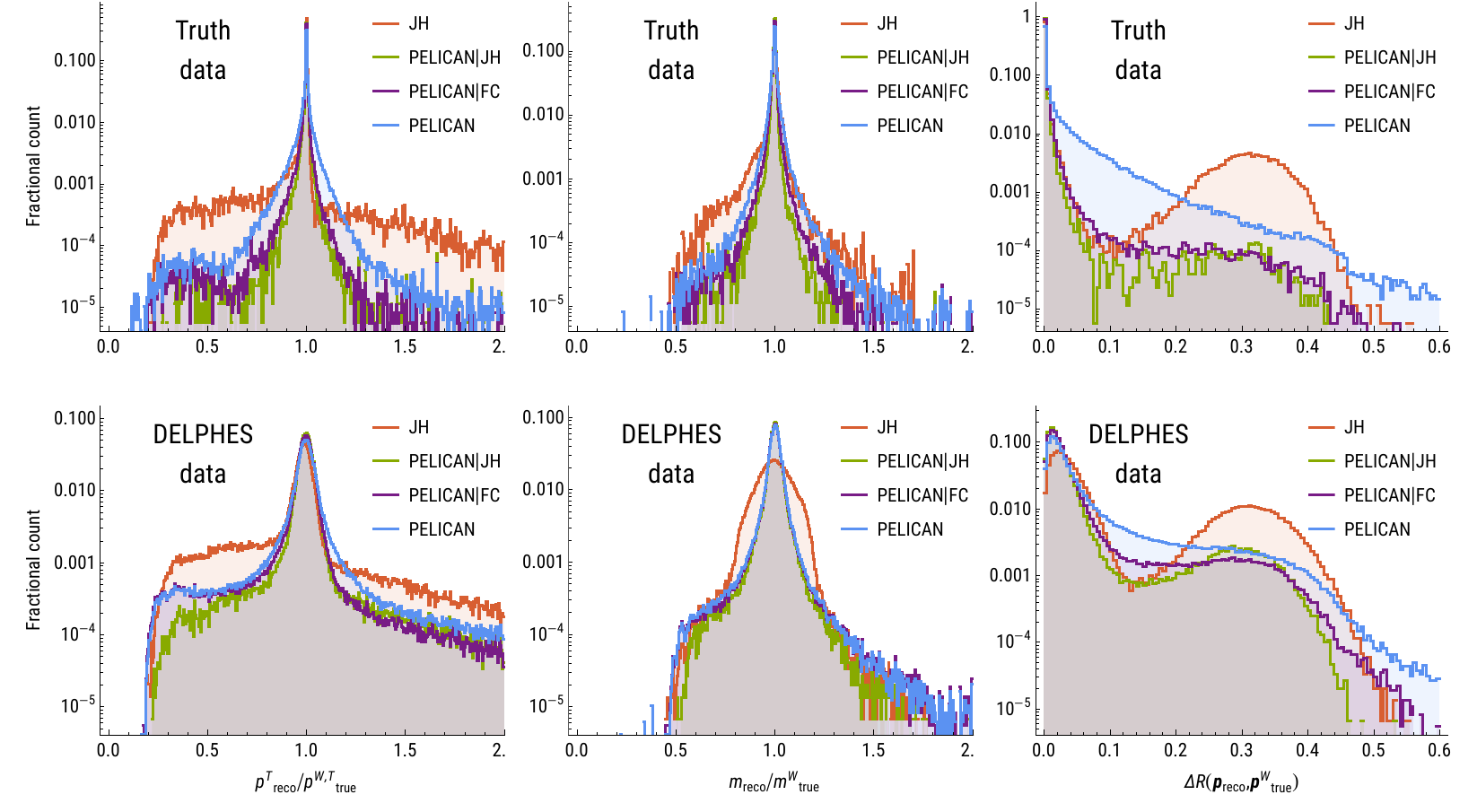}
    \caption{Full set of histograms corresponding to the entries in \tabref{btW6_W_table} (models trained targeting $p^W_{\mathrm{true}}$, truth-level and \Delphes versions).}
    \label{btW6WW_res}
    \end{center}
\end{figure}

\begin{figure}[h!]
    \begin{center}
    \includegraphics[width=\linewidth]{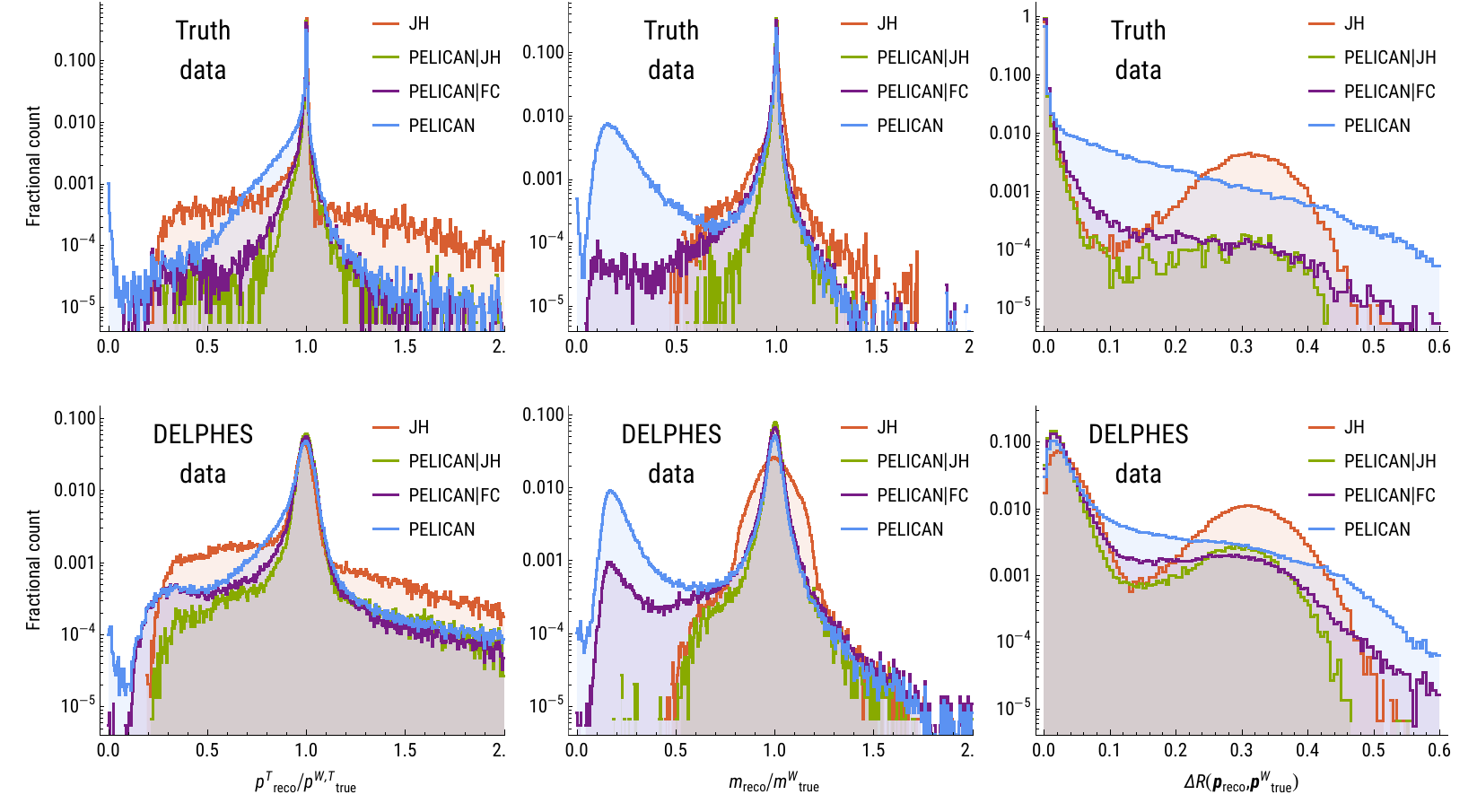}
    \caption{Full set of histograms corresponding to the entries in \tabref{btW6_D_table} (models trained targeting $p^W_{\mathrm{cont}}$, truth-level and \Delphes versions, and compared to $p^W_{\mathrm{true}}$).}
    \label{btW6DW_res}
    \end{center}
\end{figure}

\begin{figure}[h!]
    \begin{center}
    \includegraphics[width=\linewidth]{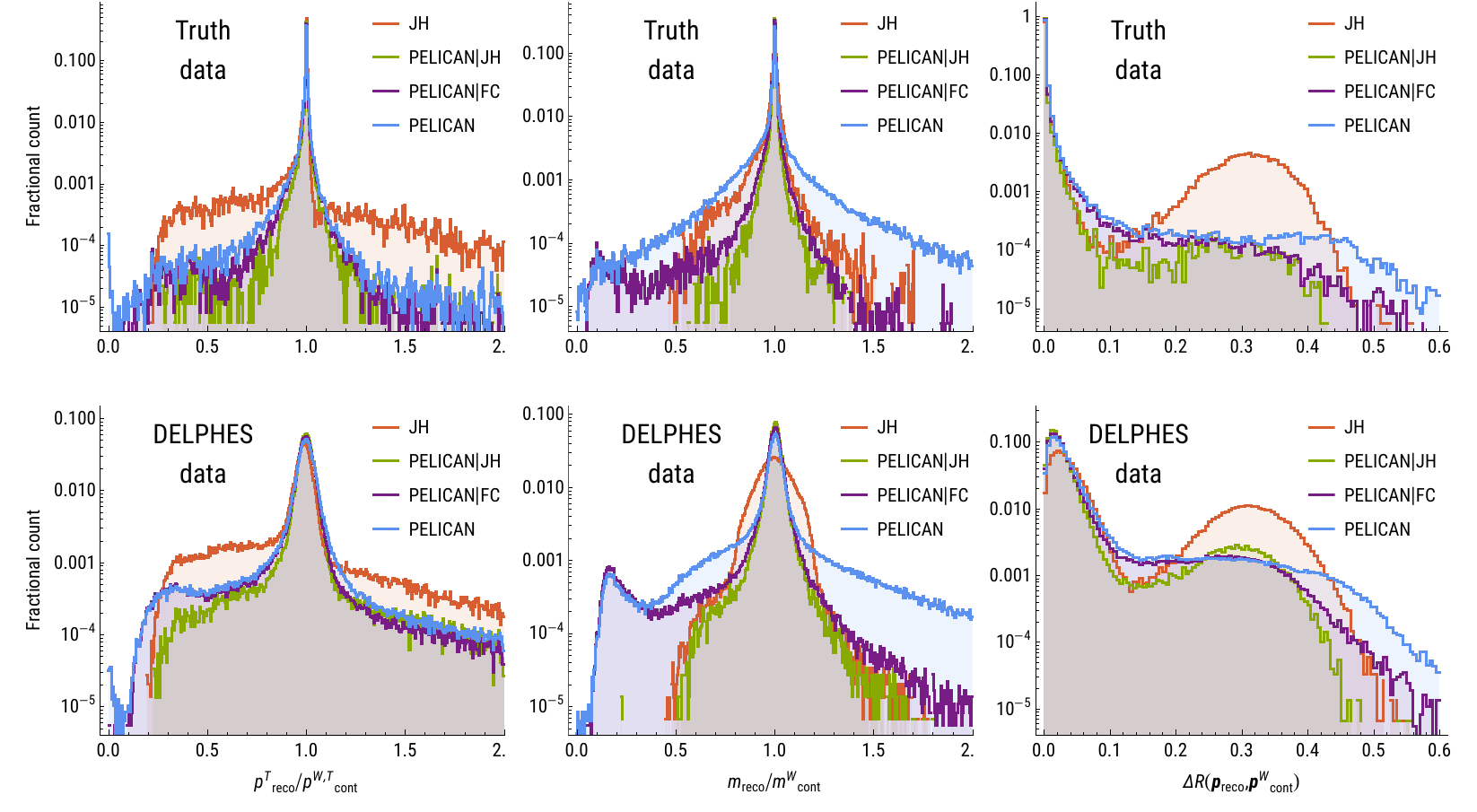}
    \caption{Full set of histograms for models trained targeting $p^W_{\mathrm{cont}}$ and compared to $p^W_{\mathrm{cont}}$.}
    \label{btW6DD_res}
    \end{center}
\end{figure}

\begin{figure}[h!]
    \begin{center}
    \includegraphics[width=\linewidth]{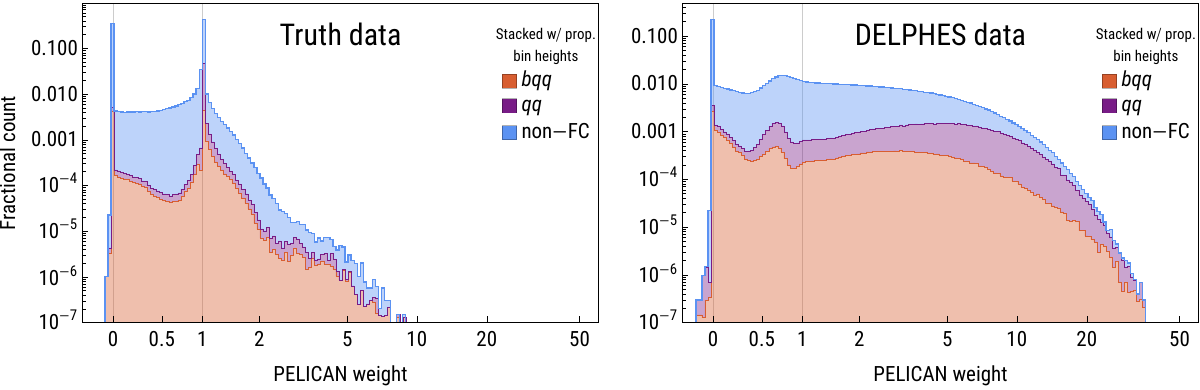}
    \caption{Distributions of PELICAN weights analogous to \figref{weights_event_6W} but for models trained targeting $p^W_{\text{cont}}$.}
    \label{weights_event_6D}
    \end{center}
\end{figure}

\begin{figure}[h!]
    \vspace{-0\intextsep}
    \begin{center}
    \includegraphics[width=0.85\linewidth]{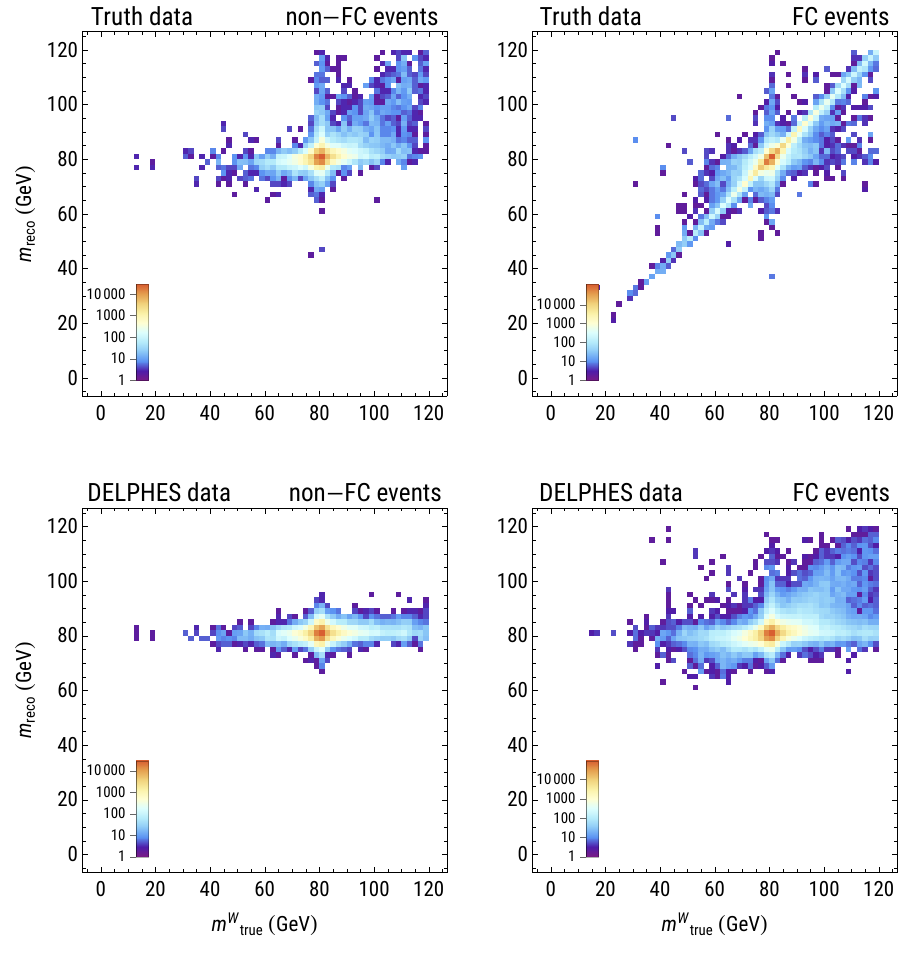}
    \caption{2D histograms of target vs.~reconstructed masses for models (top: truth data; bottom: \Delphes data) trained targeting $p^W_{\mathrm{true}}$, broken up into two populations based on containment (left: non-FC events; right: FC events).}
    \label{m_corr_WW}
    \end{center}
\end{figure}

\begin{figure}[h!]
    \vspace{-0\intextsep}
    \begin{center}
    \includegraphics[width=0.85\linewidth]{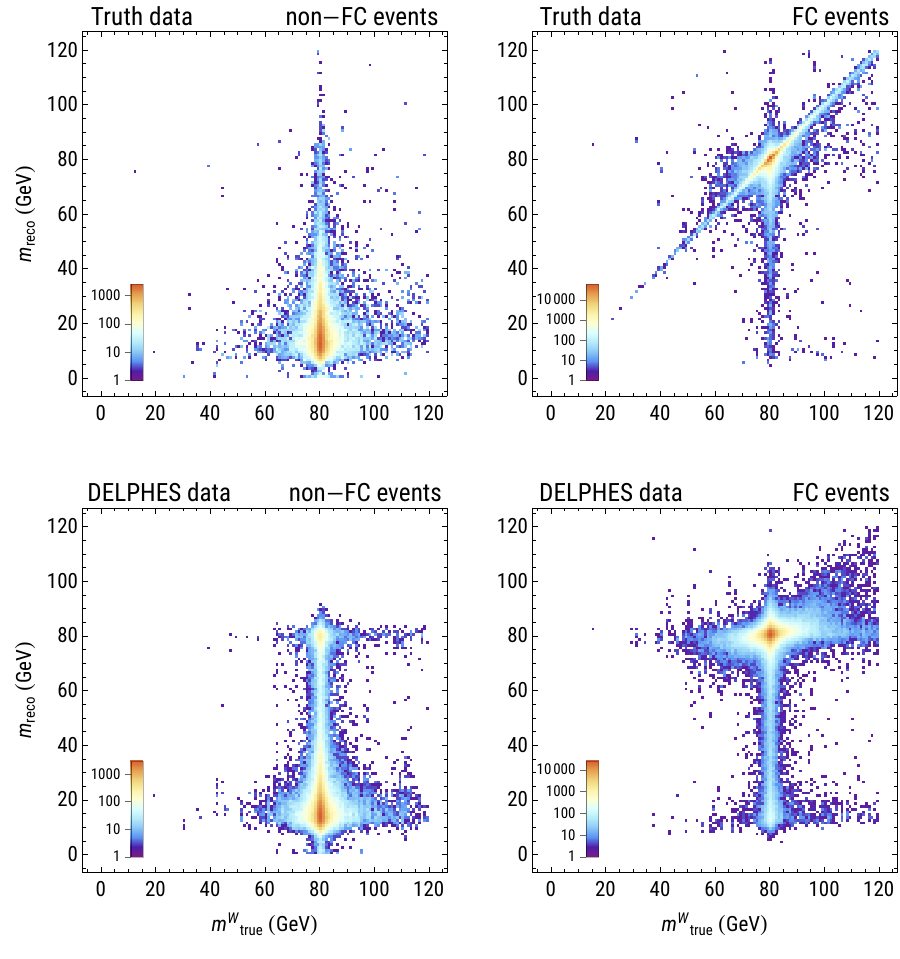}
    \caption{2D histograms of true $W$ mass vs.~reconstructed contained mass for models (top: truth data; bottom: \Delphes data) trained targeting $p^W_{\mathrm{cont}}$, broken up into two populations based on containment (left: non-FC events; right: FC events).}
    \label{m_corr_DW}
    \end{center}
\end{figure}

\begin{figure}[h!]
    \vspace{-0\intextsep}
    \begin{center}
    \includegraphics[width=0.85\linewidth]{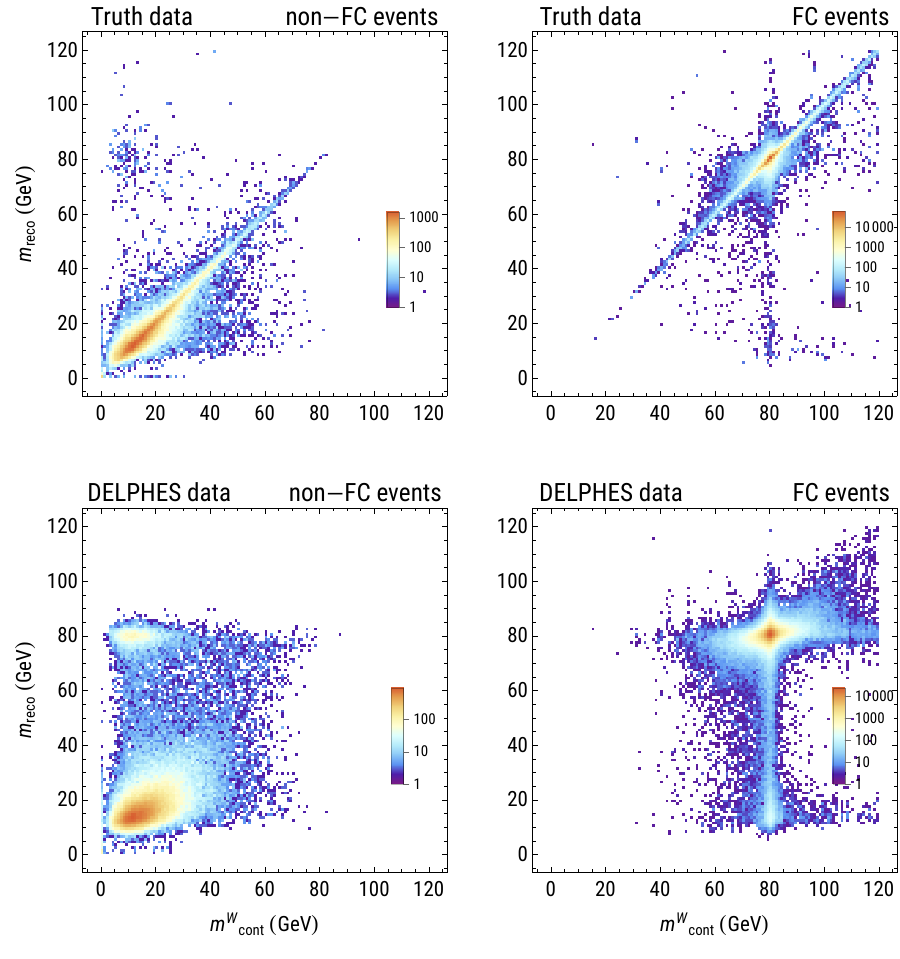}
    \caption{2D histograms of target vs.~reconstructed contained masses for models (top: truth data; bottom: \Delphes data) trained targeting $p^W_{\mathrm{cont}}$, broken up into two populations based on containment (left: non-FC events; right: FC events).}
    \label{m_corr_DD}
    \end{center}
\end{figure}

Below is the list of additional figures for the models trained on the $W$-boson 4-momentum regression dataset from \secref{Wreco}:
\begin{itemize}
    \item \Figref{btW6WW_res}: histograms corresponding to \tabref{btW6_W_table} (models trained targeting $p^W_{\mathrm{true}}$);
    \item \Figref{btW6DW_res}: histograms corresponding to \tabref{btW6_D_table} (models trained targeting $p^W_{\mathrm{cont}}$);
    \item \Figref{btW6DD_res}: histograms for models trained targeting $p^W_{\mathrm{cont}}$.
    \item \Figref{weights_event_6D} distributions of PELICAN weights for models trained targeting $p^W_{\text{cont}}$;
    \item \Figref{m_corr_WW}: target vs.~reconstructed mass correlation histograms for models trained targeting $p^W_{\mathrm{true}}$;
    \item \Figref{m_corr_DW}: true $W$-boson vs.~reconstructed mass correlation histograms for models trained targeting $p^W_{\mathrm{cont}}$;
    \item \Figref{m_corr_DD}: target vs.~reconstructed mass correlation histograms for models trained targeting $p^W_{\mathrm{cont}}$.
\end{itemize}

\subsection{$W$-boson mass measurement}

\begin{figure}[h!]
    \begin{center}
    \includegraphics[width=\linewidth]{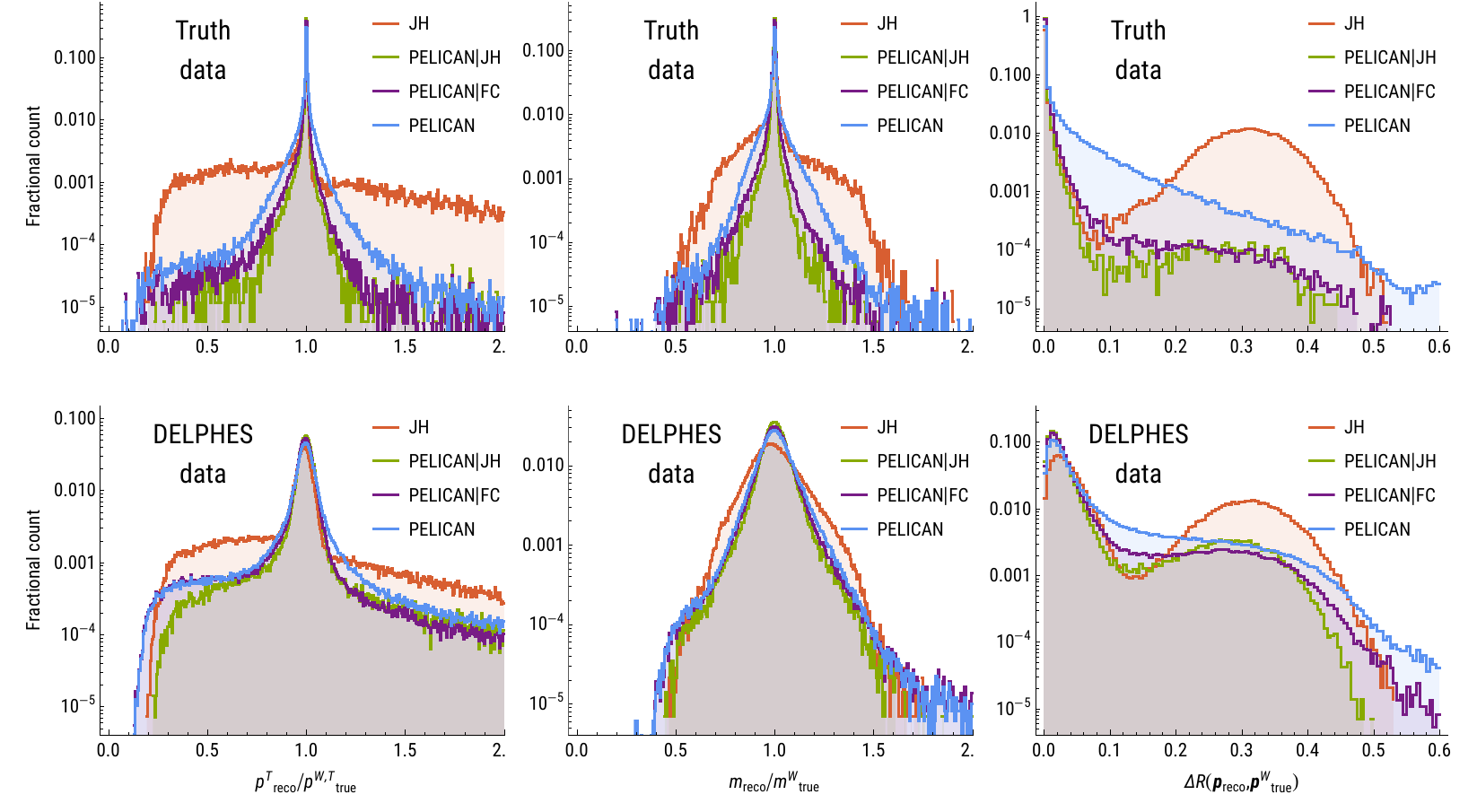}
    \caption{Full set of histograms corresponding to the entries in \tabref{btW6m_W_table} (models trained targeting $p^W_{\mathrm{true}}$, truth-level and \Delphes versions).}
    \label{btW6mWW_res}
    \end{center}
\end{figure}

\begin{figure}[h!]
    \begin{center}
    \includegraphics[width=\linewidth]{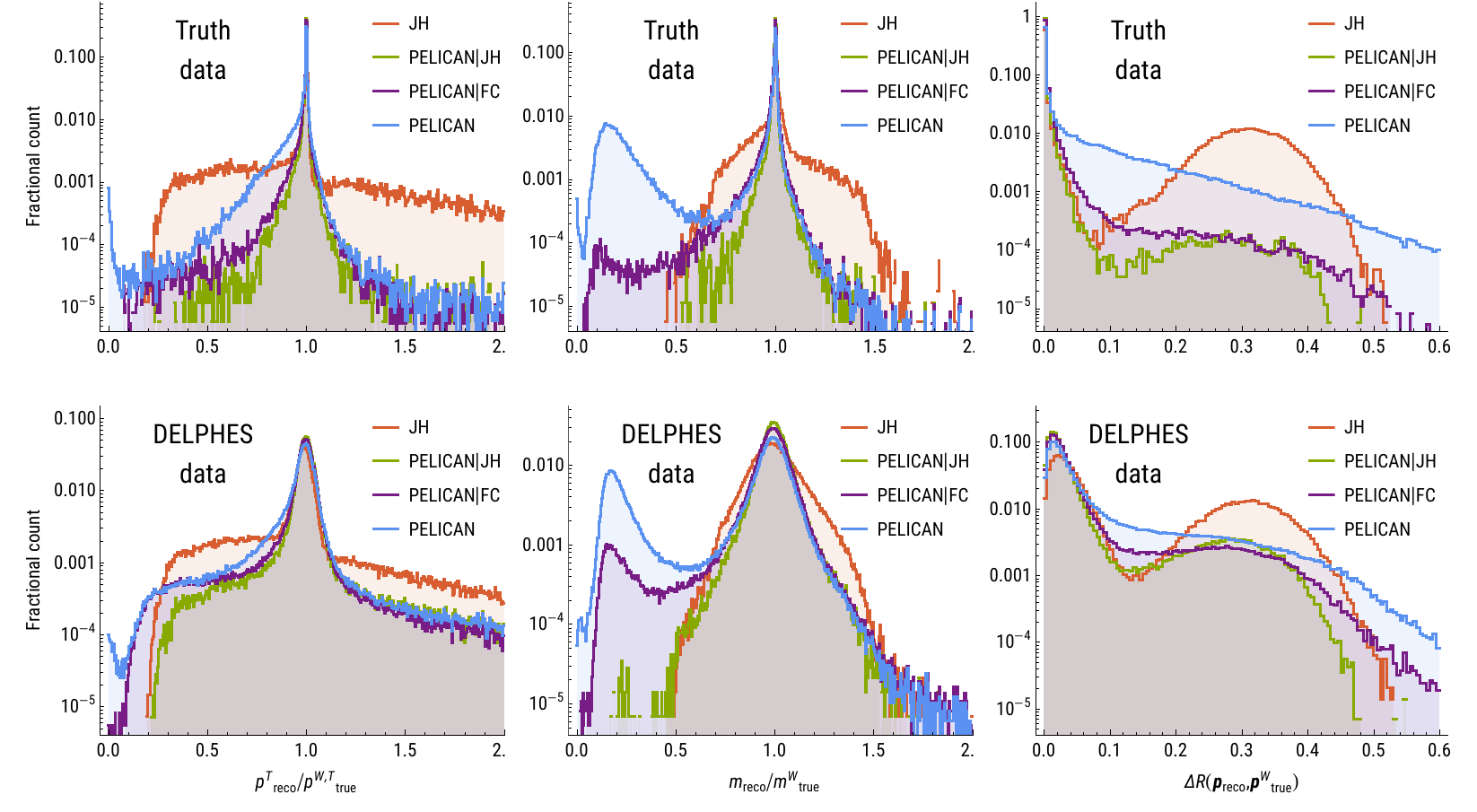}
    \caption{Full set of histograms corresponding to the entries in \tabref{btW6m_D_table} (models trained targeting $p^W_{\mathrm{cont}}$, truth-level and \Delphes versions, and compared to $p^W_{\mathrm{true}}$).}
    \label{btW6mDW_res}
    \end{center}
\end{figure}

\begin{figure}[h!]
    \begin{center}
    \includegraphics[width=\linewidth]{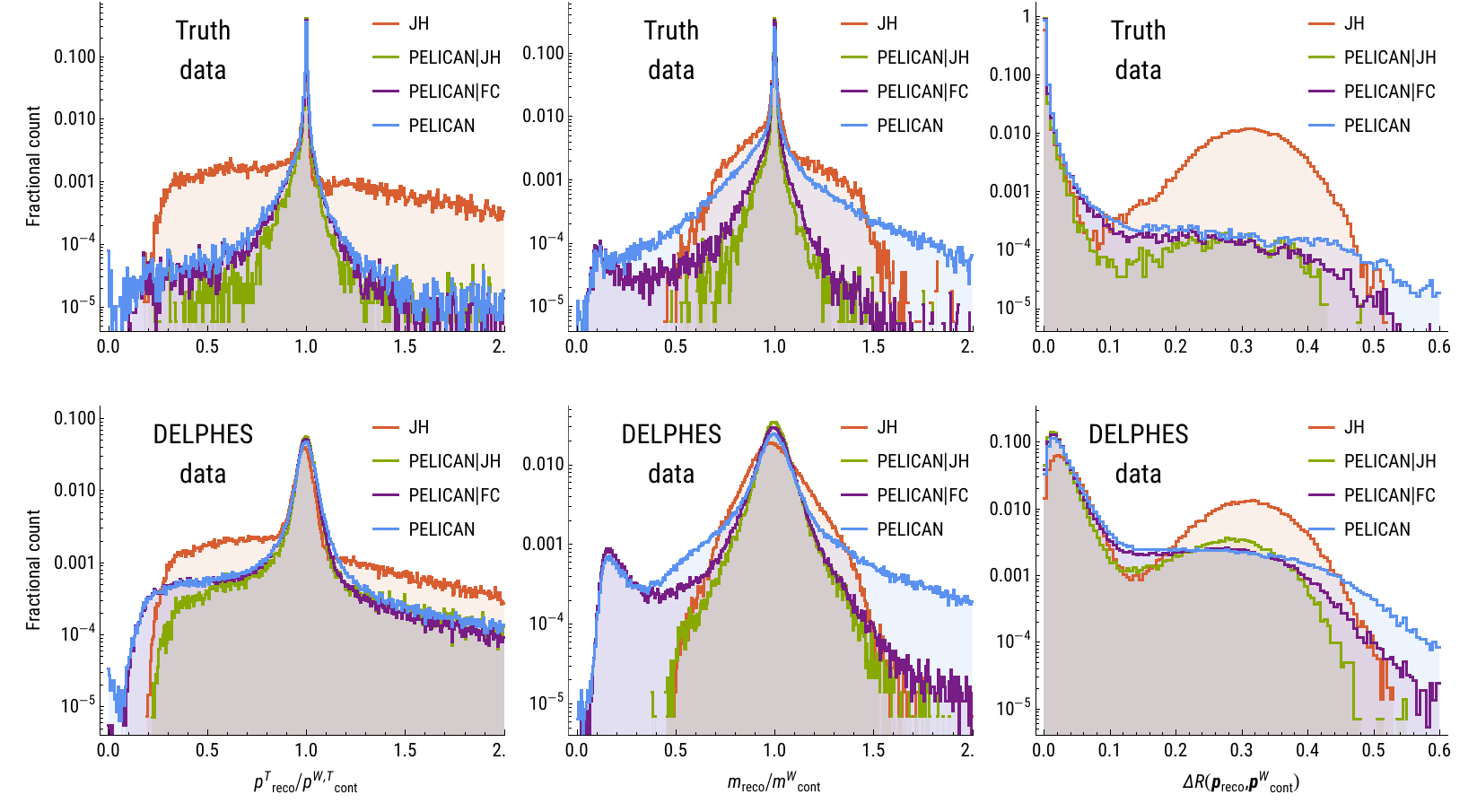}
    \caption{Full set of histograms for models trained on the variable mass dataset targeting $p^W_{\mathrm{cont}}$ and compared to $p^W_{\mathrm{cont}}$.}
    \label{btW6mDD_res}
    \end{center}
\end{figure}

\begin{figure}[h!]
    \vspace{-0\intextsep}
    \begin{center}
    \includegraphics[width=0.85\linewidth]{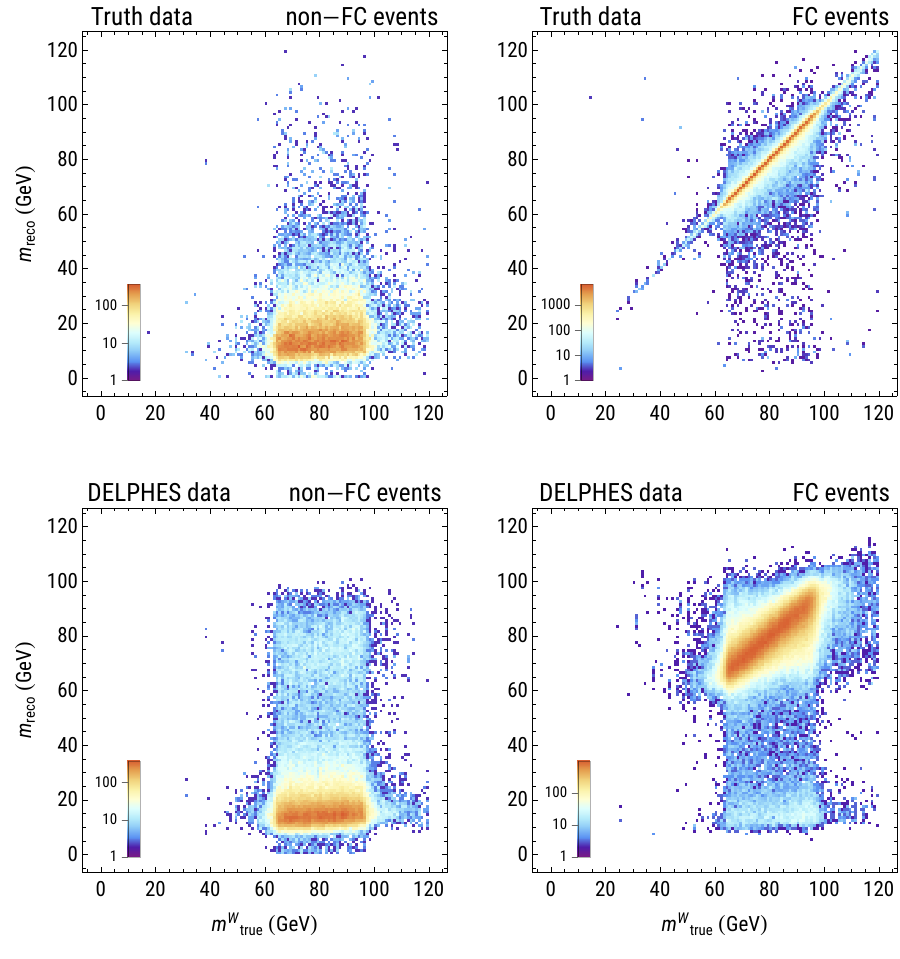}
    \caption{2D histograms of target vs.~reconstructed masses for models (top: truth data; bottom: \Delphes data) trained on the variable mass dataset targeting $p^W_{\mathrm{cont}}$, broken up into two populations based on containment (left: non-FC events; right: FC events).}
    \label{m_corr_mDW}
    \end{center}
\end{figure}

Below is the list of additional tables and figures for the models trained on the variable mass dataset from \secref{Wmass}:
\begin{itemize}
    \item \Tabref{btW6m_W_table}: resolutions for models trained targeting $p^W_{\mathrm{true}}$;
    \item \Figref{btW6mWW_res} histograms corresponding to \tabref{btW6m_W_table} (models trained targeting $p^W_{\mathrm{true}}$);
    \item \Figref{btW6mDW_res}: histograms corresponding to \tabref{btW6m_D_table} (models trained targeting $p^W_{\mathrm{cont}}$ and compared to $p^W_{\mathrm{true}}$);
    \item \Figref{btW6mDD_res}: histograms for models trained targeting $p^W_{\mathrm{cont}}$ and compared to $p^W_{\mathrm{cont}}$;
    \item \Figref{m_corr_mDW}: target vs.~reconstructed mass correlation histograms for models trained targeting $p^W_{\mathrm{cont}}$.
\end{itemize}

\begin{table}[h]
    \centering
    \begin{small}
        \begin{tabular}{ccS[table-format=3.2]<{\%}S[table-format=3.2]<{\%}S[table-format=3.3]}
        \toprule
        & Method &  \multicolumn{1}{c}{$\sigma_{p^T}$ (\%)} & \multicolumn{1}{c}{$\sigma_{m}$ (\%)} & \multicolumn{1}{c}{$\sigma_{\Delta R}$ (centirad)}\\
        \midrule
        \multirow{3}{*}{\rotatebox[origin=c]{90}{\parbox{1.3cm}{\centering Without\\ \Delphes}}} 
        & JH               & 7.98    & 4.75     & 22.180   \\
        & PELICAN$\mid$JH  & 0.26    & 0.58     & 0.111   \\
        & PELICAN$\mid$FC  & 0.31    & 0.76     & 0.142   \\
        & PELICAN          & 0.88    & 1.67    & 0.548   \\
        \midrule
        \multirow{3}{*}{\rotatebox[origin=c]{90}{\parbox{1.3cm}{\centering With\\ \Delphes}}} 
        & JH               & 16.0    & 12.0   & 25.4      \\
        & PELICAN$\mid$JH  & 4.1     & 6.3    & 3.2      \\
        & PELICAN$\mid$FC  & 4.7     & 7.3    & 3.5      \\
        & PELICAN          & 6.2     & 8.2    & 5.6      \\
        \bottomrule
        \end{tabular}
    \end{small}
    \caption{PELICAN resolutions for models trained to reconstruct $p^W_{\mathrm{true}}$ with variable $W$ mass.\label{btW6m_W_table}}
    \vspace{-0.5\intextsep}
\end{table}

\subsection{Additional event displays}

\begin{figure}[t]
    \vspace{-0\intextsep}
    \begin{center}
    \includegraphics[width=\linewidth]{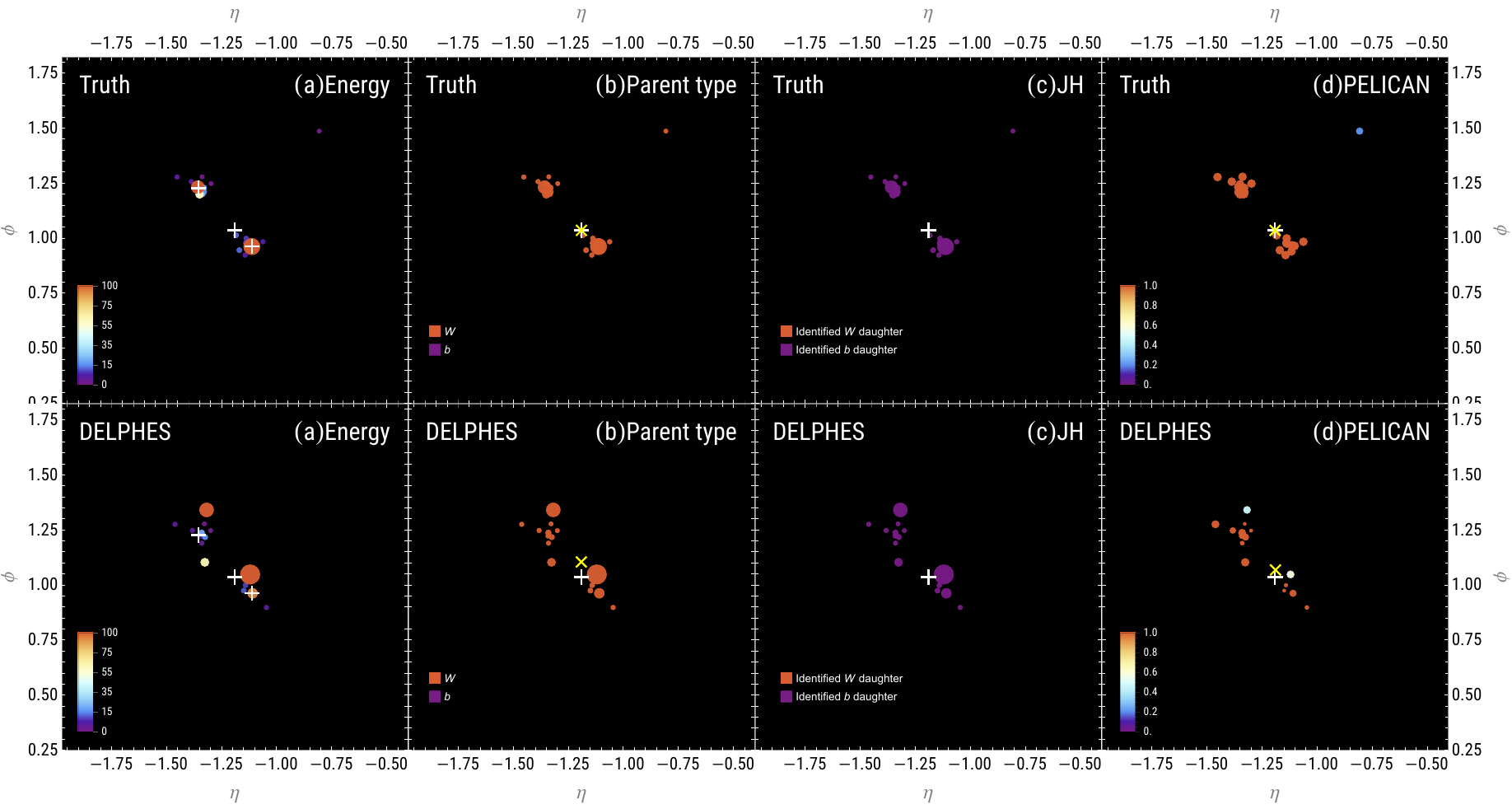}
    \caption{Event display combining the kind of information shown in \figrange{event_display}{display_energy}. This is a $qq$ event, i.e.~one where the $b$ quark from the $t\to bW$ decay fell outside of the jet radius. The JH tagger failed to tag this event as a top quark decay, whereas PELICAN correctly assigned weights close to $1$ to most of the constituents and obtained an accurate reconstruction of the $W$-boson. The legend in pane (c) loses its direct meaning here, but it can be understood as JH effectively setting all clustering weights to zero.}
    \label{event_display_2}
    \end{center}
\end{figure}

\begin{figure}[t]
    \vspace{-0\intextsep}
    \begin{center}
    \includegraphics[width=\linewidth]{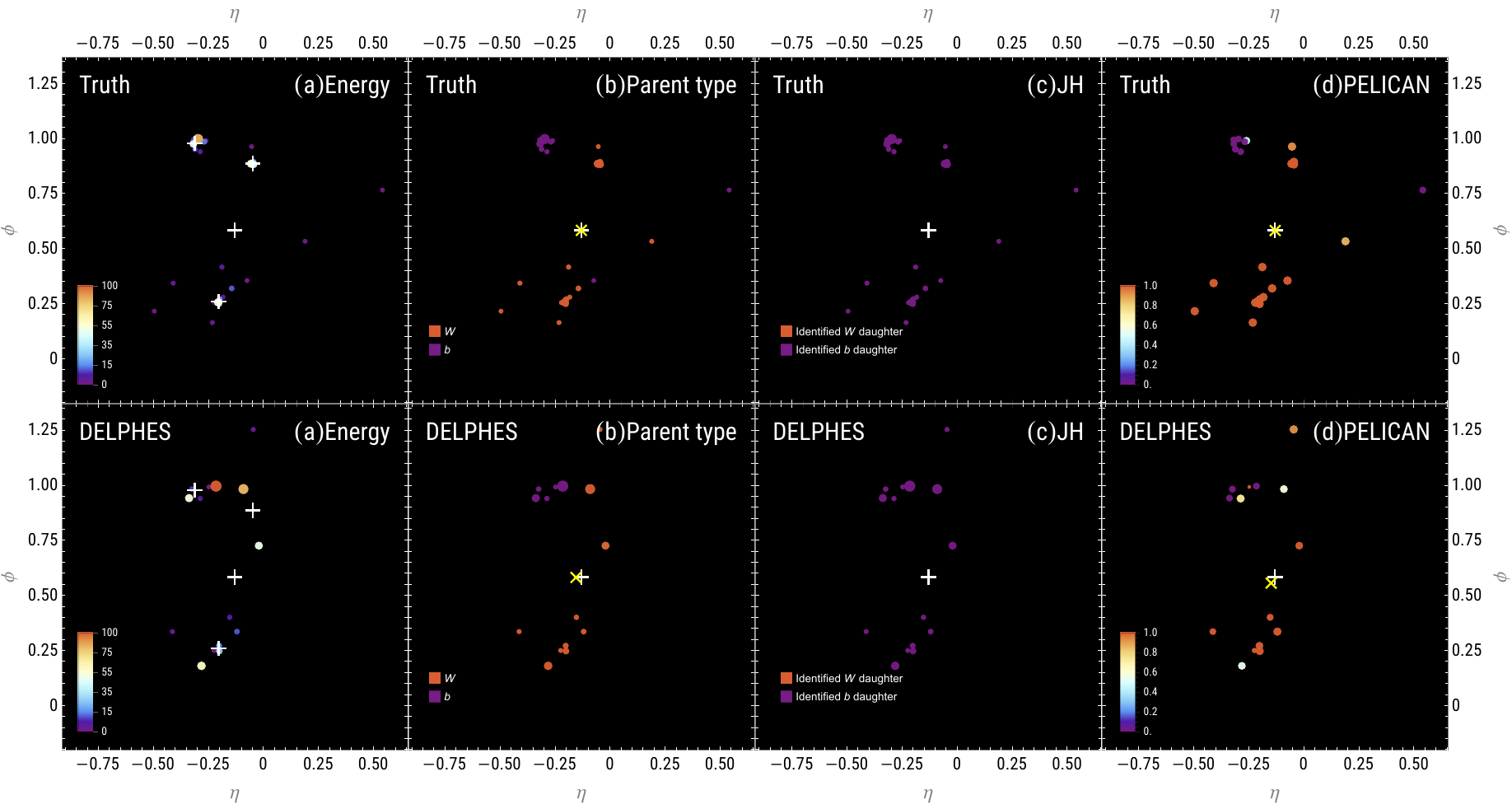}
    \caption{Yet another event display, this time a $bqq$ event that was still mistagged by the JH tagger. PELICAN correctly recognized the $b$ quark subjet and obtained a fairly accurate reconstruction of the $W$-boson.}
    \label{event_display_4}
    \end{center}
\end{figure}

Below is the list of additional event displays:
\begin{itemize}
    \item \Figref{event_display_2} an event display showing how PELICAN processes a $qq$ event that JH fails to tag;
    \item \Figref{event_display_4}: an event display showing how PELICAN processes a $bqq$ event that JH fails to tag.
\end{itemize}

\newpage
\section{IRC-safety and Lorentz symmetry}\label{appendix_irc}

Let us now try to characterize the IRC-safe Lorentz invariants of a set of jet constituents. When working with Lorentz-invariant observables, it is convenient to rewrite everything in terms of Lorentz invariant coordinates -- the dot products $d_{ij}=p_i\cdot p_j$ (note that these coordinates are in general not independent since any $5\times 5$ minor of the matrix $\{d_{ij}\}$ must vanish due to the fact that any five $4$-vectors are linearly dependent, but they will suffice for our purposes). For two massless particles, collinearity is equivalent to the vanishing of the dot product. Conversely, the dot product between two time- or light-like vectors is zero only if both are light-like and collinear. Thus the C-safety condition amounts to the equality of the gradients of $f$ with respect to any two of the rows (or columns) of $\{d_{ij}\}$ when evaluated on the corresponding coordinate hyperplane:

\begin{equation}
    \restr{\frac{\partial f}{\partial d_{ki}}}{d_{ij=0}}=\restr{\frac{\partial f}{\partial d_{kj}}}{d_{ij=0}}.\label{2525}
\end{equation}

In general this condition is difficult to solve, however it turns out to be extremely powerful in one special case. If we assume that $f$ is \textit{analytic} in the dot products $d_{ij}$ (i.e.~it can be expanded into a convergent multivariate Taylor series in some vicinity of the origin), then by the Identity theorem from complex analysis \equref{2525} must hold in an entire neighborhood of the origin, even when $d_{ij}\neq 0$. However, if this formula holds for all values of $k$ in an entire domain, then $f$ is necessarily only a function of $p_i+p_j$. Moreover, if all inputs are assumed to be massless, then this applies to all pairs of indices $i,j$, and $f$ becomes a function of only the jet mass $m_J^2$. Thus we arrive at the following rather disappointing result.

\begin{thm}\label{thmB1}
If an IRC-safe Lorentz-invariant observable with a mixture of massless and massive inputs is \textit{real analytic} in the pairwise dot products $d_{ij}$ near the origin, then it can depend on the massless inputs only through their sum. If all inputs are massless, this observable can be reduced to an analytic function of a single scalar -- the jet mass $m_J^2=\sum_{i,j}d_{ij}$.
\end{thm}

In this way, IRC-safety is a significantly more powerful restriction than one could na\"ively expect based on the definitions of IR and C-safety alone. As an example, consider the quantity $\prod_{i<j}d_{ij}=d_{12}d_{23}d_{31}$ on $N=3$ massless inputs. It is in fact C-safe because it simply vanishes whenever any two inputs are massless and collinear. However it is not IR-safe because sending $p_3\to 0$ reduces the expression to zero instead of the expected $d_{12}$. Alternatively, as we will see in the following section, the natural IR-safe extension of $d_{12}d_{23}d_{31}$ to arbitrary $N$ will not be C-safe.

Clearly, by restricting ourselves to analytic functions, i.e.~functions that can be approximated by polynomials in the variables $d_{ij}$, we have arrived at an extremely narrow set of IRC-safe observables which will be of little practical use. This is where our construction of IRC-safe PELICAN comes in handy, seeing as it clearly produces many more IRC-safe observables. In terms of the jet-frame coordinates defined in \equref{energy-def} and \equref{angle-def}, the C-safety condition reads
\[
 \restr{\frac{\partial f}{\partial\mathcal{E}_i}}{\hat{d}_{ij=0}}=\restr{\frac{\partial f}{\partial\mathcal{E}_j}}{\hat{d}_{ij=0}}.
\]
The crucial observation here is that $\mathcal{E}_i$ and $\hat{d}_{ij}$ are \textit{not} analytic functions of the original coordinates $d_{ij}$, as is evident from \equref{energy-def}. Therefore the set of analytic symmetric functions in these coordinates is very different, and in fact much larger, as we already know from our construction of IRC-safe PELICAN.

% We intend to show that any smooth Lorentz-invariant observable $f$ that is both IR-safe and C-safe in fact depends only on the jet mass $m_{\mathrm{jet}}^2=\sum_{k,l} d_{kl}=\left(\sum_k p_k\right)^2$. To that end, assume for a moment that this is not true. Then we can invoke IR-safety to summon a new input 4-vector $p_{N+1}=0$ without changing the initial values of $f$, and then start varying it. Equation (\ref{2525}) must remain true for all values of the dot products $d_{i,N+1}$ and $d_{j,N+1}$. It is certainly impossible for both sides of the equation to be independent of $d_{i,N+1}$ unless $f$ is independent of $p_{N+1}$, which would imply that $f$ is a constant by permutation symmetry. If, as we assumed, $f$ is functionally independent of $m_{\mathrm{jet}}^2$ (i.e.~$\dd f\wedge \dd m_{\mathrm{jet}}^2\neq 0$), then due to permutation symmetry the partial derivatives with respect to \textit{any} two different components of the matrix $\{d_{kl}\}$ do not always match, and the two sides of (\ref{2525}) will necessarily start diverging in value as we vary $p_{N+1}$, making $f$ not C-safe. Therefore in the C-safe case the restriction to the coordinate hyperplane in (\ref{2525}) can be lifted so that the derivatives of $f$ with respect to all dot products coincide, and hence $f$ is only a function of $m_{\mathrm{jet}}^2$.

% The theorem above becomes possible only when both of these constraints are enforced.

\subsection{Lorentz meets EFPs}

\subsubsection{Review of EFPs}

Particle data is often analyzed via IRC-safe observables such as $N$-subjettiness \cite{Thaler:2010tr}. A general polynomial basis for all IRC-safe observables was obtained in ref.~\cite{EFP}. Here we first retrace some of the  steps from the original derivation, and introduce a Lorentz-invariant analog at the end. Massless vectors can be written as $p=(E,E\hat{p})$ with a unit 3-vector $\hat{p}$. Then any (not necessarily Lorentz-invariant) smooth,\footnote{It is important to note that smoothness is generally an excessively powerful restriction, and many useful IRC-safe observables cannot be expanded in Taylor series this way. For instance, staying within the realm of Lorentz-invariant observables, the quantity $\sum_{i,j} \mathcal{E}_i \mathcal{E}_j \left(\hat{d}_{ij}\right)^\beta$ is IRC-safe for any $\beta$, but when written out in terms of $d_{ij}$ it is clearly not differentiable at the origin. We thank Andrew Larkoski for this example, prompting much of our discussion of IRC-safety as presented in this work.} symmetric observable of a set of $N$ massless vectors can be expanded at low energies as a combination of terms (omitting the constant term) of the form
\[\sum_{i_1,\ldots ,i_M=1}^N \sum_{j_1,\ldots, j_L=1}^N E_{i_1}\dots E_{i_M}\cdot f^{(N)}_{i_1,i_2,\ldots,i_M}\left(\hat{p}_1,\ldots,\hat{p}_N\right)\]
with some ``angular functions'' $f^{(N)}_{i_1,i_2,\ldots,i_M}$.  As shown in ref.~\cite{EFP}, IR-safety for such a series amounts to requiring that the angular function $f^{(N)}_{i_1,i_2,\ldots,i_M}$ in fact depends only on the particles whose indices are listed in the label of the function and is also independent of the total number $N$, i.e. it can be replaced by $f_{i_1,i_2,\ldots,i_M}\left(\hat{p}_{i_1},\ldots,\hat{p}_{i_M}\right)$. Furthermore, permutation symmetry of the entire observable descends to the total permutation symmetry of each angular function. Finally, C-safety requires that equating any two unit vectors $\hat{p}_i\to \hat{p}_j$ should lead to a function of only $E_1+E_2$. At the level of angular functions, this implies that whenever $i$ appears as an index in the subscript of the angular function, it can be replaced with $j$, and this can be done one index at a time. Ultimately this reduces the entire set of angular functions to just one $f_M=f_{1,2,\ldots,M}$, and therefore the order $M$ term in the low-energy expansion of any IRC-safe observable becomes
\[\sum_{i_1,\ldots ,i_M=1}^N E_{i_1}\dots E_{i_M}\cdot f_M\left(\hat{p}_{i_1},\ldots,\hat{p}_{i_M}\right).\]
From here, EFP's are derived by expanding the angular functions into a Taylor series around small angles $\cos \theta_{ij}=\hat{p}_i\cdot\hat{p}_j$. The resulting polynomial expansion can be broken up into a linear basis enumerated by the set of isomorphism classes of multigraphs $G$ with no loops (edges connecting a vertex to itself):
\[\text{EFP}_G=\sum_{i_1,\ldots, i_M=1}^N \prod_{j\in V(G)}E_{i_j} \cdot \prod_{(k,l)\in E(G)}\theta_{i_k i_l},\]
where $V(G)=\{1,2,\ldots,M\}$ is the set of vertices of $G$, and $E(G)$ is the set of edges. The EFP is a homogenous polynomial in the energies of degree $|V(G)|=M$ and in the angles of degree $D=|E(G)|$. Notice that the indices $i_j$ can coincide but the corresponding terms vanish since $\theta_{ii}=0$, which is why multigraphs with loops are excluded from the basis. As a trivial but important example, the total energy $E=\sum_{i=1}^N E_i$ corresponds to $G=\bullet$. If $G$ consists of multiple connected components, the resulting EFP is the product of the EFP's corresponding to the components, so the entire basis is algebraically generated by just the connected multigraphs.

It is instructive to see how IRC-safety works in EFP's. IR-safety is guaranteed by the mere presence of the energy pre-factors: sending $E_N\to 0$ will simply recover the same EFP for the remaining $N-1$ particles.  Meanwhile, C-safety is observed because if two particles, say $1$ and $2$, become collinear, then the angular factor is completely invariant under their permutations. In other words, all terms where the list $(i_1,i_2,\ldots,i_M)$ contains a fixed number, say $L$, of $1$'s \textit{or} $2$'s, can be grouped together, and after summation over those indices the energy pre-factor manifestly depends on $E_1$ and $E_2$ only through an overall factor of $(E_1+E_2)^L$. This is exactly the statement of C-safety.

\subsubsection{MFPs: Lorentz-invariant analytic EFPs}

Now let us task ourselves with identifying the subset of Lorentz-invariant IRC-safe observables. Obviously none of the EFP's are Lorentz-invariant due to the direct dependence on spatial angles, but smooth Lorentz-invariant observables can still be expanded in the EFP basis at small energies and angles. The only Lorentz-invariant of two massless 4-vectors $p_i,p_j$ is $p_i\cdot p_j=E_i E_j(1-\cos\theta_{ij})$. At small angles the approximately boost-invariant combination is then $E_iE_j\theta_{ij}^2$. The expansion of any permutation-symmetric function of such dot products into a series as above will differ from general EFP's as follows: every angle $\theta_{ij}$ must appear in the product an even number of times, and the pre-factor must consist of nothing but one factor of $E_i E_j$ for every factor of $\theta_{ij}^2$ (or $1-\cos\theta_{ij}$). The Lorentz-invariant version of the polynomial can then be obtained by replacing each $\theta_{ij}^2$ with $2\hat{d}_{ij}$. In each EFP only a subset of terms will generally satisfy these conditions, so a better representation of the Lorentz-invariant basis is warranted.

Let us start again with the most general Lorentz-invariant and permutation-invariant observable of $N$ 4-momenta that is \textit{analytic} when expressed in terms of the dot products $d_{ij}=p_i\cdot p_j$. Such a function $f$ can be expanded at small arguments into a linear combination $f\sim \sum_G w_G f_G^{(N)}$ of terms of the form
\[f^{(N)}_G=\sideset{}{'}\sum_{i_1,\ldots, i_M=1}^N \prod_{(k,l)\in E(G)} d_{i_k i_l},\label{1535}\]
where the prime indicates that the terms where any two of the indices coincide are excluded. Including them would also produce a valid basis, but we leave them out to reduce the complexity of each polynomial. Such terms simply correspond to other multigraphs obtained from $G$ by vertex identification, and those already appear in the expansion with independent coefficients. This is still an over-complete basis since the matrix of dot products $d_{ij}$ is highly degenerate for $N\geq 5$. Unlike in EFP's, we also allow multigraphs with loops since we are not restricting ourselves to purely massless inputs. However, if all inputs are massless, all graphs with loops will produce vanishing polynomials.

Now we can easily see how to enforce IRC-safety in these polynomials. IR-safety requires that sending any of the 4-momenta to zero, say $p_N\to 0$, must be equivalent to simply restricting the same expression to only the other $N-1$ particles. We observe that the only case when this is not already satisfied in the expression above is when the multigraph $G$ contains isolated vertices (vertices of degree zero). Isolated vertices lead to summations over dummy indices corresponding to those vertices, which results in an overall factor of some power of $N$. The multiplicity $N$ is not an IR-safe variable, and it is sufficient to ban graphs with isolated vertices to enforce IR-safety. We can thus drop the notation $f^{(N)}_G$ and simply write $f_G$ with the implicit understanding that \equref{1535} defines the full infinite family of observables for all $N$.

\begin{figure}[t]
    \begin{center}
        \adjustbox{trim=0cm 0.6cm 0cm 0.4cm}{
        \includegraphics[width=0.3\textwidth]{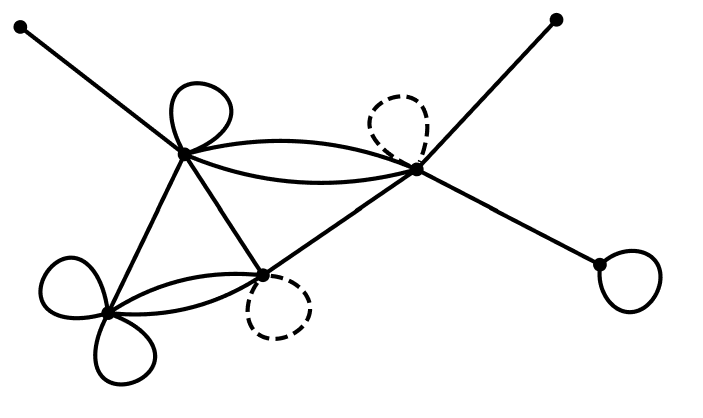}
        }   
    \end{center}
    \caption{The addition of the dashed loops makes this multigraph \textit{loop-saturated}: all non-leaf vertices come with at least one loop. Loop-saturated multigraphs enumerate all IRC-safe polynomials of $d_{ij}$.} 
\end{figure}
Now we address C-safety. Much like in EFP's, C-safety in $f_G$'s means that whenever, say, $p_1$ and $p_2$ are massless, each monomial that contains a certain number of indices equal to $1$ or $2$ must match its coefficient with any other monomial that differs only by flipping any number of the indices from $2$ to $1$ or vice versa. In the language of multigraphs, this means that each monomial that assigns a vertex of degree $n_1$ to $p_1$ and another vertex of degree $n_2$ to $p_2$ must in fact result as just one among all possible monomials obtained via vertex identification from a graph $G'$ where those two vertices have been ``blown up'' into $n_1+n_2$ vertices of degree one. We also notice that this condition applies only to vertices that do not have any loops attached to them because otherwise the corresponding terms would vanish as $p_1$ and $p_2$ become massless, thereby making any additional restrictions unnecessary. To summarize, any IRC-safe polynomial contains with equal coefficients all $f_G$'s with $G$'s obtained by any combination of vertex identifications from some multigraph $G'$ such that all of its loopless vertices are \textit{leaves} (i.e.~of degree one). We shall call such multigraphs \textit{loop-saturated}. In this process it is sufficient to allow identifications of only the leaf vertices because all other vertex identifications will result in a polynomial independently defined by another loop-saturated multigraph.

We thus define an IRC-safe Lorentz-invariant polynomial basis indexed by non-isomorphic loop-saturated multigraphs $G$ with no isolated vertices. Assuming the vertices of $G$ are identified with the set $\{1,2,\ldots,M\}$, we write
\[\text{MFP}_G=\sum_{i_1,\ldots, i_M=1}^N \prod_{(k,l)\in E(G)} d_{i_k i_l} = \sum_{i_1,\ldots, i_M=1}^N \prod_{j\in V(G)}\mathcal{E}_{i_j}^{\delta_j} \prod_{(k,l)\in E(G)} \hat{d}_{i_k i_l},\label{MFP-jet-frame}\]
where the sum is taken over all $M$-tuples and $\delta_1,\ldots,\delta_M$ are the degrees of the vertices of $G$ (each loop increases the degree by $2$). An alternative basis is
\[\text{MFP}_G'=\sideset{}{'}\sum_{i_1,\ldots, i_M=1}^N \prod_{(k,l)\in E(G)} d_{i_k i_l},\]
where the coincidence of two or more indices $i_k=i_l$ is allowed only if their corresponding vertices ($k$ and $l$) are leaves of $G$. $\text{MFP}_G$ is simply $\text{MFP}_G'$ plus a linear combination of certain $\text{MFP}_H'$'s with fewer vertices, namely such that $H$ is a graph obtained from $G$ by one or more vertex identifications each of which involves at least one non-leaf vertex of $G$. Indeed, such identifications preserve the property of being loop-saturated, and all other identifications are already included in the definition of $\text{MFP}_G'$.

Note that just like for EFP's, disconnected multigraphs correspond to products of their connected components (this holds for $\text{MFP}_G$ but not for $\text{MFP}_G'$):
\[\text{MFP}_{G\sqcup H}=\text{MFP}_G \cdot \text{MFP}_H.\]
As an example, if all inputs are known to be massless, then any connected multigraph with loops will produce a vanishing polynomial. And the only loopless loop-saturated connected multigraph is the graph consisting of just one edge, which evaluates to the jet mass: $\text{MFP}_{\bullet - \bullet}=m_J^2=\sum_{i,j} d_{ij}=\left(\sum_{i} \mathcal{E}_i\right)^2$. Therefore in the massless case MFP's generate all analytic functions of $m_J^2$, which is a disappointingly small subset of all IRC-safe Lorentz-invariant observables. As a final example, we can consider the polynomial $d_{12} d_{23}d_{31}$, which nominally seems C-safe. It can only be interpreted as an IR-safe polynomial if it corresponds to the triangle graph. However, such a graph is not loop-saturated, and therefore this polynomial is not IRC-safe despite being C-safe for $N=3$ (which can be confirmed by observing that for $N=4$ the triangle graph generates a non-C-safe polynomial), and the ``correct'' IRC-safe completion is simply $m_J^6$.

What we found is congruent with \equref{thmB1} in that MFP's represent a polynomial expansion basis for all IRC-safe observables that are analytic in the dot products $d_{ij}$, and depend on massless inputs only through their sum. Since we already know how restrictive this result is, we can now move on to a more useful generalization of EFPs.

\subsubsection{JFPs: Lorentz-invariant jet-frame EFPs}

As already discussed above, by transforming our inputs from $d_{ij}$ to the jet-frame coordinates $\mathcal{E}_i$ and $\hat{d}_{ij}$ we get a much larger space of IRC-safe analytic observables. Ideologically, the transformation to the jet-frame coordinates is extremely simple: $\mathcal{E}_i$ is nothing but the regular energy $E_i$ \textit{as measured in the rest frame of the jet}, and the same is true for the angular parts which take the form $\hat{d}_{ij}=1-\cos\Theta_{ij}$ for a pair of massless particles. This coordinate transformation itself is IRC-safe in the sense that IR and collinear splittings that don't involve constituent $p_i$ will not affect its jet-frame energy and angles with the other constituents. Therefore the notion of IRC-safe \textit{continuous} observables is the same in both coordinate systems. On the other hand, as we have seen above, the restriction to analyticity makes the two sets of IRC-safe observables very different.

All of this implies that the set of IRC-safe observables which are invariant under spatial rotations and restricted to the manifold of events whose jet momentum $J$ is purely time-like is in one-to-one correspondence with Lorentz-invariant IRC-safe observables. Indeed, given a rotationally invariant observable $f(p_1,\ldots,p_N)$, consider
\[
f_L(p_1,\ldots,p_N)=f(\Lambda_J p_1,\ldots,\Lambda_J p_N),
\]
where $\Lambda_J$ is a Lorentz transformation that maps the jet momentum $J=\sum_i p_i$ to $(m_J,0,0,0)$. Note that $\Lambda_J$ is not unique, namely $R\Lambda_J$ is also a solution for any spatial rotation $R$. However, since we have assumed $f$ to be rotationally invariant, the resulting observable $f_L$ is independent of the specific choice of $\Lambda_J$ and is \textit{fully Lorentz-invariant}. And since $\Lambda_J$ is invertible, the transformation $f\mapsto f_L$ is also invertible, which proves the one-to-one correspondence.\footnote{This principle, of course, applies not only to observables but to entire algorithms, such as jet clustering. E.g.~we defined our Lorentz-invariant analog of Soft Drop multiplicity using exactly this method.}

We conclude that a polynomial expansion basis for analytic IRC-safe Lorentz-invariant observables in the jet-frame coordinates can be trivially obtained by taking the original EFPs, replacing $\theta_{ij}^2$ with their rotationally invariant analogs $2(1-\cos\theta_{ij})$, and then replacing $E_i\mapsto \mathcal{E}_i$ and $(1-\cos\theta_{ij})\mapsto \hat{d}_{ij}$. We do have to abandon EFPs that include odd powers of any angle due to their non-analyticity, but that doesn't affect the completeness of the basis since all the angular coordinates are non-negative (however it means that every edge in our multigraphs corresponds to two edges in the analogous EFP). The only subtlety here is that the original EFPs were built only for all-massless inputs, so they don't depend on masses whereas $\hat{d}_{ij}$ do. But all the original arguments in the derivation of the EFPs still apply even when some of the inputs are massive (as long as those masses are also perturbatively small), so we still get a complete basis of IRC-safe observables that are analytic in the jet-frame coordinates. We call these polynomials the \textit{Jet Flow Polynomials} (JFPs):
\[
\mathrm{JFP}_G = \sum_{i_1,\ldots, i_M=1}^N \prod_{j\in V(G)}\mathcal{E}_{i_j} \cdot \prod_{(k,l)\in E(G)}\hat{d}_{i_k i_l},
\]
where the multigraphs $G$ have no restrictions on them and can contain loops. Comparing to \equref{MFP-jet-frame}, we see that the powers of the energies are no longer tethered to the structure of the multigraph, which is why C-safety doesn't place any constraints on the multigraph unlike in the case of MFPs. Namely, the symmetry that C-safety forces on the angular parts $\prod_{(k,l)} \hat{d}_{i_k i_l}$ (that one must be able to switch the value of an index $i_k$ from one collinear particle to another and obtain another monomial in the same polynomial) is no longer coupled to the power counting of the energy factors, which makes any multigraph $G$ permissible as long as we sum over all possible values of the indices $i_k$.

\subsection{Universality of PELICAN}

While the Deep Sets theorem \cite{ZaKoRaPoSS17} shows that permutation-invariant networks are universal for set learning problems, it is known that message passing networks are strictly non-universal for \textit{graph} learning problems, i.e.~in the presence of rank 2 data such as an adjacency matrix. Permutation-equivariant networks that take graph data as input (PELICAN's inputs can be interpreted as such) and use layers like $\mathrm{Eq}_{2\to 2}$ are known to have the expressivity of the 2-WL (Weisfeiler-Lehman) graph isomorphism test, matching the expressivity of message passing networks, see ref.~\cite{Azizian_expressivity}. By adding matrix multiplication to the list of equivariant aggregators (thus turning the linear equivariant layer into a quadratic or polynomial one), one can raise the expressivity to the ``folklore'' version of the WL test, i.e.~2-FWL for PELICAN, which is equivalent to 3-WL \cite{Maron19ProvablyPowerful}. However, in our tests we have not been able to see any performance improvements from the addition of an aggregator of this kind.

We can see the limited expressivity of PELICAN in terms of the graphs $G$ in \equref{1535} that can be generated by PELICAN when expanded at small values of the inputs (assuming a smooth activation function). Indeed, the messaging layer combined with single-index aggregation $\sum_i \bullet_{ij}$ can attach new edges ($d_{ij}$), loops ($d_{jj}$), and multi-edges ($d_{ij}^n$) to a given vertex $j$. By applying more of these layers we can only keep joining such ``starfish'' graphs at their central vertex $j$, whereas even something as simple as the triangle graph can never be generated due to the fact that it requires three free indices at an intermediate step. The only graphs with two free indices generated by PELICAN are the multi-edges $d_{ij}^n$, but joining them together will never produce a more complex graph.  At best, PELICAN can only generate multigraphs that can be broken up into subgraphs with at most two severed edges on each of them (since each severed edge requires a free index on the corresponding vertices for us to be able to ``glue'' the vertex back together via aggregation).

Finally, it is curious to compare PELICAN-like architectures to architectures that work with irreducible representations of the continuous symmetry group, like LGN \cite{Bogatskiy:2020tje}. As was shown in ref.~\cite{Maron21SO3Universality}, the latter kind of network is universal, provided that the set of the finite-dimensional irreducible representations that are stored is not limited. In particular, the proof involves showing that a universal network based on the tensor product nonlinearity will be able to generate all symmetric polynomial tensors of the form $P_{\alpha}=\sum_{i_1,\ldots,i_r} p_{i_1}^{\alpha_1}\otimes \cdots \otimes p_{i_r}^{\alpha_r}$ where $\alpha\in\mathbb{Z}^r$. It is easy to see that the Lorentz-invariant part of such a tensor produces all possible multigraphs $G$ with vertices of degrees $\alpha_1,\ldots,\alpha_r$. Therefore, indeed, the ability of a network to generate all such tensors implies its universality on the space of Lorentz-invariant observables.

However, in practice the dimensionality of these tensor representations has to be quite limited. For example, LGN kept only representations up to spin 2, that is, tensors of the form $p_i\otimes p_j$, but not higher than that (of course, these can also be multiplied by invariant scalars without increasing the dimension). This effectively limits the number of free particle indices in all latent values of the network to $2$ (right before aggregation reduces it back to 1 again), just like in PELICAN. Therefore there is ample reason to believe that the expressivity of LGN-like networks that go up to spin $k$ is not universal, but is equivalent to the expressivity of PELICAN-like networks that go up to the permutation-equivariant rank of $k$ (using $\mathrm{Eq}_{k\to k}$ blocks). Moreover, this expressivity can be stated in terms of the class of multigraphs $G$ that any analytic network of this kind can generate. We leave this problem for future work.

\subsection{Quantifying IRC-safety}

This section is an attempt to define some Lorentz-invariant quantities that can measure \textit{whether} and \textit{how} IR/C-safe any given observable is. It is mostly of theoretical interest since it does not yet have a software implementation.

Let $f^{(N)}$ be a Lorentz-invariant permutation-symmetric observable defined for varying number $N$ of inputs, for instance the output of a PELICAN instance. We define the following difference operator that measures the obstruction to $f$ being IR-safe:
%-------------------
\begin{equation}
    \Delta^{\mathrm{IR}} f^{(N)} = f^{(N+1)}\left(p_1,\ldots,p_N,0\right)- f^{(N)}\left(p_1,\ldots,p_N\right).
\end{equation}
%-------------------
IR-safety is equivalent to the vanishing of this operator:
\[\mathrm{IR-safety:}\quad \Delta^{\mathrm{IR}} f=0.\]
Furthermore, we can define \textit{IR-robustness} as the partial derivative corresponding to the injection of an infinitesimally soft but non-zero 4-momentum:
%-------------------
\begin{equation}
    \mathrm{IR-robustness:}\quad R^{\mathrm{IR}}_p f^{(N)}=\restr{\frac{\dd}{\dd\epsilon}}{\epsilon=0} f^{(N+1)}(p_1,\ldots,p_N,\epsilon p).
\end{equation}
%-------------------
All together, we have the following series expansion for a general observable near $p_{N+1}=0$:
\[f^{(N+1)}=f^{(N)}+\Delta^{\mathrm{IR}}f^{(N)}+\left(R^{\mathrm{IR}}_{p_{N+1}} f^{(N)}\right)\cdot p_{N+1}+\mathcal{O}(p_{N+1}^2).\]

An IR-safe observable must always have $\Delta^{\mathrm{IR}}f=0$, and lower values of IR-robustness indicate lower sensitivity to soft particles. In practice, we can evaluate the ratio $(\Delta^{\mathrm{IR}}f^{(N)})/f^{(N)}$ and average it over a testing dataset to quantify the IR-safety of a model.

For C-safety, we pick a massless 4-vector $p$ and define the differential operator
%-------------------
\begin{equation}
    D^{\mathrm{C}}_{p} f=\restr{\frac{\dd}{\dd\lambda}}{\lambda=0}f(p_1+\lambda p,p_2-\lambda p,p_3,\ldots).
\end{equation}
%-------------------
C-safety then is equivalent to the requirement that $D^{\mathrm{C}}_{p} f=0$ for any massless $p$ whenever $p_1$ and $p_2$ are collinear with $p$:
%-------------------
\begin{equation}
    \mathrm{C-safety: }\quad \restr{D^{\mathrm{C}}_{p} f}{p_1\parallel p_2\parallel p}=0, \quad p^2=0.
\end{equation}
%-------------------

\textit{C-robustness} can be easily defined as the second-order analog of $D^{\mathrm{C}}_{p}$. Namely, we pick a second vector $\hat{p}$ such that $p\cdot \hat{p}=0$ (which means that $\hat{p}$ is tangent to the light cone at $p$, as required by the massless constraint) and measure how quickly $D^{\mathrm{C}}_{p} f$ deviates from zero as we deform $p_1$ and $p_2$ away from collinearity:
%-------------------
\begin{equation}
    \mathrm{C-robustness: }\quad R^{\mathrm{C}}_{\hat{p},p}f=D^{\mathrm{C}}_{\hat{p}} D^{\mathrm{C}}_{p} f, \quad \hat{p}\cdot p=p^2=0, \quad p_1\parallel p_2\parallel p.
\end{equation}
%-------------------

It is instructive to express these quantities in terms of the dot products $d_{ij}$. We have
%-------------------
\begin{equation}
    D^{\mathrm{C}}_{p}f=\sum_{j=1}^N \left(\partial_{1j}f-\partial_{2j}f\right)(p_j\cdot p),
\end{equation}
%-------------------
where $\partial_{kl}$ indicates the partial derivative with respect to $d_{kl}$. Note that individual partial derivatives aren't really well-defined due to the fact that  $\{d_{ij}\}$ is not a set of independent coordinates on the manifold of $N$ 4-vectors for $N\geq 5$. However, combinations such as above are valid since they are nothing but a different way of expressing the original well-defined operator. We also immediately notice that if $N=2$ and $p_1\parallel p_2\parallel p$, then $D^{\mathrm{C}}_{p}f=0$ automatically. Indeed, every Lorentz-invariant observable for two particles is C-safe due to the fact that $d_{12}=\frac12 (p_1+p_2)^2$ for massless inputs.

C-robustness is similarly expressed by
%-------------------
\begin{equation}
    R^{\mathrm{C}}_{\hat{p},p} f=\sum_{i,j=1}^N\left(\partial^2_{1i,1j}+\partial^2_{2i,2j}-\partial^2_{1i,2j}-\partial^2_{1j,2i}\right)f\cdot (p_i \cdot \hat{p})(p_j\cdot p).
\end{equation}

\end{document}